\documentstyle[11pt,epsf]{article}
\def\be{\begin{equation}}
\def\ee{\end{equation}}
\def\espacio{\vskip13truept}

\def\headf #1 #2 {\espacio \espacio \noindent {\bf #1 #2 }\espacio}
\def\heads #1 #2 {\espacio \normp \noindent #1 #2 \espacio \norm}
\def\headt #1 #2 {\espacio \sl \noindent #1 #2}

\newcommand \e {{\rm e}}

\newcommand \D{\, d}

\newcommand \cH{{ \cal H}}

\begin{document}


{\baselineskip=11truept
\noindent{\bf STOCHASTIC EFFECTS IN PHYSICAL SYSTEMS\footnote{To be published
in {\sl Instabilities and Nonequilibrium Structures}, 
VI, E. Tirapegui and W.Zeller,
eds. Kluwer Academic Pub. (1997).}}
\espacio
\espacio
{\noindent \bf MAXI SAN MIGUEL and RA{\'UL} TORAL}\\
Departamento de F\'{\i}sica Interdisciplinar \\
Instituto Mediterr\'a‡neo de Estudios Avanzados IMEDEA (CSIC-UIB)\\
Campus Universitat de les Illes Balears\\
E-07071 Palma de Mallorca, Spain\\
http://www.imedea.uib.es/PhysDept/
}
\tableofcontents
\newpage
\section{Introduction}
The study of the effects of noise and fluctuations is a  well
established subject in several different disciplines ranging from pure
mathematics (stochastic processes) to physics (fluctuations) and electrical 
engineering (noise and radiophysics). In traditional statistical physics,
fluctuations are of thermal origin giving rise to small departures from a mean
value. They tend to zero as one approaches the thermodynamic limit in which
different statistical descriptions (different ensembles) become equivalent.
Likewise, in more applied contexts fluctuations or noise are usually regarded as
small corrections to a deterministic (noise free) behavior that degrades a
signal-to-noise ratio or can cause transmission errors. In such 
framework fluctuations are a correction that can be usually dealt with through
some sort of linearization of dynamics around a mean or noise free dynamics.
A different point of view about fluctuations emerges, for example, in the study
of critical phenomena in the 1970's. The statistical physics description of
these phenomena requires a formulation appropriate for a system dominated by
fluctuations and nonlinearities. A linear theory only identifies the existence
of a critical point by a divergence of fluctuations.

The renewed interest and activity of the last 15-20 years on stochastic
phenomena and their applications is precisely in the context of the study of
nonlinear dynamics and instabilities in systems away from equilibrium. This
activity has led to some new conceptual developments, applications, and new or
rediscovered methodology
\cite{kampen,gardiner,Risken,HL,Moss,wio94}.
Among these we would like to emphasize here two very general aspects. One is the
concept that noise need not only be a nuisance that spoils the "true" and desired
behavior of the system, but rather noise might make possible new states and forms of
behavior which do not appear in a noise-free limit. These situations might occur
when there are mechanisms of noise amplification and/or when noise interacts with
nonlinearities or other driving forces of a system. Phenomena like noise
sustained spatial structures, noise--induced transitions or stochastic resonance
go under this category. A second concept we wish to emphasize is that the 
physical relevant behavior is not
necessarily associated with some ensemble average, but rather with typical
stochastic trajectories. It is certainly a trivial mathematical statement that a
mean does not always give a typical characterization of a process, but there is
a certain tradition in physics (borrowed from equilibrium statistical physics)
of focusing on averaged values. A physical intuition or understanding of novel
stochastic driven phenomena is in fact gained by the consideration of the
individual realizations of the stochastic process. This has important
methodological consequences: one needs tools to follow trajectories instead
of just looking at averages and probability distributions.

In these lectures we follow an incomplete random walk on the phase space of some
of the current concepts, developments and applications of stochastic processes
from the point of view of the practitioner physicist and emphasizing examples of
the two key ideas outlined above. Section 2 is rather tutorial while the other
ones give more a summary and guide to different topics. Some parts of Sect. 2
are rather elementary and can be skipped by anyone with a little experience in
stochastic processes. But this section also contains a rather detailed
presentation of numerical methods for the simulation of the individual
realizations of a stochastic process. It is here important to warn against the
naive idea that including noise in the simulation of a nonlinear dynamical 
problem is
just to add some random numbers in any reasonable way. This is particularly
important when we want to learn on the physics of the problem following the
trajectories. Section 3 deals with one of the important cases of noise
amplification, namely the transient decay from unstable states. Key concepts
like trajectory dynamics, passage time statistics and mapping of linear into
nonlinear stochastic properties are discussed. As an important application of
these ideas and methods we discuss properties of laser switch-on viewed as a
process of noise amplification. Section 4 is devoted to the analysis
of the long--time, stationary, properties of physical systems in the
presence of noise. The classification of potential/non--potential
systems is succinctly reviewed and the conditions for a system to be potential
are precisely stated. We study the general form of 
the probability distribution function when noise is present. We end
with an example borrowed from fluid dynamics, the K\"uppers--Lortz 
instability, in which we illustrate the stabilization by noise of a periodic
trajectory in a system in which the deterministic dynamics has a contibution
which can not be described as relaxation dynamics in a potential.
In Section 5 we consider spatially extended systems, described either by ordinary
differential equations for the amplitudes of a few spatial modes or by
stochastic partial differential equations. Through some specific examples 
we discuss symmetry breaking by
pattern formation and symmetry restoring by noise, the issue of pattern
selection in the presence of noise and noise sustained structures in
convectively unstable situations. Section 6 reviews the concept of 
noise--induced transition, distinguishing it from that of 
noise--induced phase--transition. The difference being, mainly, that
noise--induced transitions do not break ergodicity, as we understand
a phase transition in the statistical--mechanics sense.
We present specific examples of
systems displaying one or the other and show that, in general, they
can no coexist in the same system. 

We should finally make clear that our choice of subjects included in these
lectures is rather arbitrary and sometimes dictated by personal contributions.
Among many other relevant topics of actual interest which we do not discuss here
we could mention, from the methodological
point of view,  path integral formulations of stochastic processes
\cite{wio94,pathintegral} 
and from the conceptual point of view stochastic resonance
\cite{stochres} and directed fluxes supported by noise \cite{ratchets}


\section{Stochastic Processes}
\setcounter{equation}{0}
\setcounter{figure}{0}
\subsection{Basic Concepts}
In this first subsection we want to give a quick review of what a stochastic process
is from the physical point of view. We will not be too
rigorous on the mathematical
side. The name ``stochastic process" is usually
associated with a trajectory which is random enough to demand a probabilistic
description. Of course, the paradigmatic example being that of Brownian 
motion \cite{Risken,wax,macl,brush,pathria}. 
The botanist Robert Brown discovered in 1827 that particles
of pollen in suspension execute random movements which he interpreted
initially as some sort of life. It is not so well known that L. Boltzmann
knew as early as 1896 the reason for this erratic movement when he wrote
``... very small particles in a gas execute motions which result from the
fact that the pressure on the surface of the particles may 
fluctuate" \cite{lebo}. 
However, it was A. Einstein in 1905 who successfully presented the 
correct description of the erratic movement of the Brownian 
particles \cite{gardiner}.
Instead of focusing on the trajectory of a single particle, Einstein
derived a probabilistic description valid for an ensemble of
Brownian particles. In his description, no attempt is made to follow
fully in time the (complicated) trajectory of a Brownian particle.
Instead, he introduces the concept of a coarse--grained description,
defined by a time scale $\tau$ such that different trajectories separated 
by a time $\tau$ can be considered independent. No attempt is made to 
characterize the dynamics at a time scale smaller than this coarse--grain
time $\tau$. A second concept, probabilistic in nature, introduced
by Einstein is the probability density function (pdf, for short),
$f(\vec \Delta)$, for the distance $\vec \Delta$ travelled by the Brownian 
particle in a time interval $\tau$. $f(\vec \Delta)$ is defined such that
$f(\vec \Delta)d \vec \Delta$ is the probability of having a change in position
$\vec x$ in the interval $(\vec \Delta, \vec \Delta+d\vec \Delta)$.
From the fact that $f(\vec \Delta)$ is a pdf it follows the following
properties:
\begin{equation}
\label{eq:1}
\begin{array}{rcl}
f(\vec \Delta) & \ge & 0 \\
\int f(\vec \Delta)d \vec \Delta & =  & 1 
\end{array}
\end{equation}
One could invoke the law of large numbers to predict a Gaussian form for
$f(\vec \Delta)$. However, this is not necessary and one only needs to assume
the symmetry condition:
\begin{equation}
\label{eq:2}
f(-\vec \Delta)= f(\vec \Delta)
\end{equation}
We consider an ensemble of $N$ Brownian particles. This is characterized
by the particle number 
density $n(\vec x,t)$, which is such that $n(\vec x,t)d\vec x$
is the number of particles in the volume $(\vec x,\vec x+d\vec x)$ 
at time $t$. From
the assumption that the trajectories separated a time interval $\tau$ are
independent, it follows that the number of particles at location
$\vec x$ at time $t+\tau$ will be given by the number of particles at 
location $\vec x-\vec \Delta$ at time $t$,
multiplied by the probability that the
particle jumps from $\vec x-\vec \Delta$ to $\vec x$, which is 
nothing but $f(\vec \Delta)$, and 
integrated for all the possible $\vec \Delta$ values:
\begin{equation}
\label{eq:3}
n(\vec x,t+\tau)=\int n(\vec x-\vec \Delta,t)f(\vec \Delta)d\vec \Delta 
\end{equation}
This is the basic evolution equation for the number density $n(x,t)$. By
Taylor expanding the above expression, using of the symmetry
relation eq.(\ref{eq:2}) and keeping only
the lowest non--vanishing order terms, one gets the diffusion equation:
\begin{equation}
\label{eq:4}
\frac{\partial n}{\partial t} = D \nabla^2 n
\end{equation}
where the {\sl diffusion constant} $D$ is given in terms of the second
moment of the pdf $f(\vec \Delta)$ by:
\begin{equation}
\label{eq:5}
D=\frac{1}{2\tau}\int \Delta^2 f(\vec \Delta)d\vec \Delta =
\frac{\langle \Delta^2 \rangle}{2 \tau}
\end{equation}
If the initial condition is that all particles are located at $\vec x=0$,
$n(\vec x,t=0)=N\delta(\vec x)$, the solution of the diffusion equation is:
\begin{equation}
\label{eq:6}
n(\vec x,t)=\frac{N}{(4 \pi Dt)^{3/2}}{\rm e}^{-x^2/4Dt}
\end{equation}
from where it follows that the average position of the Brownian
particle is $0$ and that the average square position 
increases linearly with time, namely:
\begin{equation}
\label{eq:7}
\begin{array}{rcl}
\langle \vec x(t) \rangle & = & 0 \\
\langle \vec x(t)^2 \rangle & = & 6 D t
\end{array}
\end{equation}
These predictions have been successfully confirmed in experiments and
contributed to the acceptance of the atomic theory. The above results
are characteristic of stochastic 
diffusion processes as the ones we will encounter
again in other sections (c.f. Sect. 5).

Even though Einstein's approach was very successful, one has to admit
that it is very phenomenological
and can not yield, for instance, an explicit expression that allows
the calculation of the diffusion constant in terms of
microscopic quantities. Langevin (1908) initiated
a different treatment which can be considered in
some way complementary of the previous
one. In his approach, Langevin focused on the trajectory of a single 
Brownian particle and wrote down Newton's equation Force=mass $\times$
acceleration. Of course, he knew that the trajectory of the Brownian
particle is highly erratic and that would demand a peculiar kind
of force. Langevin considered two types of forces acting on the Brownian
particle: usual friction forces that, according to Stokes law, would
be proportional to the velocity, and a sort of ``fluctuating" force
$\vec \xi(t)$ which represents the ``erratic" part
of the force coming from the
action of the fluid molecules on the Brownian particle. The equation
of motion becomes then:
\begin{equation}
\label{eq:8}
m\frac{d\vec v}{d t}= -6 \pi \eta a \vec v + \vec \xi
\end{equation}
$\eta$ is the viscosity coefficient and $a$ is the radius of the
(assumed spherical )
Brownian particle. Langevin made two assumptions about the fluctuating
force $\vec \xi(t)$: that it has mean $0$ and that it is uncorrelated
to the actual position of the Brownian particle:
\begin{equation}
\label{eq:9}
\begin{array}{rcl}
\langle \vec \xi(t) \rangle & = & 0 \\
\langle \vec x(t) \cdot \vec \xi(t) \rangle & = & \langle \vec x(t) \rangle
\cdot \langle \vec \xi(t) \rangle = 0 
\end{array}
\end{equation}
Multiplying eq.(\ref{eq:8}) by $\vec x$,
taking averages with respect to all realizations
of the random force $\vec \xi(t)$, and using the previous conditions
on $\vec \xi(t)$ one gets:
\begin{equation}
\label{eq:10}
\frac{m}{2}\frac{d^2\;\langle {\vec x\;}^2\rangle }{dt^2} = 
m \langle {\vec v\;}^2 \rangle-3 \pi a \eta 
\frac{d\;\langle {\vec x\;}^2\rangle }{dt}
\end{equation}
Langevin assumed that we are in the regime in which thermal equilibrium
between the Brownian particle and the surrounding
fluid has been reached. In particular,
this implies that, according to the equipartition theorem, the average
kinetic energy of the Brownian particle is $\langle m {\vec v\;}^2/2 \rangle 
= 3k_BT/2$ ($k_B$ is Boltzmann's constant and $T$ is the fluid temperature).
One can now solve very easily eq.(\ref{eq:10}) to find that the 
particle does not move on the average and that, after
some transient time, the asymptotic average square position is given
by:
\begin{equation}
\label{eq:11}
\begin{array}{rcl}
\langle \vec x(t) \rangle & = & 0 \\
\langle {\vec x(t)}^2 \rangle & = & \frac{k_BT}{\pi \eta a} t
\end{array}
\end{equation}
This is nothing but Einstein's diffusion law, but we have now an
explicit expression for the diffusion coefficient:
\begin{equation}
\label{eq:12}
D = \frac{k_BT}{6\pi\eta a}
\end{equation}
Langevin's random force $\vec \xi(t)$ and the Brownian particle
position $\vec x(t)$ are examples of stochastic processes.
It is now time we provide a more precise definition of what a stochastic
process is. It should be clear that the natural scenario 
is that of probability theory \cite{feller,papoulis}.

Let us consider a probabilistic experiment $E=(S,{\cal F},P)$, where
$S$ is a set of possible results, ${\cal F}$ is the $\sigma$--algebra
of events, i.e. those subsets of $S$ that are assigned a probability, and
the real function $P:{\cal F} \to R$ is a $\sigma$--additive probability. 
In many occasions, we are interested in a real number associated with the
experiment result. We call this a {\sl random variable} and denote
it by $\hat x$. In other words, a random variable is an application of
the set of results into the set of real numbers\footnote{This application
must satisfy some additional properties, in particular that the set
$\{u \in S~|~\hat x[u] \le x\}$ belongs to the $\sigma$--algebra ${\cal F}$,
$ \forall x$.}
\begin{equation}
\label{eq:13}
\begin{array}{rcl}
\hat x:S & \to &  R\\
u & \to & \hat x [u]
\end{array}
\end{equation}
In many occasions, the outcome $u$ of the experiment is itself
a real number, and we simply define $\hat x(u)=u$. In those cases
and by abuse of language, the experiment
result $u$ is also called a random variable. The probability density function
$f(x)$ of the random variable $\hat x$ is defined such that $f(x)dx$ is the
probability that $\hat x$ takes variables in the interval $(x,x+dx)$, namely
the probability of the set $\{u \in S~|~x \le \hat x[u] \le x+dx\}$.

We now define a stochastic process $\hat x(t)$ as a family
of random variables
depending on some continuous real parameter $t$. It is, then, a family of
applications:
\begin{equation}
\label{eq:14}
\begin{array}{rcl}
\hat x(t):S & \to &  R\\
u & \to & \hat x[t,u]
\end{array}
\end{equation}
Alternatively, for each experiment result $u$ we might now think 
of $\hat x[t,u]\equiv x_u(t)$ as
a function of the parameter $t$. This is usually the way one considers
a stochastic process $\hat x(t)$: as a collection of functions $x_u(t)$ each
one depending of the outcome $u$. In most applications, $t$ is a
physical time and $x_u(t)$ is a trajectory that depends
on the outcome of some probabilistic experiment $S$. In this
way, the trajectory
itself $x_u(t)$ acquires a probabilistic nature.

\begin{figure}
\begin{center}
\def\epsfsize#1#2{0.82\textwidth}
\leavevmode
\epsffile{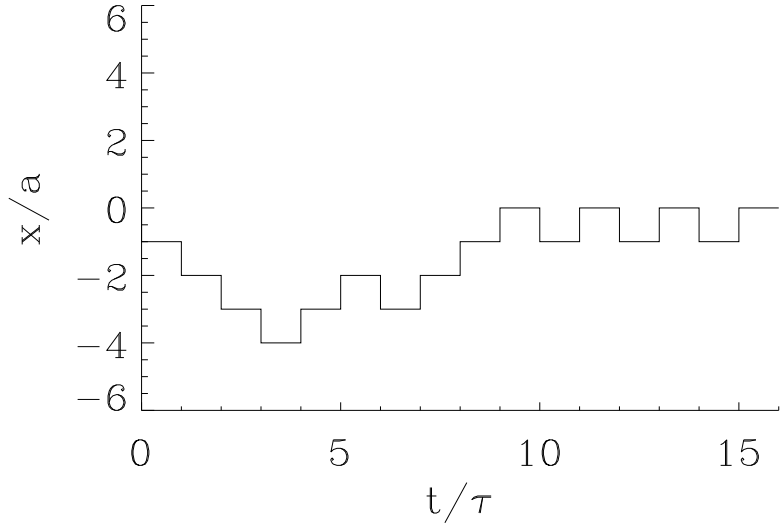}
\end{center}
\caption{\label{fig:1} Random walk trajectory $x_u(t)$ associated  to the result
$u$ of a binary experiment as discussed in the main text.}
\end{figure}

Arguably,
the most well--known example of a stochastic process is that of the
{\sl random walk} \cite{weiss}. The experiment $S$ is a series of binary results
representing, for instance, the outcome of repeatedly tossing
a coin: $u=(0,0,0,0,1,1,0,1,1,1,0,1,0,1,0,1,\dots)$ ($1$ means
``heads", $0$ means ``tails"). To this outcome we associate a $1$--dimensional
real function $x_u(t)$ 
which starts at $x_u(0)=0$ and that moves to the left (right)
at time $k \tau$ an amount $a$ if the $k$--th result of the 
tossed coin was 0 (1). Fig. (\ref{fig:1}) shows the 
``erratic" trajectory for the above result $u$.

What does one mean by characterizing a stochastic process? Since it 
is nothing but a continuous family of random variables, a stochastic process
will be completely characterized when we know the joint probability
density function for the set $\{\hat x(t_1),\hat x(t_2),\dots,
\hat x(t_m)\}$, i.e.
when we know the function $f(x_1,\dots,x_m;t_1,\dots,t_m)$ for arbitrary
$m$. This function is such that 
\begin{equation}
\label{eq:15}
f(x_1,\dots,x_m;t_1,\dots,t_m)dx_1 \dots dx_m
\end{equation}
represents the probability that the random variable $\hat x(t_1)$
takes values in the interval $(x_1,x_1+dx_1)$, the random variable
$\hat x(t_2)$ takes values in the interval $(x_2,x_2+dx_2)$, etc.
In a different language, we can say that a complete characterization of the
trajectory is obtained by giving the functional probability density function
$f([x(t)])$. One has to
realize that a complete characterization of a stochastic process implies the
knowledge of a function of an arbitrary number of parameters and is very
difficult to carry out in practice. In many occasions one is happy if one can
find simply the one--time pdf $f(x;t)$ and the two--times pdf
$f(x_1,x_2;t_1,t_2)$. In terms of those, it is possible to compute trajectory
averages: \begin{equation}
\label{eq:16}
\langle \hat x(t)^n \rangle = \int_{-\infty}^{\infty} dx \; x^n f(x;t)
\end{equation}
and time correlations:
\begin{equation}
\label{eq:17}
\langle \hat x(t_1)\hat x(t_2) \rangle = \int_{-\infty}^{\infty} 
dx_1\;dx_2\; x_1x_2 f(x_1,x_2;t_1,t_2)
\end{equation}
It is important to understand that
the averages $\langle \dots \rangle$ are taken
with respect to all the possible realizations of the stochastic process
$\hat x(t)$, i.e. with respect to all the possible outcomes $u$ of the
experiment. Every outcome $u$ gives rise to a different trajectory $x_u(t)$.
The different trajectories are usually called ``realizations" of the stochastic
process $\hat x(t)$. An alternative way of understanding the previous averages is
by performing the experiment a (preferably large) number of times $M$ to obtain
the results $u_i$, $i=1,\dots,M$, and the different trajectory realizations
$x_{u_i}(t)$. The averages can then be performed by averaging over the different
trajectories as:
\be \label{eq:17b}
\langle \hat x(t) \rangle = \frac{1}{M} \sum_{i=1}^M x_{u_i}(t)
\ee
and similar expressions for other averages.

In two very interesting cases does the knowledge of $f(x;t)$ and
$f(x_1,x_2;t_1,t_2)$ imply the knowledge of the complete pdf
$f(x_1,\dots,x_m;t_1,\dots,t_m)$ for arbitrary $m$: (i) Complete
time independence and (ii) Markov process\footnote{In fact, complete
independence is a particularly simple case of Markov processes.}. 
In a complete time--independent
process, the random variables at different times are independent and
we are able to write:
\begin{equation}
\label{eq:18}
f(x_1,\dots,x_m;t_1,\dots,t_m)=f(x_1;t_1) f(x_2;t_2) \dots f(x_m;t_m)
\end{equation}
In the case of a so--called 
Markov process, the rather general conditional probability
\begin{equation}
\label{eq:19}
f(x_m;t_m|x_1,\dots,x_{m-1};t_1,\dots,t_{m-1}) \equiv
\frac{f(x_1,\dots,x_m;t_1,\dots,t_m)}{f(x_1,\dots,x_{m-1};t_1,\dots,t_{m-1})}
\end{equation}
is equal to the two--times conditional probability
\begin{equation}
\label{eq:20}
f(x_m;t_m|x_{m-1};t_{m-1}) \equiv
\frac{f(x_{m-1},x_m;t_{m-1},t_m)}{f(x_{m-1};t_{m-1})}
\end{equation}
for all times $t_m > t_{m-1} > \dots > t_1$. Loosely speaking, the Markov 
property means that the probability of a future event depends only on the
present state of the system and not on the way it reached its present
situation. In this case one can compute the $m$--times pdf as:
\begin{equation}
\label{eq:21}
\begin{array}{rcl}
& & f(x_1,\dots,x_m;t_1,\dots,t_m)  = \\
& & f(x_m;t_m|x_{m-1};t_{m-1})
f(x_{m-1};t_{m-1}|x_{m-2};t_{m-2})\dots f(x_2;t_2|x_1;t_1)f(x_1;t_1)
\end{array}
\end{equation}
The random walk is an example of a Markov process, since the 
probability of having a particular value of the position at time $(k+1) \tau$ 
depends only on the particle location at time $k\tau$ and not on the
way it got to this location.

Another important category is that of Gaussian processes \cite{fox}
for which there is an explicit form for the $m$--times pdf, namely:
\begin{equation}
\label{eq:22}
f(x_1,\dots,x_m;t_1,\dots,t_m)= \sqrt{\frac{|S|}{(2\pi)^m}}
\exp{\left[-\frac{1}{2}\sum_{i,j=1}^m(x_i-b_i)S_{ij}(x_j-b_j)\right]}
\end{equation}
where the parameters of this expression can be related to mean values
of the random variables as:
\begin{equation}
\label{eq:23}
\begin{array}{rcl}
b_i & = & \langle \hat x(t_i)\rangle \\
(S^{-1})_{ij} & = & \langle \hat x(t_i) \hat x(t_j) \rangle-
\langle \hat x(t_i) \rangle\langle  \hat x(t_j)\rangle   
\end{array}
\end{equation}
As a consequence, one very rarely writes out the above form for the pdf, 
and rather characterizes the Gaussian process
by giving the mean value $\langle \hat x(t) \rangle$  
and the correlation function $\langle \hat x(t) \hat x(t') \rangle$.
From the many properties valid for Gaussian processes, we mention that
a linear combination of Gaussian processes is also a Gaussian process.

\subsection{Stochastic Differential Equations}
A stochastic differential equation (SDE) is a differential equation
which contains a stochastic process $\hat \xi(t)$:
\begin{equation}
\label{eq:24}
\frac{d\hat x(t)}{dt} = G(\hat x(t),t,\hat \xi(t))
\end{equation}
Let us explain a little further what is meant by the previous 
notation\footnote{The fact that this equation is a first--order differential
equation in no means represents a limitation. If $\hat x$ and $G$ are
vector functions, this equation can represent any high--order differential
equation. For simplicity, however, we will consider only 
first--order stochastic differential equations. Notice that Langevin equation
for the position of the Brownian particle is indeed a second order stochastic 
differential equation}. $G$ is a given 3--variable real function. $\hat \xi(t)$
is a stochastic process: a family of functions $\xi_u(t)$ depending on  the
outcome $u$ of some experiment $S$. As a consequence a SDE is {\sl not} a single
differential equation but rather a family of ordinary  differential equations, a
different one for each outcome $u$ of the  experiment $S$:
\begin{equation}
\label{eq:24a}
\frac{d x_u(t)}{dt} = G(x_u(t),t,\xi_u(t))
\end{equation}
Therefore, the family of solutions $x_u(t)$ of these differential equations, for
different outcomes $u$, constitute a stochastic process $\hat x(t)$. We can say
that for each realization $\xi_u(t)$ of the stochastic process $\hat \xi$,
corresponds a realization $x_u(t)$ of the stochastic process $\hat x$. The
solution $\hat x$ becomes then a functional of the process $\hat \xi$. To
``solve" a SDE means to characterize completely the stochastic process $\hat
x(t)$, i.e. to give the $m$--times pdf $f(x_1,\dots,x_m;t_1,\dots,t_m)$. Again,
this is in general a rather difficult task and sometimes one focuses only on the
evolution of the moments $\langle \hat x(t)^n \rangle$ and the correlation
function $\langle \hat x(t_1) \hat x(t_2) \rangle$. 

When the stochastic process $\hat \xi (t)$ appears linearly one talks
about a {\sl Langevin equation}. Its general form being:
\begin{equation}
\label{eq:25}
\frac{dx}{dt}=q(x,t)+g(x,t)\xi(t)
\end{equation}
(from now on, and to simplify notation, the ``hats" will be dropped from  the
stochastic processes). In this case, $\xi(t)$ is usually referred to as the
``noise" term. A word whose origin comes from the random ``noise" one can
actually hear in electric circuits. Still another notation concept: if the
function $g(x,t)$ is constant, one talks about {\sl additive} noise, otherwise,
the noise is said to be {\sl multiplicative}. Finally, $q(x,t)$ is usually
referred to as the ``drift" term,  whereas $g(x,t)$ is the ``diffusion" term (a
notation which is more commonly used in the context of the Fokker--Planck
equation, see later).  

Of course, we have already encountered a Langevin SDE, this is nothing but
Langevin's equation for the Brownian particle, equation (\ref{eq:8}).  In this
example, the stochastic noise was the random force acting upon the Brownian 
particle. The ``experiment" that gives rise to a different force is the
particular position and velocities of the fluid 
molecules surrounding the Brownian
particle. The movements of these particles are so erratic and unpredictable that
we assume a probabilistic description of their effects upon the Brownian
particle.

\begin{figure}
\begin{center}
\def\epsfsize#1#2{0.82\textwidth}
\leavevmode
\epsffile{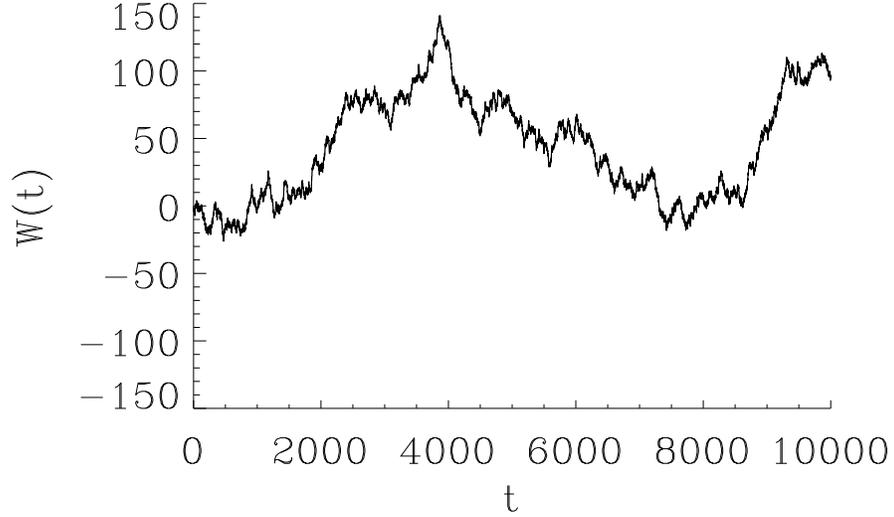}
\end{center}
\caption{\label{fig:2} A typical realization of the Wiener process done
by generating a random walk with a large number of steps.}
\end{figure}

We will now characterize the process $\xi(t)$ that appears in
the Langevin equation for the Brownian particle. We will be less
rigorous here in our approach and, in fact, we will be nothing but
rigorous at the end! but still we hope that we can give a manageable
definition of the stochastic force $\xi(t)$. We first need 
to define the stochastic Wiener process $W(t)$. This is obtained as a suitable
limit of the random walk process \cite{papoulis}. 
The probability that the walker is at location
$x=r\,a$ after time $t=n\,\tau$ can be expressed 
in terms of the binomial distribution:
\begin{equation}
\label{eq:26}
P(x(n\tau)=r a) = {n \choose {\frac{n+r}{2}}}2^{-n}
\end{equation}
From where it follows:
\begin{equation}
\label{eq:27}
\begin{array}{rcl}
\langle x(n\tau) \rangle & = & 0\\
\langle x(n\tau)^2 \rangle & = & na^2\\
\end{array}
\end{equation}
For $n \gg 1$ we can use the asymptotic result (de Moivre--Laplace theorem)
that states that the binomial distribution can be replaced by a Gaussian
distribution:
\begin{equation}
\label{eq:28}
P(x(n\tau) \le r a) = \frac{1}{2}+{\rm erf}\left(\frac{r}{\sqrt{n}}\right)
\end{equation}
(${\rm erf}(x)$ is the error function \cite{abra}).
We now take the continuum limit defined by:
\begin{equation}
\label{eq:29}
\begin{array}{lr}
n\to \infty,~ \tau \to 0 & n \tau = t\\
r\to \infty,~ a \to 0 & r a = s \\
 & a^2/\tau = 1
\end{array}
\end{equation}
with finite $t$ and $s$. In this limit the random walk
process is called the Wiener process $W(t)$ and equation (\ref{eq:28}) 
tends to:
\begin{equation}
\label{eq:30}
P(W(t) \le s) = \frac{1}{2}+{\rm erf}\left(\frac{s}{\sqrt{t}}\right)
\end{equation}
which is the probability distribution function of a Gaussian variable
of zero mean and variance $t$. 
The corresponding one--time probability density function 
for the Wiener process is:
\begin{equation}
\label{eq:31}
f(W;t)= \frac{1}{\sqrt{2\pi t}}\exp{(-\frac{W^2}{2t})}
\end{equation}
The Wiener process is a Markovian (since the random walk is Markovian)
Gaussian process. As every Gaussian process it can be fully 
characterized by giving
the one--time mean value and the two--times correlation function. These
are easily computed as:
\begin{equation}
\label{eq:32}
\begin{array}{rcl}
\langle W(t) \rangle & = & 0 \\
\langle W(t)^2 \rangle & = & t \\
\langle W(t_1) W(t_2) \rangle & = & \min(t_1,t_2) \\
\end{array}
\end{equation}
A typical realization of the Wiener process is given in Fig. (\ref{fig:2}).
The Wiener process is continuous but it does not have first derivative.
In fact it is a fractal of Hausdorff dimension $2$ \cite{falconer}.

\begin{figure}
\begin{center}
\def\epsfsize#1#2{0.82\textwidth}
\leavevmode
\epsffile{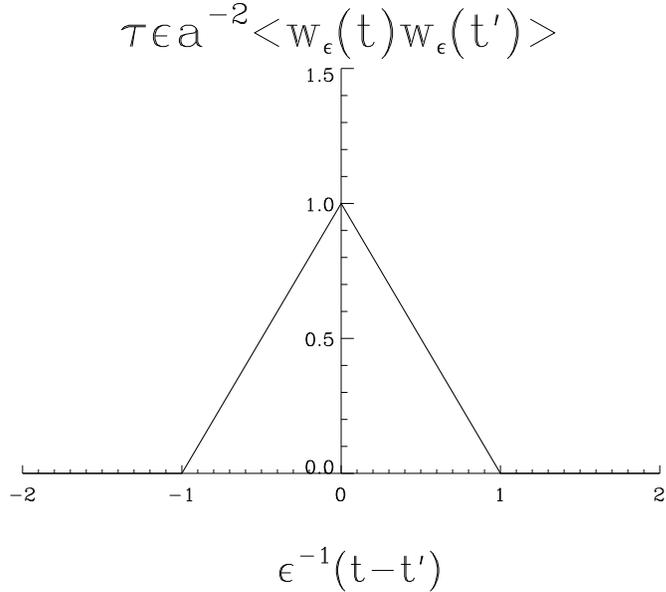}
\end{center}
\caption{\label{fig:4} Correlation function for
the derivative of the random walk process, equation (\ref{eq:34}).}
\end{figure}

We will define now the {\sl white--noise} random process as the
derivative of the Wiener process. Since we just said that the Wiener process
does not have a derivative, it is not surprising that the resulting
function is a rather peculiar function. The trick is to perform the
derivative before the continuum limit (\ref{eq:29}) is taken.
If $x(t)$ is the
random walk process, we define the stochastic process $w_{\epsilon}(t)$ as:
\begin{equation}
\label{eq:33}
w_{\epsilon}(t) = \frac{x(t+\epsilon)-x(t)}{\epsilon}
\end{equation}
$w_{\epsilon}(t)$ is a Gaussian process since it is a linear combination
of Gaussian processes. Therefore, it is sufficiently defined by its
mean value and correlations:
\begin{equation}
\label{eq:34}
\begin{array}{rcl}
\langle w_{\epsilon}(t) \rangle & = & 0 \\
\langle w_{\epsilon}(t_1) w_{\epsilon}(t_2) \rangle & = & \left\{ 
{\begin{array}{lcr}
& 0 & {\rm if}~ t_1-t_2 < - \epsilon \\
& a^2/(\tau\epsilon)(1+(t_1-t_2)/\epsilon)&{\rm if}~-\epsilon \le t_1-t_2 \le 0 \\
& a^2/(\tau\epsilon)(1-(t_1-t_2)/\epsilon)&{\rm if}~0\le t_1-t_2 \le \epsilon \\
& 0 & {\rm if}~ t_1-t_2 > \epsilon 
\end{array} }
\right.
\end{array}
\end{equation}
which is best understood by the plot in Fig. (\ref{fig:4}).
If we let now $\epsilon \to 0$ the process $w(t)=\lim_{\epsilon \to 0}
w_{\epsilon}(t)$ becomes the derivative of the random walk process.  The
correlation function (\ref{eq:34}) becomes a delta function 
$(a^2/\tau)\delta(t_1-t_2)$. If we take now the limit defined in
eqs.(\ref{eq:29}) the random walk  process tends to the Wiener process $W(t)$
and the derivative process $w(t)$ tends to $\xi_w(t)$: the {\sl white--noise}
process.  Intuitively, the white--noise represents a series of independent
pulses acting on a very small time scale but of high intensity, such that their
effect is finite. The white noise can be considered the ideal limit of a physical
stochastic process in the limit of very small correlation time $\tau$. Formally,
the white noise is  defined as a Markovian, Gaussian process  of mean value and
correlations given by: 
\begin{equation}
\label{eq:35}
\begin{array}{rcl}
\langle \xi_w(t) \rangle & = & 0 \\
\langle \xi_w(t_1) \xi_w(t_2) \rangle & = & \delta(t_1-t_2)
\end{array}
\end{equation}
and can be considered the derivative of the Wiener process:
\begin{equation}
\label{eq:36}
\xi_w(t)=\frac{dW(t)}{dt}
\end{equation}

All this lack of mathematical rigor leads to some problems of
interpretation. For instance, when in a SDE the white--noise
appears multiplicatively, the resulting process $x(t)$ will be, in 
general, a non continuous function of time. When this happens, there
is an ambiguity in some mathematical 
expressions. Giving a sense to those a priori undefined expressions
constitutes a matter of sheer definition. The most
widely used interpretations are those of It\^o and Stratonovich 
\cite{kampen,gardiner}. To 
make a long story short, we can summarize both interpretations as follows:
when in a calculation we are faced with an integral
\begin{equation}
\label{eq:36b}
\int_t^{t+h} ds\;g(x(s)) \xi_w(s) 
\end{equation}
to be computed in the limit $h \to 0$, It\^o interprets this as:
\begin{equation}
\label{eq:37}
g(x(t)) \int_t^{t+h} ds\; \xi_w(s) = g(x(t)) [W(t+h)-W(t)]
\end{equation}
and Stratonovich as:
\begin{equation}
\label{eq:38}
\frac{g(x(t))+g(x(t+h))}{2} \int_t^{t+h} ds\; \xi_w(s) = 
\frac{g(x(t))+g(x(t+h))}{2} [W(t+h)-W(t)]
\end{equation}
Although there was much arguing in the past to which is the ``correct"
interpretation, it is clear now that it is just a matter of convention.
This is to say, the Langevin SDE (\ref{eq:25}) is not completely
defined unless we define what do we interpret when we encounter expressions
such as eq.(\ref{eq:36b}). In some sense, it is related to the problem
of defining the following expression involving the Dirac--delta function:
\begin{equation}
\label{eq:39}
\int_0^{\infty} dt\;\delta(t)
\end{equation}
This can be {\sl defined} 
as equal to $1$ (equivalent to the It\^o rule) or to $1/2$ (Stratonovich). Both 
integration rules give different
answers and one should specify from the beginning which is the interpretation
one is using. The Stratonovich rule turns out to be more ``natural" for
physical problems because it is the one that commutes with the limit
(\ref{eq:29}). Unless otherwise stated, we will follow in general the
Stratonovich interpretation. Moreover, the Stratonovich interpretation allows 
us to use the familiar rules of calculus, such as change of variables
in an integration, etc. The use of the It\^o rules leads to
relations in which some of the familiar expressions of ordinary calculus are not
valid and one needs to be a little bit more cautious when not used to it. A
consequence of the It\^o interpretation is the simple result $\langle x(t)
\xi_w(t) \rangle = 0$ which is not valid in the Stratonovich interpretation
(see later Eq. (\ref{eq:51})).  However, there is a simple relation between the
results one obtains in the two interpretations. The rule is that the SDE
\begin{equation} 
\label{eq:40} 
\frac{dx}{dt}=q(x)+g(x)\xi_w(t)
\end{equation}
in the It\^o sense, is equivalent to the SDE:
\begin{equation}
\label{eq:41}
\frac{dx}{dt}=q(x)-\frac{1}{2}g(x)g'(x)+g(x)\xi_w(t)
\end{equation}
in the Stratonovich sense. This rule allows to easily translate the
results one obtains in both interpretations. Both interpretations coincide for
additive noise.

The white noise process is nothing but a physical idealization that leads
to some mathematical simplifications. For instance, it can be proven that the
solution $x(t)$ of the Langevin equation (\ref{eq:40}) is a Markov process if
$\xi(t)$ is a white--noise process. In any physical process, however, there will
be a finite correlation time $\tau$ for the noise variables. A widely used
process that incorporates the concept of a finite correlation time is the
Ornstein--Uhlenbeck noise, $\xi_{OU}$ \cite{kampen}. 
This is formally defined as a Gaussian Markov process characterized by: 
\begin{equation} 
\label{eq:41b}
\begin{array}{rcl}
\langle
\xi_{OU}(t) \rangle & = & 0 \\ 
\langle \xi_{OU}(t) \xi_{OU}(t') \rangle & = & \frac{1}{2\tau}{\rm
e}^{-|t-t'|/\tau}  
\end{array}
\end{equation}
Several comments are in order:\\
(i) The OU--noise has a non zero correlation time $\tau$ meaning that
values of the noise at different times are not independent random variables.
This is a more faithful representation of physical reality than the
white--noise limit.\\
(ii) In the limit $\tau \to 0$ the white noise limit $\xi_w(t)$ (with
the correlations given in eq.(\ref{eq:35})) is
recovered and the corresponding SDE's are to be interpreted in the
Stratonovich sense.\\
(iii) The OU--noise is, up to a change of variables, the only Gaussian,
Markov, {\sl stationary} process.\\
(iv) The OU-noise is the solution of the SDE:
\begin{equation}
\label{eq:87}
\frac{d\xi_{OU}}{dt} = -\frac{1}{\tau} \xi_{OU}(t)+\frac{1}{\tau}\xi_w(t)
\end{equation}
with the initial condition that $\xi_{OU}(0)$ is a Gaussian random
variable of mean $0$ and variance $(2\tau)^{-1}$. 

\subsection{The Fokker--Planck Equation} 
We will find now an equation for the one--time pdf for a stochastic 
process which arises as a solution of a SDE
with Gaussian white noise.
This is called the Fokker--Planck equation \cite{Risken}. It is 
the equivalent of Einstein's 
description which focuses on probabilities rather than in trajectories as
in the Langevin approach. More precisely, we want to find 
an equation for the one--time pdf $f(x,t)$ 
of a stochastic process governed by the SDE of the Langevin form:
\begin{equation}
\label{eq:42}
\frac{dx}{dt}=q(x)+g(x)\xi_w(t)
\end{equation}
with the white--noise defined in eq.(\ref{eq:35}). This is to be understood
in the Stratonovich interpretation. 
To find an equation for the probability density function $f(x;t)$ 
we will rely upon functional methods \cite{ssmkg,hanggi}. 
Let us consider first the corresponding
deterministic initial value problem:
\begin{equation}
\label{eq:43}
\begin{array}{rcl}
\dot x & = & q(x)\\
x(t=0) & = & x_0
\end{array}
\end{equation}
The solution of this equation is a (deterministic) function $x(t,x_0)$.
We can think of $x(t)$ as a random variable whose 
probability density function $\rho(x;t)$ is a delta--function:
\begin{equation}
\label{eq:44}
\rho(x;t)=\delta(x-x(t,x_0))
\end{equation}
We can now turn $x(t)$ very easily into a stochastic process by simply letting
the initial condition $x_0$ become a random variable. For each possible
value of $x_0$ we have a different solution $x(t,x_0)$, i.e. a random
process. In this case the probability density function $\rho(x,t)$ is
obtained by averaging the above pdf over the distribution of initial
conditions:
\begin{equation}
\label{eq:45}
\rho(x;t)=\langle \delta(x-x(t,x_0))\rangle_{x_0}
\end{equation}
But it is well known from mechanics that the density 
$\rho(x;t)$ satisfies Liouville's continuity equation \cite{balescu}:
\begin{equation}
\label{eq:46}
\frac{\partial \rho}{\partial t} + \frac{\partial}{\partial x}(\dot x \rho) =0
\end{equation}
or, using, eq.(\ref{eq:43}):
\begin{equation}
\label{eq:47}
\frac{\partial \rho}{\partial t} = -\frac{\partial}{\partial x}[q(x) \rho]
\end{equation}
If we consider now the full SDE (\ref{eq:42}) we can repeat the above
argument for a given realization of the noise term. The probability
density function $f(x;t)$ will be the average of $\rho(x;t)$ with 
respect to the noise distribution:
\begin{equation}
\label{eq:48}
f(x;t) = \langle \rho(x;t) \rangle_{\xi_w}=
\langle \delta(x-x(t,x_0))\rangle_{x_0,\xi_w}
\end{equation}
where $\rho$ satisfies the Liouville equation (\ref{eq:46})
and, after substitution of (\ref{eq:42}):
\begin{equation}
\label{eq:49}
\frac{\partial \rho}{\partial t} = -\frac{\partial}{\partial x}
[(q(x)+g(x)\xi(t)) \rho]
\end{equation}
By taking averages over the noise term we get:
\begin{equation}
\label{eq:50}
\frac{\partial f}{\partial t} = -\frac{\partial}{\partial x}[q(x) f]-
\frac{\partial}{\partial x}[g(x)\langle \xi(t) \rho (x,[\xi(t)])\rangle]
\end{equation}
The averages are done by using Novikov's theorem \cite{novikov}:
\begin{equation}
\label{eq:51}
\langle \xi(t)\rho([\xi(t)])\rangle = \int_0^tds\;\langle \xi(t) \xi(s)
\rangle \left\langle \frac{\delta \rho}{\delta \xi(s)}\right\rangle
\end{equation}
which applies to any functional $\rho([\xi(t)])$ of a Gaussian process $\xi(t)$ of zero mean, 
$\langle \xi(t) \rangle =0$. For the white--noise process:
\begin{equation}
\label{eq:52}
\langle \xi_w(t)\rho([\xi_w(t)])\rangle = \frac{1}{2}
\left\langle \frac{\delta \rho}{\delta \xi_w(s)}\right\rangle_{s=t}
\end{equation}
By using the functional calculus, this can be computed as:
\begin{equation}
\label{eq:53}
\left\langle \frac{\delta \rho}{\delta \xi_w(s)}\right\rangle_{s=t} =
\left\langle \frac{\delta x(t)}{\delta \xi_w(s)} \right|_{s=t}\left.
\frac{\delta \rho}{\delta x(t)}\right\rangle =
-\frac{\partial}{\partial x} 
\left\langle \frac{\delta x(t)}{\delta \xi_w(s)}\right|_{s=t}
\rho \rangle
\end{equation}
By using the formal solution of eq.(\ref{eq:42}):
\begin{equation}
\label{eq:54}
x(t)=x_0+\int_0^tds\;q(x(s)) + \int_0^tds\; g(x(s))\xi_w(s)
\end{equation}
we get:
\begin{equation}
\label{eq:55}
\left. \frac{\delta x(t)}{\delta \xi_w(s)}\right|_{s=t} = g(x(t))
\end{equation}
and
\begin{equation}
\label{eq:56}
\left\langle \frac{\delta \rho}{\delta \xi_w(s)}\right\rangle_{s=t} =
-\frac{\partial}{\partial x} [g(x)\langle \rho \rangle ] =
-\frac{\partial}{\partial x} [g(x)f]
\end{equation}
By substitution in (\ref{eq:52}) and (\ref{eq:50}) we 
get, finally, the Fokker--Planck equation for the
probability density function:
\begin{equation}
\label{eq:57}
\frac{\partial f}{\partial t} = -\frac{\partial}{\partial x}[q(x)f]
+ \frac{1}{2}\frac{\partial}{\partial x}[g(x) \frac{\partial}{\partial x}[g(x)f]]
\end{equation}

The above functional method to derive the Fokker--Planck equation from
the Langevin equation is
very powerful and can be extended to other situations such as
the multivariate case and SDE with colored (Ornstein-Uhlenbeck) noise.
For colored noise the problem is nonmarkovian and no exact Fokker-Planck
equation can be derived for the probability density. Still 
one can obtain different approximate equations for the
probability density and for the two-time probability density \cite{sms}.

Much work has been devoted to finding solutions to the Fokker--Planck equation.
For the stationary equation, $\partial P/\partial t=0$, the solution can be
always reduced to quadratures in the one--variable case. The multivariate
equation is more complicated and some aspects are discussed in Sect. 4.2.

\subsection{Numerical generation of trajectories}
In the Langevin's approach to stochastic processes, a great deal of 
relevance is given to the trajectories. It is of great importance,
therefore, when given a SDE of the form of eq.(\ref{eq:24}), 
to be able to generate representative trajectories. This is to say:
to generate the functions $x_u(t)$ for different outcomes $u$ of the experiment.
If we generate, say $M$ results $u_i$, the averages could be obtained
by performing explicitly the ensemble average as indicated 
in equation (\ref{eq:17b}).

We now explain the basic algorithms to generate trajectories starting
from a SDE of the Langevin form \cite{ssmkg,gard,GSH,kloeden}.

\subsubsection{The white noise case: Basic algorithms}

Euler's algorithm is by far the simplest one can devise to generate
trajectories. We will explain it by using a simple example. Let us
consider the SDE:
\begin{equation}
\label{eq:58}
\dot x(t) = f(t)+\xi_w(t)
\end{equation}
If fact, this equation is so simple that it can be solved exactly. For
a given realization of the noise term, the solution is:
\begin{equation}
\label{eq:59}
x(t)=x(0)+\int_0^t f(s) \D s + \int_0^t \xi_w(s) \D s \equiv
x(0)+F(t)+W(t)
\end{equation}
$W(t)$ is the Wiener process. Hence, we conclude that the stochastic
process $x(t)$ is Gaussian and is completely characterized by:
\be
\label{eq:58b}
\begin{array}{rcl} 
\langle x(t) \rangle & = & x(0)+F(t) \\
\langle x(t)x(t') \rangle & = & (x(0)+F(t))(x(0)+F(t'))+\min(t,t')
\end{array}
\ee

However, let us forget for the moment being that we can solve the SDE and
focus on a numerical solution in which we generate trajectories.
We do so by obtaining $x(t)$ at discrete
time intervals:
\begin{equation}
\label{eq:60}
\begin{array}{lll}
x(t+h) & = & x(t)+\int_t^{t+h}\dot x(s) \D s \\
       & = & x(t)+\int_t^{t+h}f(s) \D s + \int_t^{t+h}\xi_w(s) \D s \\
       & \equiv & x(t)+f_h(t)+w_h(t)
\end{array}
\end{equation}
here, $w_h(t)=W(t+h)-W(t)$ is the difference of the Wiener process at two
different times and it is, therefore, a Gaussian process. $w_h(t)$ can 
be characterized by giving the mean and correlations:
\begin{equation}
\label{eq:61}
\langle w_h(t)\rangle =\int_t^{t+h} \langle \xi_w(s)\rangle  \D s = 0
\end{equation}
\begin{equation}
\label{eq:62}
\begin{array}{lll}
\langle w_h(t)w_h(t')\rangle  & = & 
\int_t^{t+h} \D s \int_{t'}^{t'+h} \D u \langle \xi_w(s)\xi_w(u)\rangle  \\
& = & \int_t^{t+h}\D s \int_{t'}^{t'+h}\D u \delta(s-u)
\end{array}
\end{equation}
The integral is an easy exercise on $\delta$ function integration:
we can assume, without loss of generality that $t'>t$.
If $t' > t+h$ the integral is $0$ since there is no overlap in the
integration intervals and the delta function vanishes. 
If $t \leq t' < t+h$ the double integral equals the length of the overlap 
interval:
\begin{equation}
\label{eq:63}
\langle w_h(t)w_h(t')\rangle   = \int_{t'}^{t+h} \D s = t-t'+h 
\end{equation}
In particular, we notice the relation
\begin{equation}
\label{eq:64}
\langle w_h(t)^2\rangle   =  h
\end{equation}
It is important to realize that, if $t_i=i~h$, $t_j=j~h$ are the times
appearing in the recurrence relation eq.(\ref{eq:60}), we have:
\begin{equation}
\label{eq:65}
\langle w_h(t_i)w_h(t_j)\rangle  = h \delta_{ij}
\end{equation}
We introduce now a set of independent Gaussian random variables $u(t)$ 
defined only for the discrete set of recurrence times, 
$t=0$, $h$, $2h$, $\dots$,
of mean zero and variance one:
\be
\label{eq:66}
\begin{array}{rcl}
\langle u(t) \rangle  = 0 & , & \langle u(t^2) \rangle = 1  \\
\langle u(t) u(t') \rangle  = 0 & ,& t \ne t'
\end{array}
\ee
There is a vast amount of literature devoted to the question of 
generation of random numbers with a given distribution \cite{james, 
knuth, NR, TC2}. The set of
numbers $u(t)$ can be generated by any of the standard methods available.
One of the most widely used\footnote{Although not the most efficient.
See ref.  \cite{TC2} for a comparison of the timing of the different
algorithms and the description of a particularly efficient one.} 
is the Box--Muller--Wiener algorithm: if
$r_1$ and $r_2$ are random numbers uniformly distributed in the interval
$(0,1)$ the transformation:
\be
\begin{array}{rcl}
g_1 & = & \sqrt{-2 \ln (r_1)}\cos(2\pi r_2) \\
g_2 & = & \sqrt{-2 \ln (r_1)}\sin(2\pi r_2)
\end{array}
\ee
returns for $g_1$, $g_2$ two Gaussian distributed random numbers
of mean zero and variance one. This, or other appropriate algorithm,
can be used to generated the set of Gaussian variables $u(t)$. 
In terms of this set of variables we can write:
\begin{equation}
\label{eq:67}
w_h(t)=h^{1/2}\, u(t)
\end{equation}
Finally, the recurrence relation that generates trajectories of 
eq.(\ref{eq:58}) is the Euler algorithm:
\begin{equation}
\label{eq:68}
\begin{array}{rcl}
x(t=0) & = & x_0 \\
x(t+h) & = & x(t)+f_h(t)+h^{1/2}\, u(t)
\end{array}
\end{equation}
For the deterministic contribution we can approximate $f_h(t)\approx
hf(t)$ from where it follows that the deterministic contribution is
of order $h^1$ and successive contributions go as $h^2$, $h^3$, etc. 
On the other hand, the contribution coming from the
white noise term is of order $h^{1/2}$ and, in general, successive
contribution will scale as $h^1$, $h^{3/2}$, etc.

With the experience we have got by solving the previous simple example, let
us now tackle a more complicated case. Let us consider the following SDE:
\begin{equation}
\label{eq:69}
\dot x(t) = q(x) + g(x)\xi_w(t)
\end{equation}
At this step, one might ask why, given that for a particular
realization of the noise term the equation becomes an ordinary 
differential equation (ode), we need special methods to deal with this kind
of equations. The answer lies in the fact that all the methods developed
to deal with ODE's assume that the functions appearing in the equation have
some degree of ``well--behaveness". For instance, they are differentiable
to some required order. This is simply not the case for the white--noise
functions. Even for a single realization of the white--noise term, the function
is highly irregular, not differentiable and, in our non rigorous treatment,
is nothing but
a series of delta--functions spread all over the real axis. This is the
only reason why we can not use the well known predictor--corrector, 
Runge--Kutta and all the like methods without suitable modifications. 
If our SDE happened to have smooth functions as random processes
we could certainly implement all these wonderful methods and use all the
standard and very effective routines available! However, this is usually
not the case: the stochastic contribution is a non analytical function
and we must resource to new and generally more complicated
algorithms. The answer lies in integral algorithms: whereas the derivatives
of the white--noise function are not well defined, the integrals are (the
first integral is the Wiener process, which is a continuous function).
We look for a recursion relation by integration of the SDE (\ref{eq:69}):
\begin{equation}
\label{eq:70}
x(t+h)-x(t)=\int_t^{t+h}q(x(s)) \D s+ \int_t^{t+h}g(x(s)) \xi_w(s) \D s
\end{equation}
Now we assume that the functions $q(x)$ and $g(x)$ are differentiable
functions and we Taylor--expand $q(x(s))$ and $g(x(s))$ around $x=x(t)$:
\begin{eqnarray}
\label{eq:71}
q(x(s))=q(x(t))+\left. \frac{dq}{dx}\right|_{x(t)}\left( x(s)-x(t) \right) + 
O[\left( x(s)-x(t) \right)^2]\\
g(x(s))=g(x(t))+\left. \frac{dg}{dx}\right|_{x(t)}\left( x(s)-x(t) \right) + 
O[\left( x(s)-x(t) \right)^2]
\end{eqnarray}
Substitution of the lowest possible order of these expansions, i.e.,
$q(x(s))=q(x(t)),~g(x(s))=g(x(t))$ in eq.(\ref{eq:70}) yields:
\begin{equation}
\label{eq:72}
x(t+h)-x(t)=h q(x(t)) + h  O[x(s)-x(t)] + w_h(t) g(x(t))
+w_h(t) O[x(s)-x(t)] 
\end{equation}
where, as in the simple previous example,
\begin{equation}
\label{eq:73}
w_h(t)=\int_t^{t+h}\D s\; \xi_w(s) = h^{1/2}\, u(t)
\end{equation}
is of order $h^{1/2}$. To the lowest order ($h^{1/2}$) we have:
\begin{equation}
\label{eq:74}
x(s)-x(t)=g(x(t)) \int_t^s \D v\; \xi_w(v) =  O[h^{1/2}]
\end{equation}
Therefore we need to go one step further in the Taylor expansion of 
the function $g(x(s))$ and the next--order contribution to (\ref{eq:70}) is:
\be
\begin{array}{rcl}
\label{eq:75}
& & g'(x(t)) \int_t^{t+h} \D s\; (x(s)-x(t))\xi_w(s) +
O[x(s)-x(t)]^2w_h(t) = \\
& & g'(x(t)) g(x(t)) \int_t^{t+h} \D s\; \int_t^s\D v\; 
\xi_w(s) \xi_w(v) + O[h^{3/2}]
\end{array}
\ee
The double integral can be easily done by changing the integration 
order:
\begin{equation}
\label{eq:76}
\int_t^{t+h} ds \int_t^s dv\; \xi_w(s) \xi_w(v) = 
\int_t^{t+h} dv \int_v^{t+h} ds\; \xi_w(s) \xi_w(v) =
\int_t^{t+h} ds \int_s^{t+h} dv\; \xi_w(s) \xi_w(v) 
\end{equation}
where to go from the first expression to the second we have exchanged
the integration order and, to go from the second to the third, we 
have exchanged the variables names $v \leftrightarrow s$. Since the
first and third integrals are equal they also equal one half of their
sum, and the previous integrals can be replaced by:
\begin{equation}
\label{eq:77}
\frac{1}{2}\int_t^{t+h} dv \int_t^{t+h} ds\; \xi_w(s) \xi_w(v) =
\frac{1}{2}\int_t^{t+h} ds\; \xi_w(s) \int_t^{t+h} dv\;\xi_w(v) = 
\frac{1}{2}\left[w_h(t)\right]^2 
\end{equation}
Putting all the bits together, we arrive at the desired result which is
\footnote{It is instructive to realize that the same result can be
obtained by use of the Stratonovich rule Eq.(\ref{eq:38}) in equation
(\ref{eq:70})}:
\begin{equation}
\label{eq:78}
x(t+h)=x(t)+h^{1/2}g(x(t)) u(t) + h \left[ q(x(t)) + \frac{1}{2}g(x(t))
g'(x(t)) u(t)^2\right] + O[h^{3/2}]
\end{equation}
This recurrence relation is known in the literature as Milshtein 
method \cite{ssmkg,milshtein}.
If the noise is additive: $g'(x)=0$, then the resulting algorithm is 
called the Euler algorithm:
\begin{equation}
\label{eq:79}
x(t+h)=x(t)+h q(x(t))+g(x(t))h^{1/2} u(t) + O[h^{3/2}]
\end{equation}
Sometimes, though, the name ``Euler algorithm" is also given to a modification
of the Milshtein algorithm in which the $u(t)^2$ term is replaced by its
mean value: $\langle u(t)^2\rangle = 1$:
\begin{equation}
\label{eq:80}
x(t+h)=x(t)+g(x(t))h^{1/2} u(t) + 
h [q(x(t))+ \frac{1}{2}g(x(t))g'(x(t))] + O[h^{3/2}]
\end{equation}
This ``Euler algorithm" is the one appearing naturally when one does the
numerical integration of the SDE in the It\^o formalism, although at this
level it has to be considered just as an approximation, unnecessary, 
to the Milshtein method.

In the previous expressions, the correction to the recurrence algorithm
is said to be of order $O[h^{3/2}]$. Let us explain a little further what is
meant by this expression. We use the following notation: we call
$\bar x(t)$ the values obtained from the numerical integration following
the Milshtein method:
\begin{equation}
\label{eq:80a}
\bar x(t+h)=x(t)+h^{1/2}g(x(t)) u(t) + h \left[ q(x(t)) + \frac{1}{2}g(x(t))
g'(x(t)) u(t)^2\right]
\end{equation}
and we want to compare $\bar x(t+h)$ 
with the exact value $x(t+h)$ which is obtained
by exact integration of the differential equation starting from $x(t)$.
What we have proven is that the mean--square error (averaged over noise 
realizations) for the trajectories starting at $x(t)$ is of order $O[h^3]$:
\begin{equation}
\label{eq:81}
\langle (\bar x(t+h) -x(t+h))^2 \rangle = O[h^3]
\end{equation}
One says that the Milshtein algorithm has a convergence for the trajectories
in mean square of order $h^3$. This is related, but not the same, as the
order of convergence of the $n$--th order moment, $D_n(h)$,
which is defined as:
\begin{equation}
\label{eq:82}
D_n(h) \equiv \langle \bar x(t+h)^n \rangle - \langle x(t+h)^n \rangle 
\end{equation}
the averages are done starting from a given $x(t)$ and averaging over
noise realizations.
For the Milshtein algorithm one can prove \cite{GSH}:
\begin{equation}
\label{eq:83}
D_n(h)=O[h^2]
\end{equation}
Which means that, when computing moments, the Milshtein algorithm
makes an error of order $h^2$ in every integration step.
Of course, in a finite integration from $t=0$ to a time $t=k h$, the total
error will be multiplied by the number of integration steps, $k$, which
gives a contribution $k O[h^2]= (t/h)O[h^2]=O[h]$. And we can write
the following relation between the exact value $\bar x (t)^n$ and the
value $x(t)^n$ obtained when using the Milshtein approximation starting
from an initial value $x(0)$:
\begin{equation}
\label{eq:84}
\langle x(t)^n \rangle = \langle \bar x(t)^n \rangle + O[h]
\end{equation}
In practice, what one does (or rather, what one should do!) is to repeat
the numerical
integration for several time steps $h_1$, $h_2$, $\dots$ and extrapolate
the results towards $h=0$ by using a linear relation
$\langle x(t)^n \rangle = \langle \bar x(t)^n \rangle + \alpha \cdot h$

The question now is whether we can develop more precise algorithms
while preserving the structure of the Milshtein method, i.e. something
like:
\begin{equation}
\label{eq:85}
x(t+h)=x(t)+  C_1 w_h(t) + C_2 h + C_3 w_h(t)^2 + C_4 h w_h(t) + 
C_5 w_h(t)^3 + \dots
\end{equation}
The (negative) answer was given by R\"umelin \cite{rumelin}, 
who stated that higher
order algorithms necessarily will imply more random processes, say
$w_h^{(1)}$, $w_h^{(2)}$, etc. But the problem lies on the fact that
those random processes are not Gaussian and have non--zero
correlations being, in general, very difficult to generate accurately.
As a conclusion, the Milshtein
algorithm appears as the simplest alternative for integrating a SDE.
However, and as a recipe for the practitioner, Runge--Kutta type
methods offer some advantages at a small cost in the programming side.
These methods will be explained in a later section. 
As a final remark, the Milshtein method can also be used for the SDE:
\begin{equation}
\label{eq:85b}
\dot x(t) = q(t,x) + g(t,x)\xi_w(t)
\end{equation}
in which the diffusion an drift terms depend explicitly on time $t$.

\subsubsection{The Ornstein--Uhlenbeck noise}
We turn now to the numerical generation of trajectories for a SDE
with colored noise, in particular of the Ornstein--Uhlenbeck form as 
defined in equation (\ref{eq:41b}). First, we explain how to generate
realizations of the OU process itself, $\xi_{OU}(t)$, and later we will
see their use in SDE's.

Equation (\ref{eq:87}) can be solved exactly (it is a linear equation) and
the solution actually tells us how to generate trajectories of the 
OU--process. The solution is:
\begin{equation}
\label{eq:88a}
\xi_{OU}(t+h)=\xi_{OU}(t){\rm e}^{-h/\tau} + H_h(t)
\end{equation}
where we have introduced the random process $H_h(t)$ as:
\begin{equation}
\label{eq:88b}
H_h(t) = \tau^{-1}{\rm e}^{-(t+h)/\tau}
\int_t^{t+h}\D s\; \xi_w(s) {\rm e}^{s/\tau}
\end{equation}
Using the white noise properties, eq.(\ref{eq:35}), it is an easy exercise
to prove that $H_h(t)$ is a Gaussian process of mean value  zero:
\begin{equation}
\label{eq:89}
\langle H_h(t) \rangle = 0
\end{equation}
and correlations:
\begin{equation}
\label{eq:91}
\langle H_h(t) H_h(t') \rangle = \left\{ \begin{array}{lcl}
(2\tau)^{-1} \left[ {\rm e}^{-|t-t'|/\tau}- {\rm e}^{(|t-t'|-2h)/\tau}\right] 
& & {\rm if~}|t-t'|\le h \\
0 & & {\rm if~}|t-t'| > h \end{array} \right.
\end{equation}
The important part is to realize that, for the times $t_i=ih$ that
appear in the recurrence relation eq.(\ref{eq:88a}), the correlations are:
\begin{equation}
\label{eq:92}
\langle H_h(t_i) H_h(t_j) \rangle = (2\tau)^{-1} \left[1-
{\rm e}^{-2h/\tau}\right]\delta_{ij}
\end{equation}
If we introduced a set of independent Gaussian variables $u(t)$ of
zero mean and variance unity, the process $H_h(t)$ can be generated as:
\begin{equation}
\label{eq:93}
H_h(t)=\sqrt{(1-{\rm e}^{-2h/\tau})/(2\tau)} u(t)
\end{equation}
And the final recurrence relation to generate trajectories
of the Ornstein--Uhlenbeck noise is:
\begin{equation}
\label{eq:94}
\left\{
\begin{array}{rcl}
\xi_{OU}(0) & = & \sqrt{(2\tau)^{-1}} u(0) \\
\xi_{OU}(t+h) & = & \xi_{OU}(t){\rm e}^{-h/\tau} +\sqrt{(1-{\rm e}^{-2h/\tau})/ 
(2\tau)} u(t+h)
\end{array}\right.
\end{equation}

Let us consider now an SDE with OU noise. Let us start again by the
simplest example:
\begin{equation}
\label{eq:94-1}
\frac{dx(t)}{dt} = f(t)+\xi_{OU}(t)
\end{equation}
We use an integral algorithm:
\begin{equation}
\label{eq:94-2}
x(t+h)=x(t)+\int_t^{t+h} ds\;f(s) + \int_t^{t+h}ds\;\xi_{OU}(s) \equiv
x(t)+f_h(t)+g_h(t)
\end{equation}
The stochastic contribution $g_h(t)$ is a Gaussian process characterised
by the following mean value and correlations:
\begin{equation}
\label{eq:94-3}
\begin{array}{rcl}
\langle g_h(t) \rangle & = & 0 \\
\langle g_h(t)g_h(t') \rangle & = & \tau 
\left[\cosh\left(\frac{h}{\tau}\right)
-1\right]\exp\left(-\frac{|t-t'|}{\tau}\right) 
\end{array}
\end{equation}
(valid for all the times $t$, $t'$ appearing in the recursion relation 
eq.(\ref{eq:94-2}).) To order $O[h^2]$, the correlations become:
\begin{equation}
\label{eq:94-4}
\langle g_h(t)g_h(t') \rangle = \frac{h^2}{2\tau}
\exp\left(-\frac{|t-t'|}{\tau}\right)  + O[h^3]
\end{equation}
from where it follows that the 
process $g_h(t)$ is nothing but $h$ times an Ornstein--Uhlenbeck process:
$g_h(t)= h\xi_{OU}(t)$.
Summarizing, to order $h^2$ the algorithm to integrate numerically 
the SDE eq.(\ref{eq:94-1}) is:
\begin{equation}
\label{eq:94-5}
x(t+h) = x(t) + f_h(t) + h \xi_{OU}(t) + O[h^2]
\end{equation}
where $\xi_{OU}(t)$ is generated by the use of eqs. (\ref{eq:94}).

If one needs more precision in the stochastic part, 
one step further in the integration of the equation can be achieved by
generating {\sl exactly} the process $g_h(t)$ \cite{erm}. We define the process: 
\begin{equation}
\label{eq:94-7}
G(t) = \int_0^t ds\; \xi_{OU}(s)
\end{equation}
in terms of which:
\begin{equation}
\label{eq:94-8}
g_h(t) = G(t+h)-G(t)
\end{equation}
Since $dG(t)/dt= \xi_{OU}(t)$ and $\xi_{OU}(t)$ satisfies the differential
equation eq.(\ref{eq:87}) we can write down the following equation for $G(t)$:
\begin{equation}
\label{eq:94-9}
\begin{array}{rcl}
\frac{d^2G(t)}{dt^2}+\frac{1}{\tau}\frac{dG(t)}{dt} & = & \frac{1}{\tau}
\xi_w(t) \\
G(0) & = & 0\\
\frac{dG(t)}{dt}|_{t=0} & = & \xi_{OU}(0) \equiv \xi_0
\end{array}
\end{equation}
whose solution is:
\begin{equation}
\label{eq:94-10}
G(t)=\tau\xi_0-\tau\xi_0{\rm e}^{-t/\tau} + \int_0^t ds\; \xi_w(s)
-{\rm e}^{-t/\tau}\int_0^tds\; {\rm e}^{s/\tau}\xi_w(s)
\end{equation}
From where it follows the recursion relation:
\begin{equation}
\label{eq:94-12}
g_h(t+h) = p g_h(t) - p f_1(t) +f_1(t+h)-f_2(t)+f_2(t+h)
\end{equation}
The initial condition is that $g_h(0)$ is a Gaussian variable of
mean and variance given by equation (\ref{eq:94-3}) for $t=t'=0$. This can be
written as:
\be
\label{eq:94-12b}
g_h(0)=\sqrt{\tau\left[ \cosh\left(\frac{h}{\tau}\right)-1\right]} u
\ee
where $u$ is a Gaussian random number of zero mean and unit variance.
In equation (\ref{eq:94-12}) we have introduced the following definitions:
\begin{equation}
\label{eq:94-11}
\begin{array}{rcl}
p & = & {\rm e}^{-h/\tau} \\
f_1(t) & = & \int_t^{t+h} ds\; \xi_w(s) \\
f_2(t) & = & -p{\rm e}^{-t/\tau}\int_t^{t+h}ds\; {\rm e}^{s/\tau}\xi_w(s)
\end{array}
\end{equation}
The processes $f_1(t)$ and $f_2(t)$ are correlated Gaussian processes,
whose properties, for the times $t_i=ih$ appearing in the recursion
relations, are given by:
\begin{equation}
\label{eq:94-12a}
\begin{array}{rcl}
\langle f_1(t_i) \rangle & = & 0 \\
\langle f_2(t_i) \rangle & = & 0 \\
\langle f_1(t_i)f_1(t_j) \rangle & = & h \delta_{ij} \\
\langle f_2(t_i)f_2(t_j) \rangle & = & \frac{\tau(1-p^2)}{2} \delta_{ij} \\
\langle f_1(t_i)f_2(t_j) \rangle & = & -\tau(1-p) \delta_{ij} 
\end{array}
\end{equation}
It is possible to generate the processes $f_1(t)$ and $f_2(t)$ satisfying
the above correlations by writing them in terms of
two sets of independent Gaussian
distributed random numbers of zero mean and unit variance, $u_1(t)$, $u_2(t)$:
\begin{equation}
\label{eq:94-13}
\begin{array}{rcl}
f_1(t) & = & \alpha_1 u_1(t) \\
f_2(t) & = & \beta_1 u_1(t) + \beta_2 u_2(t) \\
\end{array}
\end{equation}
where the constants $\alpha_1$, $\beta_1$ and $\beta_2$ are chosen in order
to satisfy the correlations (\ref{eq:94-12a}):
\begin{equation}
\label{eq:94-14}
\begin{array}{rcl}
\alpha_1 & = & \sqrt{h} \\
\beta_1 & = & -\frac{\tau(1-p)}{\sqrt{h}} \\
\beta_2 & = & \sqrt{\frac{\tau(1-p)}{2}\left[1-\frac{2\tau}{h} +
p\left(1+\frac{2\tau}{h}\right) \right]}
\end{array}
\end{equation}
In summary, the process $g_h(t)$ is generated by the recursion relation
(\ref{eq:94-12}) with the initial condition (\ref{eq:94-12b}) and the 
processes $f_1(t)$ and $f_2(t)$ obtained from the relations (\ref{eq:94-13}).

We now consider a more general equation:
\begin{equation}
\label{eq:94-17}
\frac{dx}{dt} = q(t,x) + g(t,x)\xi_{OU}(t)
\end{equation}
We start again by an integral recursion relation:
\begin{equation}
\label{eq:94-18}
x(t+h) = x(t) + \int_t^{t+h} ds\;q(s,x(s)) + \int_t^{t+h} ds\; g(s,x(s))
\xi_{OU}(s)
\end{equation}
By Taylor expanding functions $q$ and $g$ one can verify that at lowest 
order:
\begin{equation}
\label{eq:94-19}
x(t+h) = x(t) + h q(t,x(t)) + g_h(t) g(t,x(t)) + O[h^2]
\end{equation}
where $g_h(t)$ is the process introduced before. As explained,
$g_h(t)$ can be generated exactly. 
Alternatively, and in order to make the resulting
algorithm somewhat simpler, one can replace
$g_h(t)$ by $h \xi_{OU}(t) + O[h^2]$ without altering the order of
convergence of the algorithm.  However, these algorithms suffer from 
the fact that they do not reproduce adequately the white--noise
Milshtein method. This is to say: always the integration step $h$ 
has to be kept 
smaller than the correlation time $\tau$. If one is interested in the limit
of small values for the correlation time $\tau$ (in particular, if one
wants to consider the white noise limit $\tau \to 0$), it is better to
turn to the Runge--Kutta methods which smoothly extrapolate to the
Milshtein method without requiring an extremely small time step.

\subsubsection{ Runge--Kutta type methods}

We focus again in the SDE with white noise:
\begin{equation}
\label{eq:94-20}
\dot x(t) = q(t,x) + g(t,x)\xi_w(t)
\end{equation}
We will develop now a method similar to the second--order Runge--Kutta (RK)
method for solving numerically
ordinary differential equations. As we said before, the particular
features of the white--noise process prevent us from simply taking
the standard R-K methods and we need to develop new ones. 

Let us recall briefly how a Runge--Kutta method works
for an ordinary differential equation: 
\begin{equation}
\label{eq:95}
\frac{dx(t)}{dt} = q(t,x)
\end{equation}
Euler method:
\begin{equation}
\label{eq:96}
x(t+h)=x(t)+hq(t,x(t))+O[h^2]
\end{equation}
can be modified as:
\begin{equation}
\label{eq:97}
x(t+h)=x(t)+\frac{h}{2}[ q(t,x(t))+ q(t+h,x(t+h))] 
\end{equation}
Of course, this is now an implicit equation for $x(t+h)$ which appears
on both sides of the equation. RK methods replace $x(t+h)$ on the
right hand side by the predictor given by the Euler method, eq.(\ref{eq:96})
to obtain an algorithm of order $O[h^3]$: 
\begin{equation}
\label{eq:98}
x(t+h)=x(t)+\frac{h}{2}[ q(t,x(t))+ q(t+h,x(t)+hq(t,x(t)))]+O[h^3]
\end{equation}
This is usually written as:
\begin{equation}
\label{eq:99}
\begin{array}{rcl}
k & = & hq(t,x(t)) \\
x(t+h) & = & x(t)+\frac{h}{2}[ q(t,x(t))+ q(t+h,x(t)+k)]
\end{array}
\end{equation}
The same idea can be applied to the SDE (\ref{eq:94-20}). Let us modify Euler's
method, eq.(\ref{eq:79}) (which we know is a bad approximation 
in the case of multiplicative noise anyway) to:
\begin{equation}
\label{eq:100}
\begin{array}{rcl}
x(t+h) & = & x(t)+\frac{h}{2}[q(t,x(t))+q(t+h,x(t+h)]+\\
& & \frac{h^{1/2}u(t)}{2} [g(t,x(t))+ g(t+h,x(t+h))] 
\end{array}
\end{equation}
And now replace $x(t+h)$ on the right--hand--side by the predictor of
the Euler method, eq.(\ref{eq:79}) again. The resulting algorithm:
\begin{equation}
\label{eq:101}
\begin{array}{rcl}
k & = & hq(t,x(t)) \\
l & = & h^{1/2} u(t) g(t,x(t)) \\
x(t+h) & = & x(t)+\frac{h}{2}[ q(t,x(t))+ q(t+h,x(t)+l+k)] +\\
&  &  \frac{h^{1/2}u(t)}{2} [g(t,x(t))+g(t+h,x(t)+k+l)]
\end{array}
\end{equation}
is known as the Heun method \cite{gard}. 
To study the order of convergence of this 
method one can Taylor expand functions $q(t,x)$ and $g(t,x)$ to see
that one reproduces the stochastic Milshtein algorithm up to order $h$.
Therefore, from the stochastic point of view, the Heun method is of 
no advantage with respect to the Milshtein method. 
The real advantage of the Heun method is that it treats better the
deterministic part (the convergence of the deterministic part is of
order $h^3$) and, as a consequence, avoids 
some instabilities typical of the Euler method.

Similar ideas can be applied to the integration of the SDE (\ref{eq:94-17}) 
with colored noise (\ref{eq:41b}). It can be easily shown that the
RK type algorithm:
\begin{equation}
\label{eq:99b}
\begin{array}{rcl}
k & = & hq(t,x(t)) \\
l & = & g_h(t)g(t,x(t)) \\
x(t+h) & = & x(t)+\frac{h}{2}[ q(t,x(t))+ q(t+h,x(t)+k+l)] +\\
& & \frac{g_h(t)}{2}[g(t,x(t))+g(t+h,x(t)+k+l)]
\end{array}
\end{equation}
correctly reproduces the algorithm (\ref{eq:94-19}). Moreover, when the
stochastic process $g_h(t)$
is generated exactly as explained before, this algorithm
tends smoothly when $\tau \to 0$ to the Milshtein algorithm for white noise
without requiring an arbitrarily small integration step $h$.

\subsubsection{Numerical solution of Partial Stochastic Differential Equations}

We end this section with the basic algorithms for the generation of trajectories
for partial stochastic differential equations (PSDE). We will encounter several
examples in Sects. 5 and 6. In general, one has a field $A(\vec r,t)$,
function of time $t$ and space $\vec r$, that satisfies a PSDE of the 
form:
\be
\label{eq:99c}
\frac{\partial A}{\partial t} = G[A,\vec \nabla A, \nabla^2 A, \dots; 
\xi(\vec r,t)]
\ee
Where $G$ is a given function of the field $A$ and its space derivatives.
For the stochastic field $\xi(\vec r,t)$ usually a white noise
approximation is used, i.e. a Gaussian process of zero mean and
delta--correlated both in time and space:
\be
\label{eq:99d}
\langle \xi(\vec r,t) \xi(\vec r\;',t') \rangle = \delta(\vec r - \vec r\;')
\delta(t-t')
\ee
The numerical solution of (\ref{eq:99c}) usually proceeds as 
follows: one discretizes space $\vec r \to r_i$ by an appropriate
grid of size $\Delta r$. The index $i$ runs over the $N$ lattice sites.
Usually, but not always, one considers a $d$--dimensional regular lattice
of side $L$, such that $N=(L/\Delta r)^d$. 
The elementary volume of a single cell
of this lattice is $(\Delta r)^d$.
Next, one replaces the fields
$A(\vec r,t)$, $\xi(\vec r,t)$ by a discrete set of variables.  
For the field $A(\vec r_i,t)$ we simply replace it by 
$A(\vec r_i,t) \to A_i(t)$. For the white noise, we have to consider the
delta--correlation in space and use the substitution:
\be
\label{eq:99e}
\xi(\vec r_i,t) \to (\Delta r)^{-d/2} \xi_i(t)
\ee
which comes from the relation between the Dirac $\delta(\vec r)$ and
the Kronecker $\delta_i$ functions. In this expression
$\xi_i(t)$ are a discrete set of independent stochastic white 
processes, i.e. Gaussian variables of zero mean and correlations:
\be
\label{eq:99e1}
\langle \xi_i(t) \xi_j(t') \rangle = \delta_{ij}\delta(t-t')
\ee
In the simplest algorithms, 
the field derivatives are replaced by finite differences\footnote{Alternatively,
one could use Fourier methods to compute the Laplacian or other
derivatives of the field; see, for instance, \cite{ss}.}.
For instance: if the $d$--dimensional regular 
lattice is used for the set $\vec r_i$,
the Laplacian $\nabla^2 A(\vec r_i,t)$ can be approximated by the
lattice Laplacian:
\be
\label{eq:99f}
\nabla^2 A(\vec r_i,t) \approx (\Delta r)^{-2}\sum_{j\in n(i)} 
[A_j(t)-A_i(t)]
\ee
Where the sum runs over the set of $2d$ nearest neighbors of site $\vec r_i$.
With these substitutions the PSDE (\ref{eq:99c}) becomes a set of coupled
ordinary differential equations:
\be
\label{eq:99g}
\frac{dA_i(t)}{dt}= G_i(A_1(t),\dots, A_N(t); \xi_1(t), \dots, \xi_N(t)), 
\hspace{1.0cm}i=1,\dots,N
\ee
In most occasions, these equations are of the generalized Langevin form:
\be
\label{eq:99h}
\frac{dA_i(t)}{dt}= q_i([A]) +\sum_j g_{ij}([A])\xi_j(t)
\ee
$[A]$ denotes the set $[A]=(A_1,\dots,A_N)$ and $q_i([A])$, $g_{ij}([A])$
are given functions, maybe depending explicitly on time. 
The numerical integration of (\ref{eq:99h}) proceeds, as in the single
variable case, by developing integral algorithms \cite{GSH,kloeden}.
It turns out, however, that in the most general case it is very difficult
to accurately generate the necessary stochastic variables appearing in the
algorithms. That is the reason why one rarely goes beyond Euler's
modification of Milshtein's method (eq. (\ref{eq:80}), which reads 
\cite{rpshm}:
\be
\label{eq:99i2}
A_i(t+h) = A_i(t) + h^{1/2} \sum_j g_{ij}([A(t)]) u_j(t) +
h\left[
q_i([A(t)]) + \frac{1}{2} \sum_{j,k} g_{jk}([A(t)])
\frac{\partial g_{ik}([A(t)])}{\partial A_j(t)} \right]
\ee
$u_i(t)$ are a set of independent random variables defined
for the time $0$, $h$, $2h$ of the recurrence relation, with zero mean and
variance one:
\be
\label{eq:99j}
\begin{array}{rcl}
\langle u_i(t) \rangle  = 0 & , & \langle u_i(t)u_j(t) \rangle = \delta_{ij} \\
\langle u_i(t) u_j(t') \rangle  = 0 & ,& t \ne t'
\end{array}
\ee
which can be generated by the Box--Muller--Wiener or an alternative
algorithm. We stress the fact that the functions 
$g_{ij}([A])$ are of order $(\Delta r)^{-d/2}$
due to the substitution (\ref{eq:99e}). For small $\Delta r$,
this usually demands a small time--integration step for the convergence of the
solution \cite{RED,CTG,LSHMT}.

An important case in which
one can use straightforward generalizations of the Milshtein and Heun
methods is that of {\sl diagonal}\hspace{0.1cm} noise, i.e. one in which 
the noise term does not couple different field variables, namely:
\be
\label{eq:99h2}
\frac{dA_i(t)}{dt} = q_i([A]) +g_i(A_i)\xi_i(t)
\ee
In this case, the Milshtein method reads:
\be
\label{eq:99i}
A_i(t+h) = A_i(t) + g_i(A_i(t)) h^{1/2}u_i(t) +
h\left[ q_i([A(t)]) + \frac{1}{2} g_i(A_i(t)){g'}_i(A_i(t)) u_i(t)^2\right]
\ee
The Heun method is also
easily applied in the case of diagonal noise:
\begin{equation}
\label{eq:101a}
\begin{array}{rcl}
k_i & = & hq_i([A(t)]) \\
l_i & = & h^{1/2} u_i(t) g_i([A(t)]) \\
A_i(t+h) & = & A_i(t)+\frac{h}{2}[ q_i([A(t)])+ q_i([A(t)+l+k])] +\\
&  &  \frac{h^{1/2}u_i(t)}{2} [g_i(A_i(t))+g_i(A_i(t)+k_i+l_i)]
\end{array}
\end{equation}

\subsection{A trivial (?) example: The linear equation with 
multiplicative noise}

\begin{figure}[h]
\begin{center}
\def\epsfsize#1#2{0.82\textwidth}
\leavevmode
\epsffile{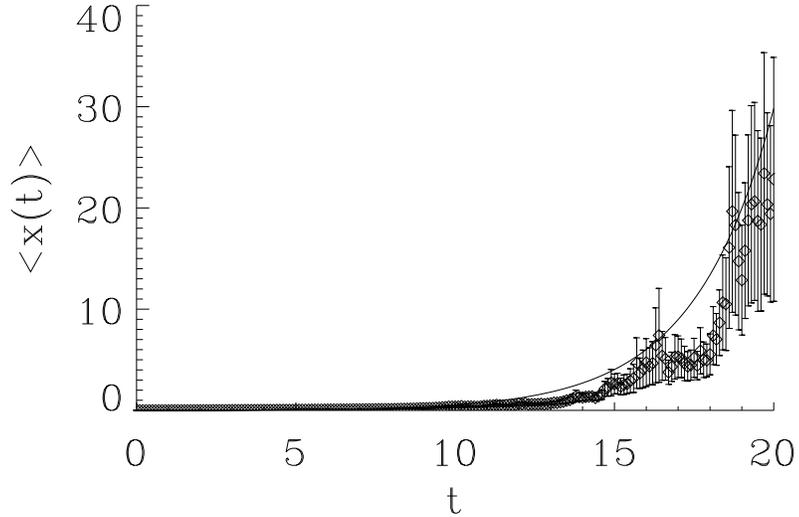}
\end{center}
\caption{\label{fig:10} 
Evolution of the mean value for the linear equation (\ref{eq:102})
for $a=-0.1$, $\sigma=1$, $x_0=0.01$. The solid line is the exact result 
eq. (\ref{eq:106}) and the symbols and 
error bars come from  a numerical solution of
the equation. Notice that although the numerical solution follow closely
the analytical result, the error bars greatly increase with time. This is
a clear signature of very strong fluctuations, see the main text.}
\end{figure}

In previous sections, we have shown how it is possible to generate
trajectories from a given stochastic differential equation and how
these trajectories can help us to perform the necessary averages. 
Of course, if one has computed the probability density function
(by solving, may be numerically, the Fokker--Planck equation \cite{Risken}) one
can also perform the same averages. However, there are cases in which
much information 
can be learnt about the solution of an stochastic differential
equation by looking at individual trajectories. Information which is
not so easy to extract from the probability density function.
To illustrate
this point we will analyse in some detail the apparently simple SDE:
\begin{equation}
\label{eq:102}
\frac{dx}{dt} = [a+\sigma \xi_w(t)]x(t)
\end{equation}
with $\xi_w(t)$ a white noise process with correlation given by 
eqs.(\ref{eq:35}).
In fact, this linear stochastic differential equation is so simple that it
can be solved explicitly:
\begin{equation}
\label{eq:103}
x(t) = x(0){\rm e}^{at+\sigma W(t)}
\end{equation}
where $W(t)$ is the Wiener process.
From the explicit solution and using the known properties of the Wiener
process we can compute, for instance, the evolution of the mean value of $x(t)$:
\begin{equation}
\label{eq:106}
\langle x(t) \rangle = \langle x(0){\rm e}^{at+\sigma W(t)} \rangle =  
x(0){\rm e}^{(a+\sigma^2/2)t}
\end{equation}
where we have used the result 
\begin{equation}
\label{eq:107}
\langle {\rm e}^z \rangle = {\rm e}^{\langle z^2 \rangle/2}
\end{equation}
valid for a Gaussian variable $z$ of zero mean. 
From equation (\ref{eq:106}) it follows
that the mean value of $x(t)$ grows towards infinity for $a > -\sigma^2/2$ and
decays to zero for $a < -\sigma^2/2$. Fig. (\ref{fig:10}) shows this 
exponential behaviour of the mean value 
together with some results obtained by
numerically integrating the SDE (\ref{eq:102}).
This exponential growth is at variance to what happens
in the deterministic case, $\sigma=0$, for which $x(t)$ decays to $0$ for
$a<0$ and grows for $a>0$. One would say, in this case, that there has
been a shift in the critical value of the parameter $a$ to induce a
transition from the state $x=0$ to the state $x \ne 0$,
representing perhaps the transition from a disordered to an ordered state.
Of course, it is of no concern the fact that $\langle x(t) \rangle$ tends
to infinity. In a real system, they will always be saturating terms that
will stop growth of $x(t)$ and will saturate $\langle x(t) \rangle$ to
a finite value. For instance, a realistic equation could be one with
a saturating cubic term:
\begin{equation}
\label{eq:108}
\frac{dx}{dt} = [a+\sigma \xi_w(t)]x(t)-x(t)^3
\end{equation}
The conclusions of our simple linear analysis would then say that the
stationary state of the non--linear equation (\ref{eq:108}) is $0$ for
$a < a_c$ and non--zero for $a > a_c$, being $a_c=-\sigma^2/2$ a ``critical"
value for the control parameter $a$. Obvious conclusion ... or is it? 
Well, let us have a closer look.

\begin{figure}[h]
\begin{center}
\def\epsfsize#1#2{0.82\textwidth}
\leavevmode
\epsffile{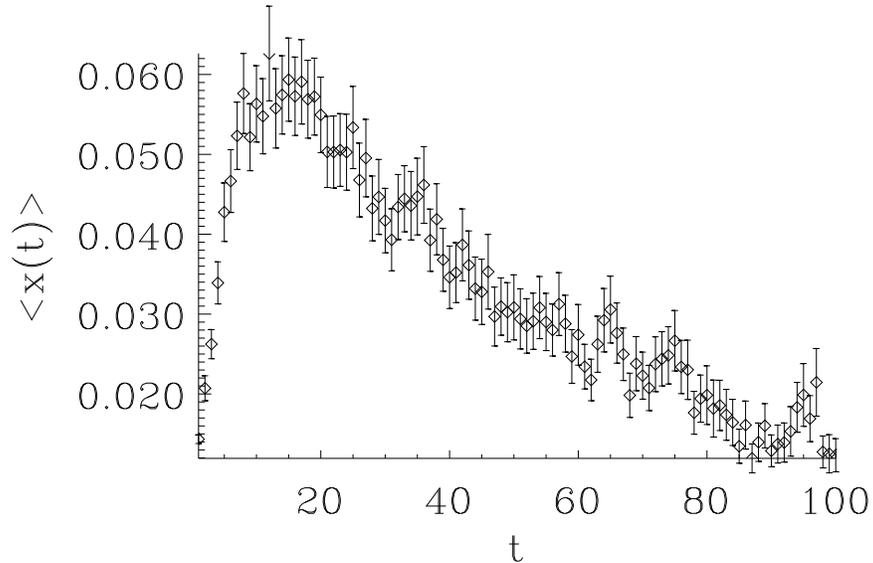}
\end{center}
\caption{\label{fig:12}
Evolution of the mean value for the non--linear equation (\ref{eq:108})
with the same values of the parameters than in Fig. (\ref{fig:10}). The 
results, with the error bars, come from a numerical simulation of 
eq.(\ref{eq:108}).}
\end{figure}
A first signature that something is missing in the previous analysis is that
we could repeat it as well for the evolution of the $n$--th moment
with the result:
\begin{equation}
\label{eq:109}
\langle x(t)^n \rangle = \left \langle (x(0){\rm e}^{at+W(t)} )^n
\right \rangle =  x(0)^n{\rm e}^{n(a+n\sigma^2/2)t}
\end{equation}
We see that the $n$--th moment of the linear equation (\ref{eq:102})
diverges at a critical value $a_{c,n} = -n\sigma^2/2$ that depends on the order
$n$ of the moment. 
By repeating the arguments sketched above, one would conclude that
the asymptotic value of the $n$--th moment of the non--linear 
equation (\ref{eq:108})
would change from zero to a non--zero value at $a_{c,n}$ and hence, the 
location of the putative transition from order to disorder would 
depend on $n$, which does not make much sense.

The solution of the associated Fokker--Planck equation associated to
the non--linear 
Langevin equation (\ref{eq:108}) and  
the complete calculation of the moments $\langle x(t)^n \rangle$ 
is given in  \cite{SB,graham} with the
conclusion that all moments tend to $0$ for $a < 0$. This is illustrated in
Fig. (\ref{fig:12}) in which we plot the time evolution for the
mean value $\langle x(t)\rangle$ for a situation in which the 
linear equation explodes. Since the mathematics
used  in this proof do not provide much physical intuition, it is instructive to rederive the
same result by studying the behavior of individual trajectories. This
will also shed light on the relation between the behavior of the moments
in the linear and the non--linear equations.

\begin{figure}
\begin{center}
\def\epsfsize#1#2{0.82\textwidth}
\leavevmode
\epsffile{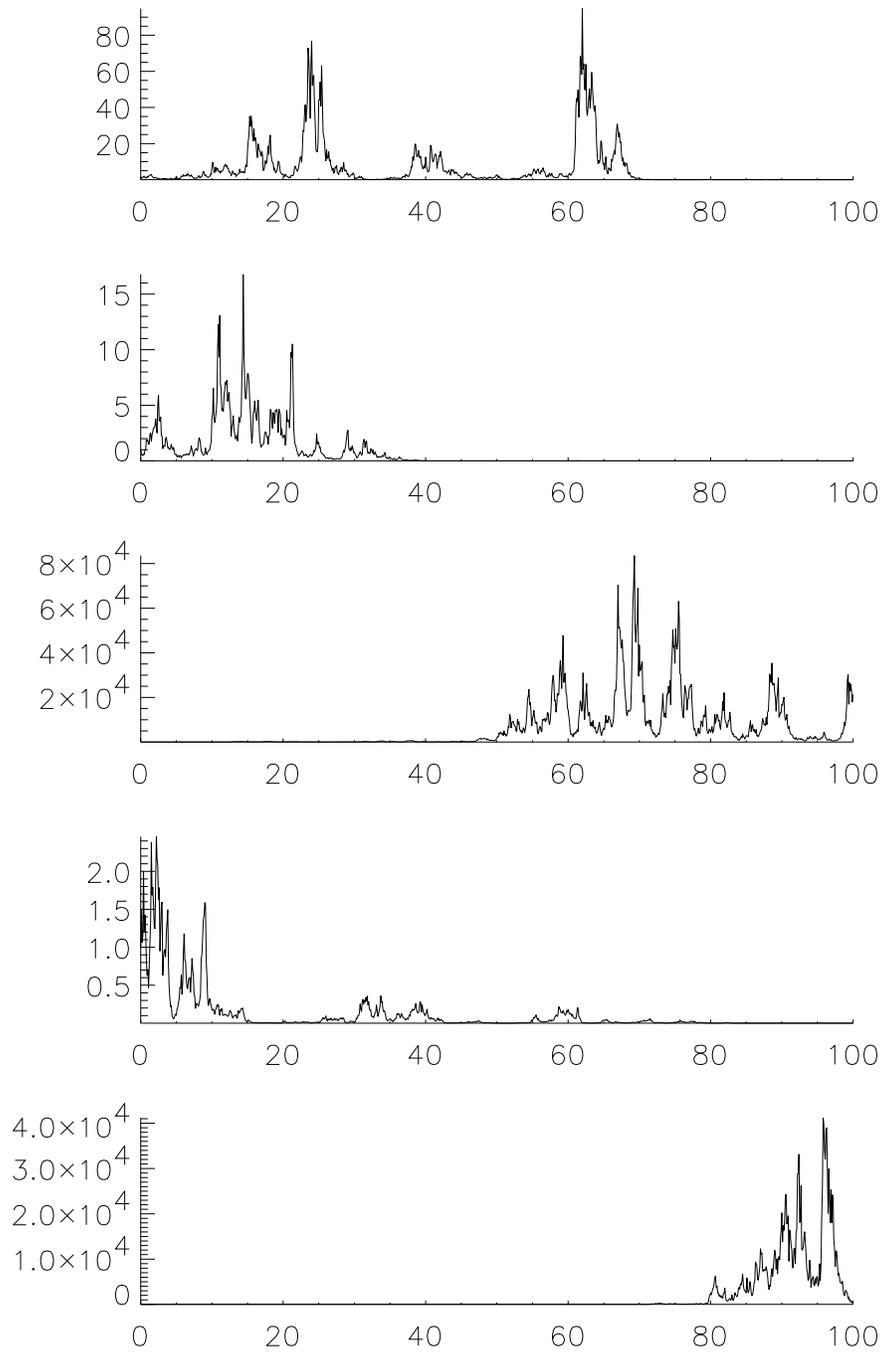}
\end{center}
\caption{\label{fig:9}
Series of five trajectories $x(t)$
for the linear multiplicative problem
to show that there are large amplitude fluctuations.  
Same parameters than in Fig. (\ref{fig:10}).}
\end{figure}

If we look at the solution, eq.(\ref{eq:103}), of the linear equation,
we can say that the
deterministic contribution $at$ will always dominate for large time $t$ 
over the stochastic contribution, $W(t)$, which, according to eq.(\ref{eq:32})
is of order $t^{1/2}$. Hence, for large $t$, and for $a<0$ every
trajectory will go to zero, and consequently $\langle x(t)^n \rangle =0$,
in contradiction with previous results, in particular
equation (\ref{eq:109}). 
If fact, the statement that $at$ dominates over $W(t)$ and hence
$x(t)$ decays to zero is not very 
precise since $at$ is a number and $W(t)$ is a stochastic process. 
A precise statement is  that the probability that $x(t)$ decays
to zero (i.e. that it takes values less than any number $\epsilon$) tends
to one as time tends to infinity: 
\be
\forall \epsilon, \lim_{t \to \infty} {\rm Probability}(x(t) < \epsilon) = 1
\ee
Remember that $W(t)$ is a stochastic
Gaussian variable and, hence, it has a finite probability
that it overcomes the deterministic contribution $at$ at {\sl any} time.
What the above relation tells us is that,
as time tends to infinity, this probability tends to zero, and we can
say, properly speaking, that every trajectory will tend to zero with
probability one. However, for any finite time there is always
a finite (however small) probability that the stochastic term $W(t)$
overcomes the deterministic contribution $at$ by any large amount. 
This is illustrated in Fig. (\ref{fig:9}) in which we plot some typical
trajectories for the linear equation. In these figures it is important
to look at the vertical scale to notice that, although every trajectory 
decays to zero as time increases, there are very large indeed fluctuations
for any given trajectory.

In brief, for the linear equation (\ref{eq:102})
there are unlikely trajectories (with decreasing probability as
time increases) that become arbitrarily large. It is the contribution
of those small probability, large amplitude trajectories, which 
make the moments diverge. 

\begin{figure}
\begin{center}
\def\epsfsize#1#2{0.82\textwidth}
\leavevmode
\epsffile{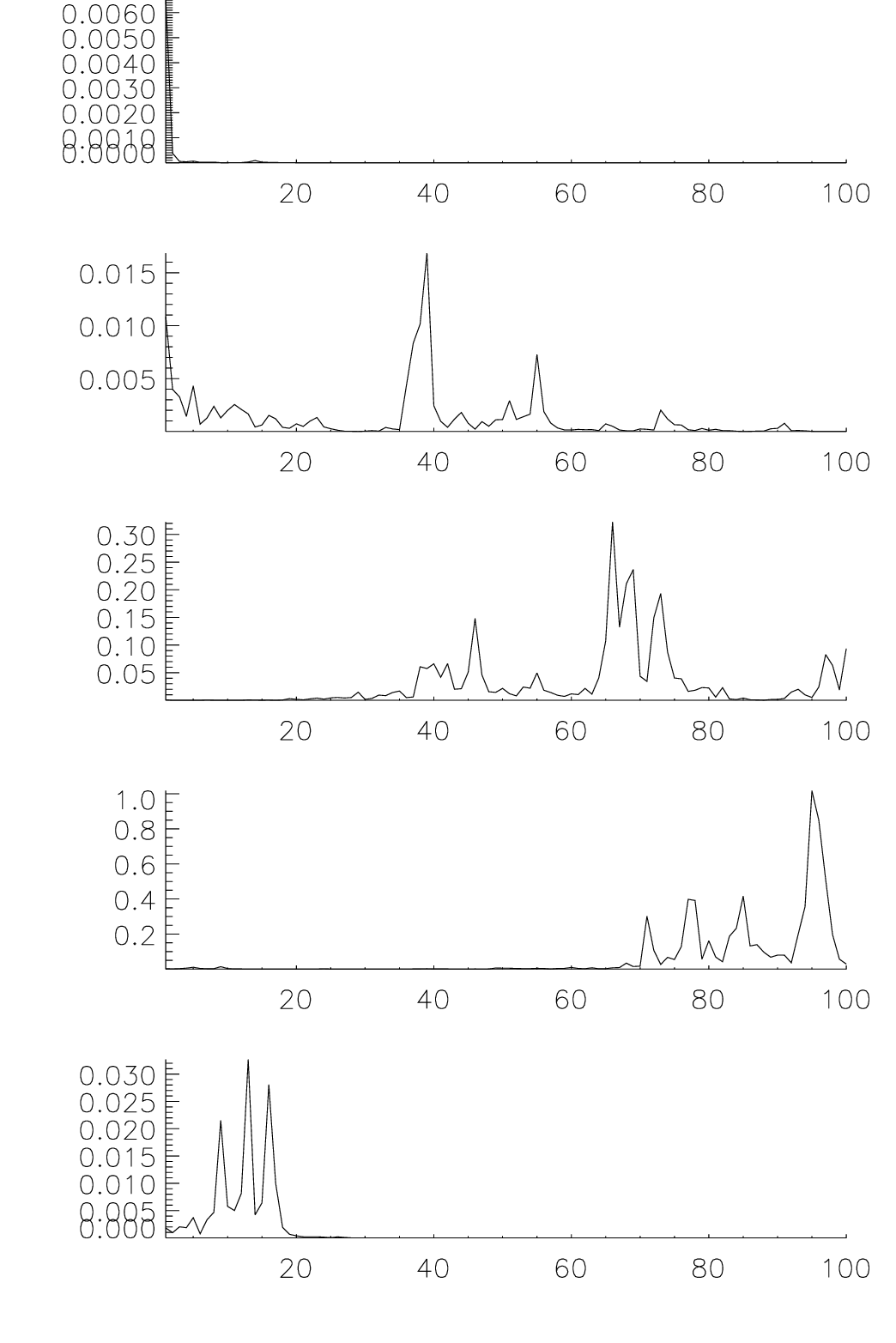}
\end{center}
\caption{\label{fig:11}
Series of five trajectories $x(t)$ 
for the non--linear multiplicative problem eq.(\ref{eq:108})
to show that the large amplitude fluctuations are suppressed by the 
non--linear terms for each trajectory. Same parameters than in Fig.
(\ref{fig:10}.)}
\end{figure}

Now we can look at the non--linear problem with other eyes. The non--linear
term will suppress those large fluctuations in the trajectories {\sl
one by one}, not just on the average \cite{kra95}. We conclude, then, that the
mean value of any moment $\langle x(t)^n \rangle$ will tend to zero 
for $a < 0$, in agreement, of course, with the exact result in \cite{SB,graham}.
We can say that the presence of the multiplicative noise
makes no shift in the transition from ordered to disordered states.
Fig. (\ref{fig:11}) shows some trajectories in the non--linear case
to see that, effectively, there are no large fluctuations as one can see
comparing the vertical scale of Figs. (2.6) and (2.7).
In Sect. 6 we will analyze other situations in which noise can actually
induce an order/disorder transition. 


\section{Transient stochastic dynamics}
\setcounter{equation}{0}
\setcounter{figure}{0}
In many physical situations noise causes only small fluctuations around a
reference state. Of course, noise effects are more interesting when this is not the
case, as for example when there is a mechanism of noise amplification. A typical
situation of this type is the decay of an unstable state \cite{Suzuki81}: a given system 
is forced to switch 
from one state to another by changing a control parameter. After the change of
 the parameter,
the system, previously in a stable state, finds itself in a state which is an unstable fixed point of the
deterministic dynamics, and it is driven away
from it by fluctuations. Noise starts the decay process and fluctuations are amplified 
during the transient dynamics. Examples of such situations include spinodal decomposition 
in the dynamics of phase transitions \cite{Gunton83} and the process of laser switch-on
\cite{ArecchiTorino,SanMiguelSPIE}. The latter essentially consists in the amplification of spontaneous emission noise. We will discuss
laser switch-on in the second part of this section after some general methodology is presented.
A related interesting situation, which we will not discuss here, occurs when a system
is periodically swept through an unstable state. The periodic amplification
of fluctuations often leads to important consequences 
\cite{SanMiguel89,DePasquale84,Hohenberg91}.

A standard way of characterizing transient fluctuations of the stochastic
process $x(t)$
is by the time dependent moments
$\langle x^n(t) \rangle$. This characterization
 gives the statistics of
the random variable $x$ at a fixed given time $t$. An
alternative characterization is given by considering $t$ as
a function of $x$. One then looks for the statistics of the
random variable $t$ at which the process reaches for the
first time a fixed given value $x$. The distribution of
such times is the First Passage Time Distribution
(FPTD). This alternative characterization emphasizes the
role of the individual realizations of the process $x(t)$.
It is particularly well suited to answer questions related
to time scales of evolution. For example, the lifetime of
a given state can be defined as the Mean FPT (MFPT) to
leave the vicinity of that state. The value of the associated variance
of the PTD identifies whether that lifetime is a meaningful
quantity.
We will follow here the approach of first finding an
approximation for the individual stochastic paths of the process, and then
extracting the PT statistics from this description of the
paths. In many cases the individual stochastic paths are easily approximated 
in some early regime of evolution, from which FPT statistics can be calculated.
It is also often the case that statistical properties at a late stage of
evolution can be calculated by some simple transformation of early time statistics.
For example, we will see that there exists a linear relation between the random switch-on 
times of a laser calculated in a linear regime and the random heights of laser pulses
which occur well in the nonlinear regime of approach to the steady state.

\subsection{Relaxational dynamics at a pitchfork bifurcation}
The normal form of the dynamical equation associated with
a pitchfork bifurcation is 
\begin{equation}
d_t x=ax+bx^3 -cx^5 +\sqrt{\epsilon} \xi(t) = - {\partial V \over 
\partial x} + \sqrt{\epsilon} \xi (t)   ,\hskip1truecm \ c>0 
\end{equation}
where $x(t)$ follows relaxational gradient dynamics in the potential $V$ (see Section 4) and
where we have added a noise term $\xi(t)$ with noise
intensity $\epsilon$. The noise $\xi(t)$ is assumed to be
Gaussian white noise of zero mean and correlation 

\begin{equation}
\langle\xi(t)\xi(t^{\prime})\rangle=2\delta(t-t^{\prime}).
\end{equation}

\begin{figure}
\begin{center}
\def\epsfsize#1#2{0.60\textwidth}
\leavevmode
\epsffile{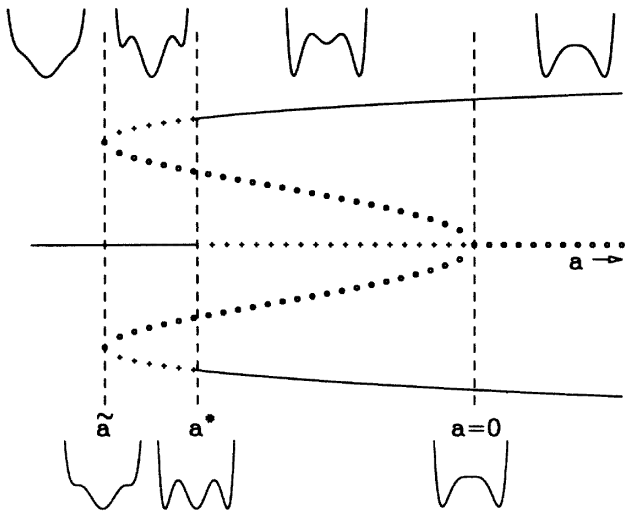}
\end{center}
\caption{Bifurcation diagram for the subcritical bifurcation showing stationary
solutions of $x$
vs. the control parameter $a$. The form of the potential $V$ for different values
of $a$ is
also shown. Solid lines denote stable states. Dots and crosses stand,
respectively, for unstable
and metastable states, where $\tilde{a}=-{b^2\over 4c}$, $a^* = -{3b^2\over16c}$.}
\label{bifsubcritico}
\end{figure}

\noindent We are here interested in describing the relaxation
triggered by noise from the state $x=0$. In a
{\it supercritical bifurcation} $b<0$ and we consider the
relaxation for $a>0$.
In this case the quintic term in (3.1) is irrelevant and
we can set $c=0$. In a {\it subcritical bifurcation} $b>0$. When $-{3b^2\over16c}
<a<0$, $x=0$ is a metastable state and it becomes unstable for $a>0$. Changing
the control parameter $a$ from negative to positive values there is a
crossover in the relaxation mechanism from relaxation via activation to
relaxation from an unstable state. The bifurcation diagram for the subcritical bifurcation is shown
in Fig. (3.1). We will first consider the relaxation from $x=0$ in the 
supercritical case ($a>0$,$b<0$, $c=0$) and
secondly the relaxation in the critical case $a=0$ of the subcritical 
bifurcation ($b>0$, $c<0$). In the
supercritical case the state $x=0$ is an unstable state and the relaxation process follows,
initially, Gaussian statistics for $x$. However, the state $x=0$ has marginal 
stability in the case of a subcritical bifurcation for 
relaxation with $a=0$, and in this case Gaussian statistics, or equivalently, linear 
theory, does
not hold at any time of  the dynamical evolution. 

A usual approach to the statistical description of (3.1) features the 
Fokker Planck equation (see Sect. 2.3)
for the probability density $P(x,t)$ of the process, (see Fig. (3.2))
\begin{equation}
\partial_{t} P (x, t) = \partial_{x} (\partial_{x} V) 
P + \epsilon \partial^{2}_{x} P \equiv H P (x, t),
\end{equation}
\noindent where $H$ is the Fokker-Planck operator. Equations for the time dependent 
moments of the
process $x(t)$ are easily obtained from (3.3). The calculation of PT statistics is also
standard \cite{kampen,gardiner,Stratonovich67} for a Markov process $x(t)$: the
probability that at time $t$, $x(t)$ is still in an interval
$(x_1,x_2)$ given an initial condition $x_0$ at $t=0$ is known as the
survival probability $F({x_0},t)$. It obeys the
equation
\begin{equation}
\partial_t F({x_0},t)=H^+_D F({x_0},t) 
\end{equation}
\noindent where $H_D^+$ is the adjoint of $H_D$. The latter operator
differs from $H$ in that it avoids the probability of
reentering the interval $(x_1,x_2)$ after having left it. The
FPTD is given by $f({x_0},t)=-\partial_{t}F({x_0},t)$, and as
a consequence the MFPT, $T_{x_0}$, obeys the Dynkin
equation 
\begin{equation}
H_D^+ T_{x_0} =-1.
\end{equation} 
The lifetime $T$ of the state $x=0$ is given by the MFPT for $x^2$ to reach a value
$R_0^2$ starting at $x=0$. From (3.5) we obtain
\begin{equation}
T ={2\over\epsilon}\int_0^{R_0}dx{_1}e^{V(x{_1})/
\epsilon}\int_0^{x_1}e^{V(x{_2})/\epsilon}dx_2. 
\end{equation}
\begin{figure}[h]
\begin{center}
\def\epsfsize#1#2{0.80\textwidth}
\leavevmode
\epsffile{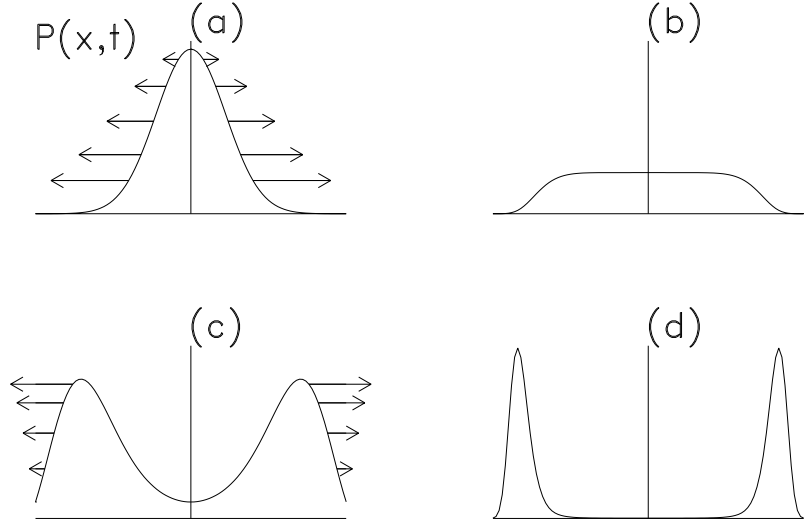}
\end{center}
\caption{Schematic evolution of the probability density $P(x,t)$ for the decay
from $x=0$ in a supercritical bifurcation: An initial distribution centered in
$x=0$ broadens and develops two peaks around the deterministic steady state
solutions $x= \pm (-a/b)^{1/2}$. Time evolution is (a) $\rightarrow$ (b)
$\rightarrow$ (c) $\rightarrow$ (d).}
\end{figure}
This relation is generally valid. In the supercritical case, and for
asymptotically small $\epsilon$, we obtain 
\begin{equation}
T \sim{1\over2a} \ln{a R^2_0\over\epsilon}
-{\psi(1)\over2a}\sim{1 \over 2a} \ln \epsilon^{-1}, 
\end{equation}
where $\psi$ is the digamma function \cite{abra}. The variance of the FPT distribution 
is given by 
\begin{equation}
(\Delta T)^{2}  = {1 \over 4 a^{2}} \psi' (1). 
\end{equation}
We note that while in a deterministic
treatment, $x(t)$ starts to grow exponentially as $x^2 \sim 
e^{2at}$, the lifetime of the state is not simply $a^{-1}$
because such lifetime is determined by fluctuations and it diverges logarithmically with $\epsilon$.
On the other hand, $(\Delta T)^{2}$ turns out to be independent of $\epsilon$ in the limit of small noise
intensity. The same type of calculation for the subcritical bifurcation \cite{Colet91,SanMiguel91}
gives a formula for $T$ which interpolates between the Kramers result for
relaxation from a metastable state, obtained for $a<0$ and
$\vert k \vert \gg 1$, and (3.7) for
relaxation from an unstable state, obtained for $a>0$ and $\vert k \vert \gg 1$.
The parameter $k\equiv a/(b\epsilon)^{1/2}$ measures the distance to the situation of marginality
$a=0$.

As we stressed in the previous section much physical insight can be gained following the 
actual stochastic trajectories $x(t)$ (Fig. (3.3)). 
We will now use this approach to describe the
relaxation process and to reobtain the results quoted above for the distribution of
passage times. A first intuitive idea of an individual trajectory in the supercritical
 case 
distinguishes two dynamical regimes in the relaxation process: There is first 
a linear regime of evolution in which the solution of (3.1) can be written ($b=c=0$) as
\begin{equation}
x(t) = h (t) e^{at}, \quad  h(t) = \sqrt{\epsilon} \int^{t}_{0} dt' e^{- at'} \xi (t'),
\end{equation}
where $h(t)$ is a Gaussian process which plays the role of a stochastic time 
dependent initial 
condition which is exponentially amplified by the deterministic motion.
There is a second time regime in which noise along the path is not important, and the 
trajectory follows 
nonlinear deterministic dynamics from a nonzero initial condition $x(0)$. The deterministic 
solution of (3.1) ($\epsilon=0$) is
\begin{equation}
 x (t) = {x (0) e^{at} \over [1 - {b \over a} x^{2} (0) ( e^{2at} - 
 1) ]^{1/2} }.
\end{equation}
\begin{figure}
\begin{center}
\def\epsfsize#1#2{0.80\textwidth}
\leavevmode
\epsffile{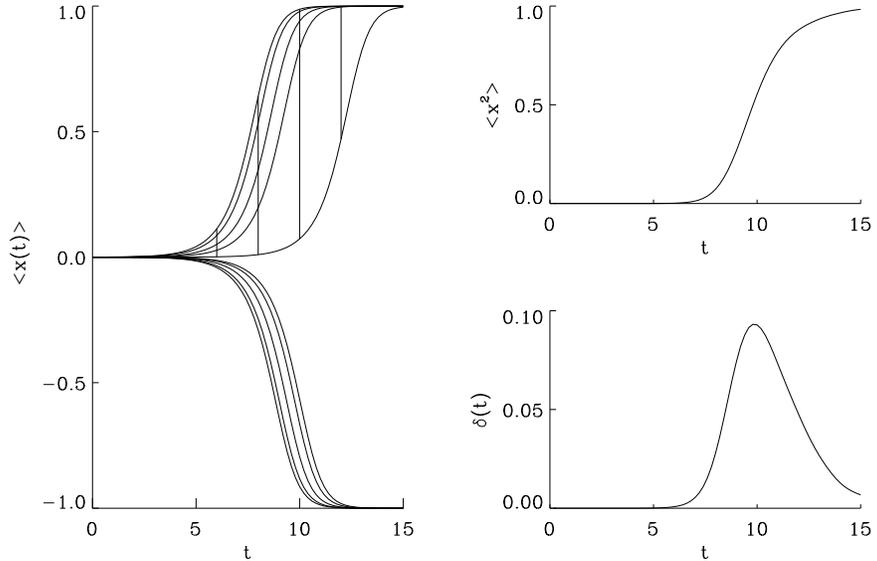}
\end{center}
\caption{A sample of individual trajectories for the decay process obtained 
from a numerical solution of (3.1)
with $x(0)=0$, $a=1$, $b=-1$, $c=0, \epsilon=10^{-8}$.,
and different realizations of the noise.
The vertical bars indicate the
difference between the most separated trajectories at a fixed time.
The evolution of an average trajectory
$\langle x^2(t) \rangle $ and a statistical spread $ \delta(t)
=\langle x^2(t) \rangle -\langle |x(t)| \rangle ^2$
is also shown. The form of $\delta(t)$ is associated with transient anomalous
fluctuations.} 
\end{figure}

A theory for the relaxation process is based on this separation of dynamical
 regimes observed in each trajectory. 
In this theory the complete stochastic evolution is replaced by the nonlinear 
deterministic mapping of the initial Gaussian
process $h(t)$: The stochastic trajectory is approximated by replacing the initial condition
$x(0)$ in (3.10) by $h(t)$ so that $x(t)$, becomes the following functional of the process
$h(t)$: \begin{equation}
 x ( [h (t) ], t) = {h (t) e^{at} \over [1 - {b \over a} h^{2} (t) 
(e^{2at} - 1) ]^{1/2} }.
\end{equation}
Eq. (3.11) implies a dynamical scaling result for the
process $x(t)$, in the sense that $x(t)$ is given by a
time dependent nonlinear transformation of another
stochastic process, namely the Gaussian process $h(t)$. This result for $x(t)$ is also  often
called the quasideterministic theory (QDT) \cite{DePasquale79} and has alternative
presentations and extensions \cite{Suzuki81,Haake81}. The fact that
Eq. (3.11) gives an accurate representation of the
individual trajectories of (3.1) (except for the small fluctuations around
the final steady state $x{_f^2}=-a/b$) justifies the approach. 

In the approximation (3.11) the PT are calculated as random times to reach a given value $R_0$
within the linear stochastic regime. In fact, $T$ sets the upper limit of validity of the linear
approximation. The process $h(t)$ has a second moment
$\langle h^{2} (t) \rangle = {\epsilon \over a} (1 - e^{-2at})$ which saturates to a time independent
value far from criticality, $at \gg 1$. In this limit the process $h(t)$ can be replaced
by a Gaussian random variable $h(\infty)$ with $\langle h(\infty)\rangle=0$ and 
$\langle h^{2} (\infty)\rangle={\epsilon \over a}$. This permits to solve (3.9) for the time $t^*$
at which $R_0$ is reached
\begin{equation}
t^* = {1 \over 2a}  \ln {R^{2}_0 \over h^{2} (\infty)}.
\end{equation}
This result gives the PT statistics as a transformation of the random variable
$h=h(\infty)$. The statistical properties are, for example, completely determined by 
the generating function
\begin{equation}
W (\lambda) = \langle e^{- \lambda t^*}\rangle = \int dh P (h) e^{- \lambda t^* (h)},
\end{equation}
where $P(h)$ is the known Gaussian probability distribution of $h(\infty)$. Moments
of the PT distribution are calculated by differentiation with respect to $\lambda$ at
$\lambda=0$. In this way one recovers (3.7) and (3.8). We note that this calculation assumes a
separation of time scales between $a^{-1}$ and $T$. When this separation fails there is no
regime of linear evolution. The decay process is then dominated by fluctuations and
nonlinearities (see an example in Sect. 5.2).

Time dependent moments of $x(t)$ can also be calculated from the approximated trajectory
(3.11) by averaging over the Gaussian probability distribution $P(h,t)$ of the process $h(t)$:
\begin{equation}
\langle x^{n} (t) \rangle = \int dh P(h, t) x^{n} ([h (t)], t) 
\end{equation}
It turns out that in the limit $e^{at} \gg 1$ there is dynamical scaling in the sense that the 
moments only depend on time through their dependence in the parameter $\tau = {\epsilon \over a} e^{2at}$.
The result for $\langle x^{n} (t) \rangle$, given in terms of hypergeometric functions, can then be
expanded in a power series in $\tau$ \cite{Suzuki81}. For example for the second order moment
one obtains:
\begin{equation}
\langle x^{2}(t)\rangle = \langle x^{2}({\infty})\rangle \sum^{\infty}_{n = 1} (-1)^{n-1} (2n-1)!! \tau^{n}.
\end{equation}
This result indicates that our approximation to the trajectory corresponds to a summation
of the perturbative series in the noise intensity $\epsilon$ which diverges term by term
with time. It also gives an interpretation of the MFPT as the time $t$ for which the scaling
parameter $\tau \sim 1$. For times of the order of the MFPT the amplification of initial 
fluctuations gives rise to transient anomalous fluctuations of order
$\epsilon^0$  as compared with the initial or final
fluctuations of order $\epsilon$ as shown in Fig. (3.3): At early and late times
of the decay process different trajectories are close to each other at a fixed
time, resulting in a small dispersion as measured by 
$\delta(t)$.
However, at intermediate times the trajectories are largely separated as a
consequence of the amplification of initial fluctuations and $\delta(t)$ 
shows a characteristic peak associated with large transient
fluctuations.

The scaling theory discussed above can be put in a more systematic basis which can also be used
in the subcritical case: In order to approximate the individual paths of the
relaxation process defined by (3.1) and $x(0)=0$, we write
$x(t)$ as the ratio of two stochastic processes
\begin{equation}
x(t)=z(t)/y^{1/2}(t)
\end{equation}

\noindent Then (3.1) (with $c=0$) is equivalent to the set of equations

\begin{equation}
d_t z(t)=az(t)+\sqrt{\epsilon}y^{1/2}(t)\xi(t)
\end{equation}
\begin{equation}
d_t y(t)=-2bz^2(t),
\end{equation}
with $z(0)=x(0)=0$, $y(0)=1$. Eqs. (3.17)-(3.18) can be solved
iteratively from the initial conditions. In the zero--th
order iteration
\begin{equation}
x(t){\sim}z(t)=h(t)e^{at},
\end{equation}
where 
\begin{equation}
h(t) =\sqrt{\epsilon} \int^t_0 e^{-as}\xi (s)ds,
\end{equation}
is a Gaussian stochastic process. In the first order iteration
\begin{equation}
x(t)={e^{at}h(t)\over[1-2b\int_0^t e^{2as}h^2(s)ds]^{1/2}}.
\end{equation}
In this approximation the decomposition (3.16) is interpreted as
follows. The process $z(t)$ coincides with the linear
approximation ($b=0$,$c=0$) to (3.1). The process $y(t)$ introduces
saturation effects killing the exponential growth of
$z(t)$. The scaling theory approximation (3.11) is recovered from this
approach whenever $at \gg 1$ so that
\begin{equation}
\int_0^t e^{2as}h^2(s)ds \sim {1\over {2a}}h^2(t)(e^{2at}-1)
\end{equation}
This indicates that the regime in which the scaling theory is valid is rapidly achieved
far from the condition of criticality $a\sim 0$.

We next consider the relaxation from $x=0$ with $a=0$ in a subcritical bifurcation $b>0$
\cite{Colet90}. As we did before we look for a
decomposition of the form (3.16). Since we are only
interested in the escape process we set $c=0$ in (3.1). Eq.
(3.1) with $a=c=0$ is equivalent to the set (3.17)-(3.18) with
$a=0$. In the zero--th order iteration

\begin{equation} 
x(t) \sim z(t) = \sqrt{\epsilon}W(t),\quad
W(t)=\int_0^t \xi (s)ds. 
\end{equation}
The process $x(t)$ coincides with the Wiener process
giving diffusion in the locally flat potential $V$. In the
first order iteration we find

\begin{equation}
x(t)={{\sqrt{\epsilon}W(t)}\over {[1-2\epsilon
b\int_0^t W^2 (s) ds]^{1/2}}}. 
\end{equation}

\noindent Two important differences between (3.24) and (3.21) should
be stressed. The first one is that in the first stage of evolution given by
(3.23) there is no escape from $x=0$. The nonlinearities introduced by the
process $y(t)$ are essential for the escape process to
occur. This means that, contrary to the case of (3.21), there is no
regime of the escape process in which Gaussian statistics
holds. The second difference is that $x(t)$ given by (3.24)
does not have a scaling form, since it is not a transformation of
a single stochastic process. Indeed, $x(t)$ in (3.24)
depends on two non independent processes, namely $W(t)$
and $\int_0^t W^2 (s)ds$. A naive approach to the problem
would be to assume, by analogy with (3.11), a scaling
form in which $x(t)$ is given by the deterministic
nonlinear mapping of the fluctuating initial process
$\sqrt{\epsilon}W(t)$, i.e., $ x(t)$ given by the
deterministic solution of (3.1) with $x(0)$ replaced by
$\sqrt{\epsilon}W(t)$. This would amount to take
$y(t)=1-2\epsilon btW^2 (t)$. This scaling representation
is qualitatively incorrect since it leads to a diverging
MPT, $T =\infty$.
The accuracy of the representation (3.24) is shown in
Fig. (3.4) where an individual trajectory given by (3.24) is
compared with the corresponding one of the exact process
given by (3.1). In this figure we observe that $x(t)$
coincides initially with the Wiener process, it later
departs from it and at a time rather sharply defined it departs
from the vicinity of $x=0$. In fact, the strong
nonlinearity implies that the solution of (3.1) with $c=0$
and $x(0)\neq 0$ reaches $\vert x\vert =\infty$ in a finite
time. It is then natural to identify the PT for the escape from $x=0$ as a random 
time
$t^*$ for which $|x(t^*)|=\infty$ in (3.24), or equivalently $y(t^*)=0$. From (3.24) we
find
\begin{equation}
1=2b\epsilon \int_0^{t^*} W^2 (s) ds.
\end{equation} 
Taking into account that $W^2 (t^* s)=t^* W^2 (s)$, (3.25) can be solved 
for $t^*$ as 
\begin{equation}
t^* = \left[ 1\over {2b\epsilon
\Omega}\right]^{1/2},
\end{equation}
with
\begin{equation}
\Omega \equiv \int_0^1 W^2 (s)ds.
\end{equation}
Eq. (3.26) gives the statistics of $t^*$ as a transformation
of the statistics of another random variable $\Omega$.
This scaling result for $t^*$ has the same basic contents
than (3.12). In (3.12) the result appeared for $at\gg 1$ as a
consequence of Gaussian statistics while here the
transformation (3.26) appears as an exceptional scaling at
the critical point $a=0$, and $\Omega$ has non-Gaussian
statistics. A discussion of the crossover between these two regimes is given
in \cite{Colet91}.

\begin{figure}[h]
\begin{center}
\def\epsfsize#1#2{0.60\textwidth}
\leavevmode
\epsffile{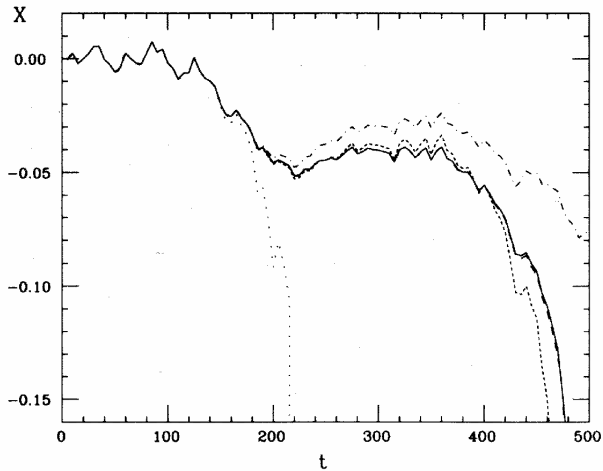}
\end{center}
\caption{Stochastic paths in the decay from $x=0$ for $a=0$ in the subcritical bifurcation with
$b=c=1$ and $\epsilon=10^{-6}$.
Different paths correspond to different approximations for the same realization of the noise.
The solid line corresponds to a simulation of the exact process given by (3.1). Dotdashed line
corresponds to a Wiener process. Dots correspond to the scaling approach similar to (3.11)
(see text) and the dashed line corresponds to the approximation (3.24).}
\label{traysubcritico}
\end{figure}
The calculation of the statistics of $t^*$ from (3.26)
requires the knowledge of the statistics of $\Omega$. The
latter is completely determined by the generating function
$G(\lambda)$, for which an exact result is available \cite{Colet90},

\begin{equation}
G(\lambda)\equiv \langle e^{-\lambda \Omega}\rangle= (\cosh (2\lambda^{1/2}))^{-1/2}.
\end{equation}
The moments of the PTD are obtained from (3.26) in terms
of $G(\lambda)$ as

\begin{equation}
\langle t^{*2n}\rangle={{(2b\epsilon )^{-n}}\over {\Gamma(n)}} \int_0^\infty
d\lambda\quad\lambda^{n-1} G(\lambda).
\end{equation}
The transient moments
$\langle x^n (t)\rangle$ can also be  obtained in terms of the statistics
of the PT \cite{Colet90}: Given the strong nonlinearity of the escape
process a good approximation to calculate ensemble
averages is to represent $x(t)$ by

\begin{equation}
x^2 (t)=x^2 (\infty) \theta (t-t^*), 
\end{equation}
where $x (\infty)$ is the final stable state (local minima of the
potential V) and $\theta (t-t^*)$ is the Heaviside step
function. The individual path is approximated by a jump from
$x=0$ to $x=x (\infty)$ at a random time $t^*$. The transient
moments are then easily calculated as averages over the
distribution of the times $t^*$. 

The methodology described here to characterize relaxation processes by focusing on the statistics
of the passage times can be generalized to a variety of different situations which include
colored noise \cite{Sancho89}, time dependent control parameter $a(t)$ sweeping through the instability
at a finite rate \cite{Torrent88,Torrent89}, competition between noise driven decay and 
decay induced
by a weak external signal \cite{Balle90}, role of multiplicative noise in transient fluctuations
\cite{DePasquale86}, saddle-node bifurcation \cite{Colet89}, etc. We have limited here ourselves
to transient dynamics in stochastic processes defined by a SDE of the Langevin type, 
but the PT characterization is also
useful, for example, in the framework of Master Equations describing transport in disordered
systems \cite{SanMiguel91,HernandezGarcia91}

\subsection{Statistics of laser switch-on}
Among the applications to different physical systems of the method described above, the analysis
of the stochastic events in the switch-on of a laser is particularly interesting for several 
reasons. Historically (see the review in \cite{ArecchiTorino}) the concept of transient anomalous
fluctuations already appeared in this context in the late 60's, and from there on a detailed
experimental characterization exists for different cases and situations. The idea of noise amplification
is also very clear here, and for example the term "statistical microscope" \cite{Arecchi89} has been used
to describe the idea of probing the initial spontaneous emission noise through the switch-on 
amplification process. From the methodological point of view the idea of following the individual
stochastic trajectory has in this context a clear physical reality. While in more traditional equilibrium
statistical physics it is often argued that measurements correspond to 
ensemble averages, 
every laser switch-on is an experimental measured event which corresponds to an individual
stochastic trajectory. Also here the PT description has been
widely used experimentally \cite{passagetimes}. Finally, from an applied point of view, the
variance of the PT distribution for the switch-on of a semiconductor laser sets limitations in
the maximum transmission rate in high speed optical communication systems
\cite{SpanoBlanes,Spano89}. We will limit our quantitative discussion here to semiconductor
lasers. A review of other situations is given in  \cite{SanMiguelSPIE}.
 
The dynamics of a single mode semiconductor laser can be described by the following equations for
the slowly varying complex amplitude of the electric field $E$ and the carrier number $N$ 
\cite{SpanoBlanes,Agrawal,Balle91,Balle92,Balle93}:
\begin{equation}
\partial_{t}E = {1 + i \alpha \over 2} [ G - \gamma ] E + \sqrt{\beta 
N} \xi_{E} (t),
\end{equation}

\begin{equation}
\partial_{t} N = C - \gamma_{e} N - G \vert E \vert^{2} - \sqrt{4 
\beta N} [ E^{*} \xi_{E} (t) + E \xi_{E}^{*} (t) ].
\end{equation}
Terms proportional to the gain coefficient $G$ account for stimulated emission
 and, to a first 
approximation, $G=g(N-N_0)$, where $g$ is a gain parameter and $N_0$ the value of
carrier number at transparency. The control parameter of the system is here the injection current
$C$. Spontaneous emission is modeled by a complex Gaussian white noise $\xi_{E} (t)$
of zero mean and correlation
\begin{equation}
\langle \xi_{E} (t) \xi_{E}^{*} (t')\rangle = 2 \delta (t - t'). 
\end{equation}
Noise is here multiplicative and its intensity is measured by the spontaneous emission rate $\beta$.
The noise term in the equation for $N$ can be usually neglected \footnote {When
the noise term in (3.32) is neglected, the It\^o and Stratonovich interpretations
are equivalent for (3.31)}. The variables $E$
and
$N$ have decay rates $\gamma$ and $\gamma_e$ such that
$\gamma \sim 10^3 \gamma_{e}$. This implies that the slow variable is $N$, and that when the
laser is switched-on the approach to the final state is not monotonous. Rather, a light pulse is
emitted followed by damped oscillations called relaxation oscillations. This also occurs for
$CO_2$ and solid state lasers, but there are other lasers such as He-Ne (``class
Alasers") in
which $E$ is the slow variable, the approach is monotonous and a simpler description
with a single equation for $E$ holds. Such description is similar to the one in 
Sect. 3.1
\cite{SanMiguelSPIE}. Finally, the $\alpha$-parameter, or linewidth enhancement factor, gives
the coupling of phase $\phi$ and laser intensity $I$ ($E = I^{1 / 2} e^{i \phi}$). 
\begin{equation}
\partial_{t} I = ( G - \gamma ) I + 4 \beta N + \sqrt{4 \beta N I} 
\xi_{I}(t),
\end{equation}
\begin{equation}
\partial_{t} \phi = {\alpha \over 2} (G - \gamma ) + 
\sqrt{4 \beta N \over I} \xi_{\phi} (t), 
\end{equation}
where $\xi_{I}$ and $\xi_{\phi}$ are the intensity and phase noise components 
of $\xi_{E}$. These
equations are conventionally written as It\^o SDE. 
The instantaneous frequency of the laser is the time derivative of the phase.
Transient frequency fluctuations for class A lasers were discussed in \cite{Ciuchi}.

To switch-on the laser, the injection current $C$ is switched, at  a time $t=0$, from below to
above its threshold  value $C_{th} = ({\gamma \over g} + N_{o} ) \gamma_{e}$. Below threshold
the laser is off with $I$ fluctuating around $I=0$. The steady state value above threshold
is given by  $I = {1 \over \gamma}( C - C_{th} ) + 0 (\beta )$ and $N = N_{th} = {C_{th} \over
\gamma_{e}}$. An example of the early time dynamics in the approach to this steady state is
shown in Fig. (3.5) where two switch-on events corresponding to two different stochastic
trajectories are shown. We observe that the laser switches-on when $N$ reaches a maximum value.
This happens at a random time which gives the delay from the time at which
$C$ is switched. The dispersion in the
delay times is known as ``jitter". We will characterize the switch-on statistics by the PT
distribution for $I$ to reach a given  reference value. We also observe a
statistical
spread in the height $I_M$ of the light pulses (maximum output power) for different
switch-on events, being larger for longer delay times. The frequency of the laser shows huge
fluctuations while it is off, and it drifts from a maximum to a minimum frequency during the
emission of the light pulse due to the phase-amplitude coupling caused by 
the $\alpha$-parameter.
This excursion in frequency is known as ``frequency chirp", and again there is a
statistical distribution
of chirp for different switch-on events. Relevant questions here are 
the calculation of
switch-on times, maximum laser intensity and chirp statistics, as well as the relation among
them.

The calculation of switch-on time statistics follows the same basic ideas than in Sect. 3.1
\cite{Balle91,Balle93}:
We consider the early linearized regime formally obtained setting $G=0$
in the equation for $N$. 
When the solution of this equation is substituted in (3.31) we still have a linear
equation for $E$ but with time dependent coefficients, so that,
\begin{equation}
E(t) = e^{A (t) / 2} h (t),
\end{equation}

\begin{equation}
A(t) = (1 + i \alpha ) \int^{t}_{\bar t} d t' [g (N (t') - N_{o}) - 
\gamma ], 
\end{equation}

\begin{equation}
h (t) = \int^{t}_{\bar t} dt' \sqrt{ \beta N (t')} \xi (t') 
e^{-A(t')/2}, 
\end{equation}
\begin{figure}[h]
\begin{center}
\def\epsfsize#1#2{0.50\textwidth}
\leavevmode
\epsffile{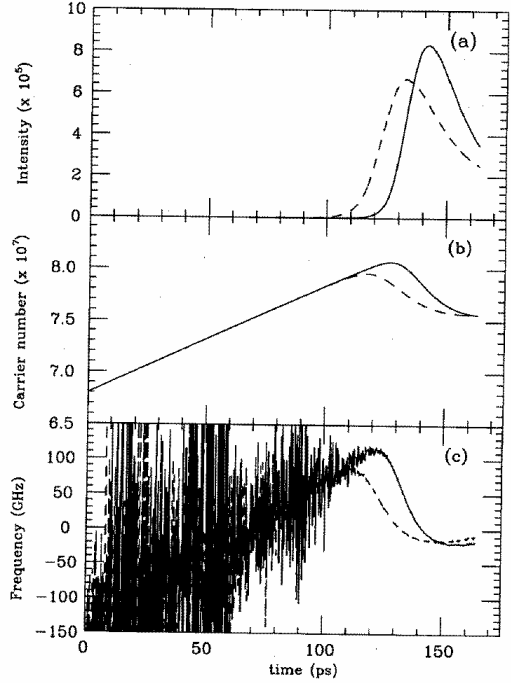}
\end{center}
\caption{Time evolution of (a) laser intensity $I=|E|^2$, (b) carrier number $N$ and
(c) frequency $f= (2 \pi)^{-1} {\Delta \phi \over \Delta t}$ as obtained from the numerical
integration of Eqs. (3.31)-(3.32) for typical parameter values and for two different switch-on
transients: solid and dashed line. Times are given in picoseconds.}
\label{twoevents}
\end{figure}
where we have chosen a reference time ${\bar t}$, $N (\bar t) = N_{th}$. We note that
for ordinary parameter values $N(t)$ grows linearly with $t$ around
${\bar t}$ (as it is seen in Fig. (3.5)), so that $A(t)\sim (t - \bar t)^{2}$. The important fact is again that the 
complex Gaussian process $h(t)$ saturates for times of interest to a complex Gaussian random
variable $h (\infty)$ of known statistics. We can then solve (3.36) for the time $t^*$ at which
the reference value $I_r=|E_r|^2$ of $I$ is reached
\begin{equation}
(t^{*} - \bar t)^{2} \sim \ln {I_{r} \over \vert h (\infty) \vert^{2}}.
\end{equation}
The statistics of $t^*$ are now easily calculated as a transformation of that of $h(\infty)$.
In particular we obtain $(\Delta T) \ll T$, with
\begin{equation}
T= \langle t^{*}\rangle = \bar t + 
\sqrt{{2 \tau \over g (C-C_{th})}} [1 - {\psi (1) \over 2 \tau}
], \quad \tau = \ln {I_r \over b} 
\end{equation}

\begin{equation}
(\Delta T)^2 = {\psi' (1) \over 2 \tau g (C-C_{th})} 
\end{equation}
where $b$ is proportional to $\beta$. 

We next address the calculation of the statistics of the maximum laser intensity $I_M$
\cite{Balle91,Balle94}. Fig (3.6) shows a superposition of different switch-on events, obtained
from a numerical integration of (3.31)-(3.32), in which the statistical dispersion in the values of $I_M$
is evidentiated. The laser intensity peak value is reached well in the nonlinear regime and we
adopt, as in Sect 3.1, the strategy of mapping the initial stochastic linear regime with the
nonlinear deterministic dynamics. For each switch-on event, the deterministic equations should
be solved with initial conditions at the time $t^*$ which identifies the upper limit of the linear
regime: $E=E(t^*)=E_r$ and $N=N(t^*)$. The value of $N(t^*)$ is calculated linearly and takes
random values for different switch-on events. Eliminating the  parameter $t$ the solution can be
written in an implicit form as 
\begin{equation} E = E(N, E_{r}, N(t^{*})).
\end{equation}
\begin{figure}
\begin{center}
\def\epsfsize#1#2{0.60\textwidth}
\leavevmode
\epsffile{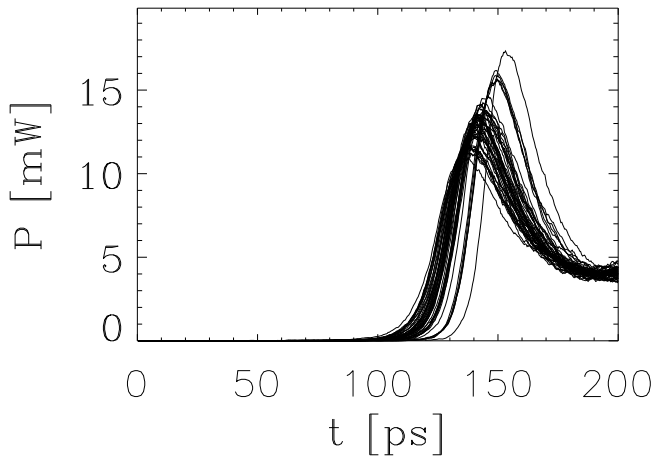}
\end{center}
\caption{Superposition of 50 switch-on events (50 individual stochastic trajectories) for the
laser intensity as obtained from the numerical
integration of Eqs.(3.31)-(3.32). The laser power $P$ is proportional to the
laser intensity I.}
\label{eyediagram}
\end{figure}
Any arbitrary function $Y$ of the dynamic variables in which one might be interested
can then be written as
\begin{equation}
Y (t) = Y (N (t), E(t) ) = Y (t, N (t^{*}), E_{r})  = Y (N, E_{r},  N (t^{*})). 
\end{equation} 
To proceed further one generally needs an explicit form for the solution of the equations.
However, important results  can be already obtained if one is interested in extreme values of
$Y$. The variable $Y$ takes an extreme value $Y_e$ at a time $t_e$ such that 
$\partial_t Y(t) =
0$. This gives an additional implicit relation $ N (t_e) = N_e (E_{r}, N (t^{*}))$ so that,
\begin{equation}
Y (t_e) = Y_e (E_{r}, N (t^{*})).
\end{equation}
This equation gives the desired mapping of the statistical properties of the switch-on times $t^*$
into those of an extreme value of a dynamical variable: For each switch-on event there is a value
of $t^*$ and a corresponding value $Y_e(t^*)$. Taking advantage of the fact 
that $(\Delta T) \ll T$
we can expand $Y_e(t^*)$ around $t^*=T$:
\begin{equation}
Y_e (t^*) = Y_e (T) + {\partial Y_e \over \partial N (t^{*})} {\left. {dN \over 
dt^{*}}\right \vert}_{T} (t^{*} - T) + 0(\Delta T)^2.
\end{equation}
Applying this relation to the maximum output power, $Y_e = I_M$, it predicts a straight line for
a plot of  the values of $I_M$ vs. switch-on times obtained from many independent switch-on
experiments. Such linear dependence can be already seen in Fig. (3.6). 
This prediction has been
corroborated experimentally for $CO_2$ \cite{Balle94,MeucciPalma},  semiconductor
\cite{LDslope} and solid state lasers \cite{Balle94}. The slope of this linear relation can be
estimated in different cases. For semiconductor lasers \cite{Balle91} it is easily seen from
the equations above, that the slope is positive and proportional to the distance to threshold
$C-C_{th}$. However in  other cases there is a predominant contribution of the initial
fluctuations of $N(t=0)$ and the slope becomes negative \cite{Balle94}.

Turning now to transient frequency fluctuations, a first delicate issue is the proper
definition of an instantaneous frequency, since $\partial_{t} \phi$ in (3.35) contains a white
noise noise contribution, and therefore, in principle,
an infinite variance. A meaningful operational
definition of the instantaneous frequency can be given in terms of a short time interval average
of $\partial_{t} \phi$, or equivalently in terms of a time resolved field Fourier spectrum
 \cite{Balle92,Ciuchi96}.
From (3.35) we also see that there are two sources of frequency fluctuations. One is the noise
along the path given by the phase noise  $\xi_{\phi}$, and the second is the stochastic nature of the gain
term $G$ for different switch-on events, due to the random nature of $t^*$.
The first dominates steady state properties such as laser  linewidth, but the second dominates
transient properties such as the fluctuations in the chirp range. Since the chirp range is
defined by a difference between two extreme values of the frequency the general framework
discussed above applies. In fact one finds that the maximum and minimum instantaneous
frequencies during the transient are linear functions of the switch-on time $t^*$
\cite{Balle93}. Chirp noise statistics is again calculated by a simple transformation of the
switch-on time statistics.

We finally mention that the operation of a semiconductor laser in a system of optical communications
involves a repetition of switch-on events (large signal modulation) to transmit pulses at a given
 bit-rate. Communication
errors are likely to happen when $\Delta T$ is large or when memory effects are present. 
The effect of memory occurs when a given switch-on
event is not independent of the previous one in very fast modulation. The determination of "memory-free"
regions in parameter space for a given modulation rate involves, as a crucial time scale of the
problem, the mean switch-on time $T$ \cite{Colet93}. Other errors are due to pulse dispersion in
the optical fibre because of large chirp noise \cite{Balle95}.
 

\section{Noise in Dynamical Systems}
\setcounter{equation}{0}
\setcounter{figure}{0}
\subsection{Classification of Dynamical Systems: Potentials and Lyapunov
Functions}
In this section, we will study the effect of noise in the long time
behavior of dynamical systems. First of all, though, we will briefly
review the basic concepts about the stability of dynamical systems \cite{GH}.

\begin{figure}
\begin{center}
\def\epsfsize#1#2{0.82\textwidth}
\leavevmode
\epsffile{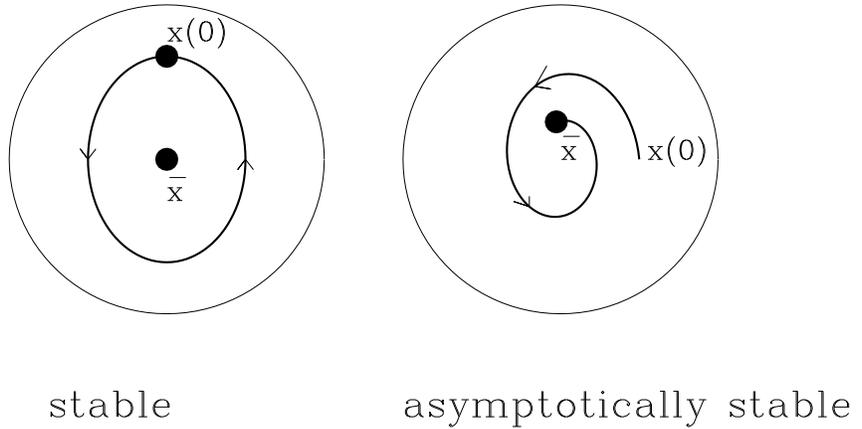}
\end{center}
\caption{\label{fig:4-1}
Illustration of the concepts of stable and asymptotically stable fixed points.}
\end{figure}

Let $x\equiv(x_1,\dots,x_N)$ be a set of real dynamical variables. They 
follow the evolution equations:
\begin{equation}
\label{eq4:1}
\frac{dx_i}{dt} = f_i(x_1,\dots,x_N), ~~~~~~ i=1,\dots,N
\end{equation}
One is usually interested in finding the fixed points $\bar x$ of the
dynamical system as those points which do not evolve in time,
i.e. those satisfying:
\begin{equation}
\label{eq4:2}
\left. \frac{dx_i}{dt}\right \vert_{x=\bar x}  = 0, ~~~~~~ i=1,\dots,N
\end{equation}
or, equivalently,
\begin{equation}
\label{eq4:3}
f_i(\bar x_1,\dots,\bar x_N) =0, ~~~~~~ i=1,\dots,N
\end{equation}
Knowing the fixed points is as important as knowing their stability.
Loosely speaking, a fixed point $\bar x$ is stable if any initial condition
$x(0)$ sufficiently close to $\bar x$ remains close to
$\bar x$ as time evolves. The fixed point is said to be asymptotically stable if
any initial condition $x(0)$ sufficiently close to $\bar x$ tends to 
$\bar x$ as time evolves. Fig. (\ref{fig:4-1}) illustrates the concepts
of stability and asymptotic stability.

To determine the stability of a fixed point, Lyapunov's stability theorem
is of great help. The theorem states that if we can find a {\sl
Lyapunov function}, $L(x) = L(x_1,\dots,x_N)$ which is such that it has a local
minimum at $x=\bar x$ and monotonically decreases with time, i.e.:
\begin{equation}
\label{eq4:3a}
\begin{array}{lll}
(a) &  L(x) > L(\bar x) & {\rm for~}x \ne \bar x\\
(b) & {\displaystyle \frac{dL}{dt} = \sum_{i=1}^N
\frac{\partial L}{\partial x_i} 
\frac{dx_i}{dt} \le 0} & {\rm for~}x \ne \bar x 
\end{array}
\end{equation}
then $\bar x$ is a stable fixed point. If condition (b) is replaced by
\begin{equation}
\label{eq4:3b}
(b') \hspace{0.1cm} \frac{dL}{dt}  <  0 \hspace{1.0cm} {\rm for~}x \ne \bar x
\end{equation}
then $\bar x$ is an asymptotically stable fixed point. $L(x)$ is also 
called a {\sl Lyapunov potential} or, simply, the potential.

\begin{figure}[h]
\begin{center}
\def\epsfsize#1#2{0.82\textwidth}
\leavevmode
\epsffile{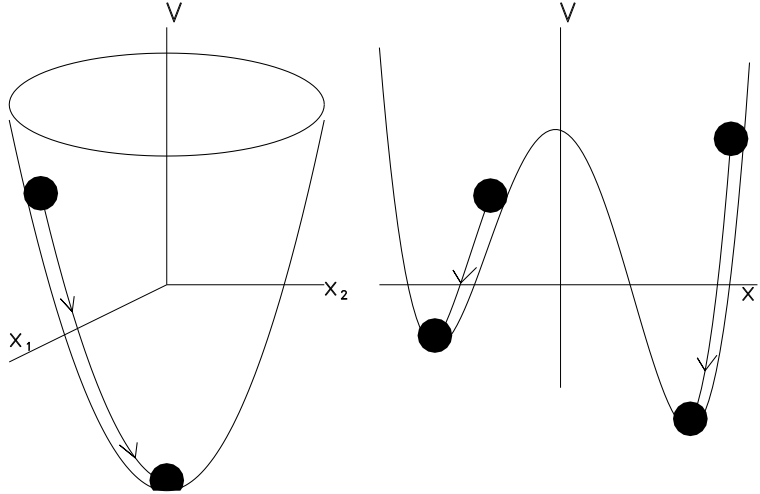}
\end{center}
\caption{\label{fig:4-2}
Schematic trajectories for the case of relaxational gradient flow in two
cases: a Lyapunov function with a single minimum (left picture) and
a Lyapunov function with two minima (right picture). The system evolves
towards one of the two minima, according to initial conditions, following
the lines of maximum slope.}
\end{figure}

The description of the behaviour of a dynamical system
is greatly simplified if we know of the existence of a Lyapunov
potential: the system evolves towards the minima of the Lyapunov
and once there it stays nearby when small perturbations act upon the 
system.
The Lyapunov potential not only tells us about local stability, but also 
gives information about global stability. If two fixed points $\bar x^{(1)}$
and $\bar x^{(2)}$ are such that $L(\bar x^{(2)}) < L(\bar x^{(1)}) $ then
we can infer that $\bar x^{(2)}$ is the stable fixed point and that
$\bar x^{(1)}$ is a metastable fixed point. This means that a sufficiently
strong perturbation might take the system out of $\bar x^{(1)}$ to
$\bar x^{(2)}$. Therefore, to understand the
asymptotic dynamics it is of great importance
to find out whether a Lyapunov potential can be found
for the dynamical system under study. It is not an easy task to (i)
prove that the Lyapunov potential does exist, (ii) find it. 
The systems for which a Lyapunov function exists are called in the
literature {\sl potential systems} and the name {\sl non--potential
systems} is therefore reserved for those systems which do not 
have a Lyapunov function (notice that this is different from the
fact that there is no {\sl known} Lyapunov potential). 
According to these definitions, and following the ideas of Graham
on nonequilibrium potential \cite{gra1,gra2,gra3,grahamsitges}, the following
classification of dynamical systems can be
established \cite{smmahg,montagne96}:\\
\noindent 1.- Relaxational Gradient Flow \\
\noindent 2.- Relaxational Non--Gradient Flow \\
\noindent 3.- Non--Relaxational Potential Flow \\
\noindent 4.- Non--Potential Flow \\
The first three categories are called potential flows. We now describe
each of them in some detail.
\espacio
{\noindent \bf (1) Relaxational Gradient Flow.}\\
Those are systems for which 
there exists a function (the ``potential") $V(x)=V(x_1,\dots,x_N)$ in terms
of which the dynamics is written as:
\begin{equation}
\frac{dx_i}{dt} = -\frac{\partial V}{\partial x_i}
\end{equation}
The fixed points of this dynamical system are the extrema of the 
potential $V(x)$. The trajectories lead to any of the minimum values of 
$V(x)$ following the lines of maximum slope (steepest descent) as
indicated schematically in Fig. (\ref{fig:4-2}). 
In this case the potential $V$ is a Lyapunov function. 
The proof of this statement is very simple indeed (we need
the additional condition that $V(x)$ is bounded from below):
\begin{equation}
\frac{dV}{dt} = \sum_{i=1}^N \frac{\partial V}{\partial x_i}
\frac{dx_i}{dt}=
 - \sum_{i=1}^N\left(\frac{\partial V}{\partial x_i}\right)^2 \le 0
\end{equation}

\begin{figure}[h]
\begin{center}
\def\epsfsize#1#2{0.82\textwidth}
\leavevmode
\epsffile{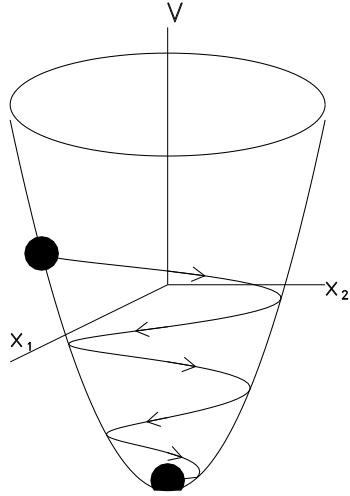}
\end{center}
\caption{\label{fig:4-3}
Schematic trajectories for the case of relaxational non--gradient flow. The
system evolves towards the minimum of the potential, but not necessarily
through the lines of maximum slope.}
\end{figure}

An example of this class of systems is the 
real Ginzburg--Landau equation (see later), also known in the 
theory of critical dynamics as model $A$ in the classification of
\cite{HH} (see Section 6).
\espacio
{\noindent \bf (2) Relaxational non--Gradient Flow.}\\
There exists again a potential function, $V(x)$, but now the dynamics
is not governed just by $V$, but it is given by:
\begin{equation}
\frac{dx_i}{dt} = -\sum_{j=1}^NS_{ij}\frac{\partial V}{\partial x_j}
\end{equation}
where $S_{ij}(x)$ is a real, symmetric, positive definite matrix. The fixed 
points of the dynamics are still given by the extrema of $V$ and the
trajectories lead to the minima of $V$ but not necessarily through the lines
of maximum slope. In this sense, we can say that the transient dynamics
is not governed just by the potential. Fig. (\ref{fig:4-3}) 
shows schematically
the situation. A typical example of this dynamics is the so--called
Cahn--Hilliard \cite{CH} model for spinodal decomposition \cite{Gunton83}
or model
$B$ in the context of critical dynamics \cite{HH}. In this model the dynamical
variables represent a concentration field that follows a conservation
law:
\begin{equation}
\frac{d~}{dt}\sum_{i=1}^N x_i = 0
\end{equation}
This constrain is satisfied by choosing $S=-\nabla^2$, the lattice Laplacian
as defined, for instance, in (\ref{eq:99f}).

\begin{figure}
\begin{center}
\def\epsfsize#1#2{0.82\textwidth}
\leavevmode
\epsffile{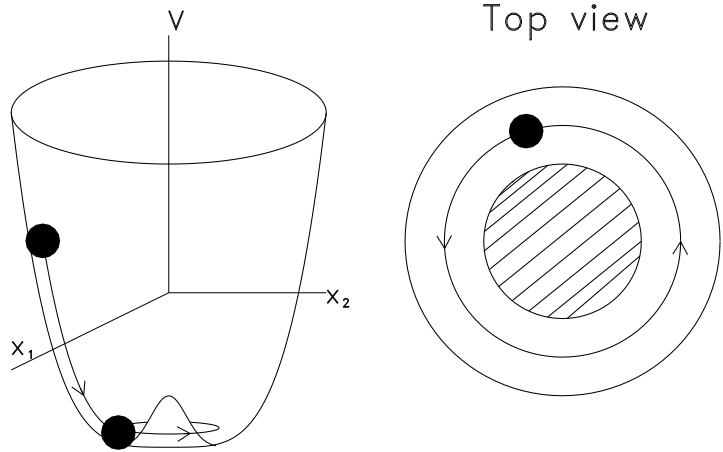}
\end{center}
\caption{\label{fig:4-4} 
Schematic trajectories for the case of non--relaxational potential flow.
The system evolves towards the (degenerate) minima of the potential 
$V$ and, once it has reached there, there is a residual movement
on an iso--potential region.}
\end{figure}

In this case, the potential $V$ is still a Lyapunov functional:
\begin{equation}
\frac{dV}{dt} = \sum_{i=1}^N \frac{\partial V}{\partial x_i}
\frac{dx_i}{dt}=-\sum_{i,j=1}^NS_{ij}\frac{\partial V}{\partial x_i}
\frac{\partial V}{\partial x_j} \le 0
\end{equation}
since, by definition, $S$ is a positive definite matrix.
\espacio
{\noindent \bf (3) Non--Relaxational Potential flow.}\\
These systems are characterized because the asymptotic behavior is not
determined simply by the minima of the potential, but there is a residual
dynamics once those minima have been reached. A first category
within this class is given by:
\begin{equation}
\frac{dx_i}{dt} = -\sum_{j=1}^ND_{ij}\frac{\partial V}{\partial x_j}
\end{equation}
Here, $D_{ij}(x)$ is an arbitrary matrix that can be split into symmetric
and antisymmetric parts:
\begin{equation}
D= S+A \left\{ \begin{array}{rcl}
S & = & \frac{1}{2}(D+D^T)\\
A & = & \frac{1}{2}(D-D^T)\end{array} \right.
\end{equation}
and we demand that $S$ is positive definite matrix. Again, the fixed point
of the dynamics are determined by the extrema of the potential $V$ and
$V$ is a Lyapunov potential:
\begin{equation}
\frac{dV}{dt} = -\sum_{i,j=1}^NS_{ij}\frac{\partial V}{\partial x_i}
\frac{\partial V}{\partial x_j}- 
\sum_{i,j=1}^NA_{ij}\frac{\partial V}{\partial x_i}
\frac{\partial V}{\partial x_j} \le 0
\end{equation}
The second sum of this expression is zero due to the antisymmetry of matrix
$A$. We are then led to the situation sketched in Fig. (\ref{fig:4-4}):
the system evolves towards the minima of $V$ but once it reaches there,
there is a residual movement produced by the antisymmetric part of 
the dynamical coefficients that can make the system flow without cost
of the Lyapunov function.
The dynamics of the nematodynamics equations commonly used to
describe the dynamics of liquid crystals in the nematic phase
belongs to this class, as shown in \cite{SanMiguel87}.

The second, more general, category of  non--relaxational potential flow 
will be one in which the 
dynamical equations can be split into two parts, namely:
\begin{equation}
\label{eq4:10}
\frac{dx_i}{dt} = f_i \equiv -\sum_{j=1}^NS_{ij}\frac{\partial V}{\partial x_j}
+v_i
\end{equation}
Here $S(x)$ is a symmetric, positive definite matrix and
$v_i(x)$ represents the residual dynamics after the
relaxational part has acted. Since we want $V$ to 
be a Lyapunov potential the residual
dynamics must not contribute to its decay:
\begin{equation}
\label{eq4:10b}
\frac{dV}{dt} = -\sum_{i,j=1}^NS_{ij}\frac{\partial V}{\partial x_i}
\frac{\partial V}{\partial x_j} + \sum_{i=1}^N v_i \frac{\partial V}
{\partial x_i} \le 0
\end{equation}
or, since the first term of the r.h.s. is always negative, a sufficient
condition is:
\begin{equation}
\label{eq4:10a}
\sum_{i=1}^N v_i \frac{\partial V}{\partial x_i}=0
\end{equation}
This is the so--called {\sl orthogonality condition}. By using
(\ref{eq4:10}) can be written as:
\begin{equation}
\label{eq4:12}
\sum_{i=1}^N\left(f_i+\sum_{j=1}^N S_{ij}\frac{\partial V}{\partial x_j}\right)
\frac{\partial V}{\partial x_i} =0
\end{equation}
or, in compact vector notation:
\begin{equation}
(\vec f + S \vec \nabla V)\cdot \vec \nabla V = 0
\end{equation}
It can be shown that this equation has the structure of a Hamilton--Jacobi
equation and in general it is very difficult to solve to find the potential
$V$, although 
important results have been obtained in the literature \cite{gra1,gra3,grahamsitges}.
\espacio
{\noindent \bf (4) Non--Potential flow.}\\
Finally, a non--potential flow is one
for which the splitting (\ref{eq4:10}), satisfying (\ref{eq4:10b}),
admits only the trivial solution
$V=0$, $v_i=f_i$. Notice that for a system to be classified as non--potential,
we have to prove the non--existence of non--trivial solutions of 
(\ref{eq4:12}) which, of course, it is not the same that not being able 
to find non--trivial solutions of (\ref{eq4:12}).
\espacio
Let us consider in detail an example of the previous classification scheme:
the complex Ginzburg--Landau equation (CGLE) \cite{montagne96,gra4,gra5}.

Let $A(\vec r,t)$ be a complex field satisfying the 
equation of motion:
\begin{equation}
\frac{\partial A}{\partial t} = \mu A + \gamma \nabla^2A - \beta |A|^2A
\end{equation}
$\mu=\mu_R+i\mu_I$, $\beta=\beta_R+i\beta_I$ and $\gamma=\gamma_R+i\gamma_I$ 
are complex numbers. 
This is the so--called complex Ginzburg--Landau equation \cite{saa}. It appears
in a great variety of situations and it is  a generic equation for
the slow varying amplitude of the most unstable model near a Hopf 
bifurcation in an extended system (see Sect. 5.3).  This might be put
in what is called in the literature a ``potential" form:
\begin{equation}
\frac{\partial A}{\partial t} = - \frac{\delta V}{\delta A^*}
\end{equation}
where the ``potential" $V$ is given by:
\begin{equation}
V\{A,A^*\} = \int d\vec r \left[ -\mu |A|^2+\gamma |\vec \nabla A|^2
+\frac{\beta}{2}|A|^4 \right]
\end{equation}
This, being of course true, is somewhat misleading because the function
$V$ does not have most of the properties associated with a potential
function as defined in the previous sections. In particular it can not
be a Lyapunov function because it generally takes complex values.

We pose now the problem in a different form. Considering the evolution
of the real and imaginary components of $A(\vec r,t) = X(\vec r,t)+
iY(\vec r,t)$, is it possible to find a real function $\tilde V(X,Y)$ such that
the flow is relaxational gradient? i.e.
\begin{equation}
\left.
\begin{array}{rcl}
\displaystyle{
\frac{\partial X}{\partial t}} & = & 
\displaystyle{
- \frac{\delta \tilde V}{\delta X}} \\
\displaystyle{
\frac{\partial Y}{\partial t}} & = & 
\displaystyle{
- \frac{\delta \tilde V}{\delta Y} }
\end{array}
\right\}
\end{equation}
This would imply:
\begin{equation}
\frac{\delta~}{\delta Y} \left(\frac{\partial X}{\partial t}\right) = 
\frac{\delta~}{\delta X} \left(\frac{\partial Y}{\partial t}\right)
\end{equation}
which leads to 
\begin{equation}
\mu=\mu^* \hspace{1.0cm} \gamma=\gamma^* \hspace{1.0cm} \beta=\beta^*
\end{equation}
i.e. a complex Ginzburg--Landau equation with real coefficients.

In general, though, when the coefficients are complex there is no obvious
Lyapunov function. However, there is an interesting case in which the
dynamics can be put in the category (3) described above: there exists
a split of the dynamics in terms of a potential term plus an extra
term which, being orthogonal to the first, does not contribute to the
variation of the Lyapunov potential. 
The easiest way of showing this is noticing that
if we write $V(A,A^*)= V_R(X,Y)+iV_I(X,Y)$ the variation of the
real part is:
\begin{equation}
\begin{array}{rcl}
\frac{dV_R}{dt} & = & -\frac{1}{2}\int d\vec r\left[ 
\left(\frac{\delta V_R} {\delta X}\right)^2 + 
\left(\frac{\delta V_I} {\delta Y}\right)^2 \right] + \\
& & 2(\gamma_I\beta_R-\beta_I\gamma_R)\int d \vec r (X^2+Y^2)
\vec \nabla (Y\vec \nabla X - X \vec \nabla Y)
\end{array}
\end{equation}
It is clear now that if the following condition is fulfilled
\be
\label{eq4:30}
\gamma_I\beta_R-\beta_I\gamma_R=0
\ee
$V_R$ acts
as a Lyapunov. Introducing $\hat V = V/2$ the dynamical system can be 
written as:
\begin{equation}
\pmatrix{\frac{\partial X}{\partial t} \cr \frac{\partial X}{\partial t}} = - 
\pmatrix{\frac{\delta \hat V_R}{\delta X} \cr \frac{\delta \hat V_R}{\delta Y}}+
\pmatrix{\frac{\delta \hat V_I}{\delta Y} \cr -\frac{\delta \hat V_I}{\delta X}}
\end{equation}
and the orthogonality condition is:
\begin{equation}
\left(\frac{\delta \hat V_I}{\delta Y},-\frac{\delta \hat V_I}{\delta X}\right)
\pmatrix{\frac{\delta \hat V_R}{\delta X} \cr \frac{\delta \hat V_R}{\delta Y}}
=0
\end{equation}
which is satisfied by condition (\ref{eq4:30}).
In this case a change of variables of the form $A(\vec r,t) \rightarrow 
A(\vec r,t) {\rm e}^{i\alpha t}$ for a convenient value of $\alpha$ leads to 
\begin{equation}
\frac{\partial A}{\partial t} = - (1+i) \frac{\delta V_R}{\delta A^*}
\end{equation}
The $1$ in the r.h.s. is the real CGLE and the $i$ in the r.h.s. gives rise
to a form of the Nonlinear Schr\"odinger equation.

We finally mention that although for general complex coefficients
$\mu$, $\beta$ and $\gamma$ there is no simple form for a Lyapunov 
functional, this does not mean that there is no solution for the
orthogonality condition (\ref{eq4:12}). In fact, Graham and 
collaborators \cite{gra4,gra5} have found an approximate Lyapunov function
for the CGLE by a gradient expansion solution of the orthogonality condition
(\ref{eq4:12}). A numerical test of its Lyapunov properties and a 
discussion of such Lyapunov function for the CGLE is given in \cite{montagne96}.

\subsection{Potentials and Stationary Distributions}
\begin{figure}
\begin{center}
\def\epsfsize#1#2{0.82\textwidth}
\leavevmode
\epsffile{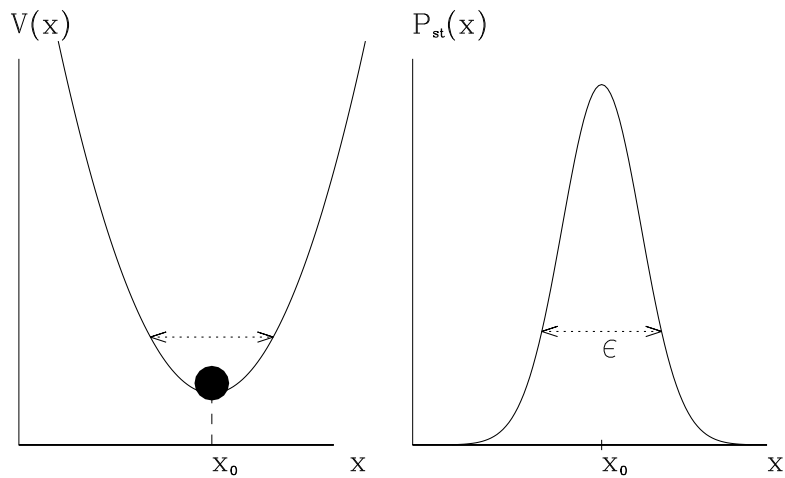}
\end{center}
\caption{\label{fig:4-5} 
Potential and stationary probability density function}
\end{figure}

We will see now which is the effect of noise on the dynamics of the above
systems. To this end, we consider the general dynamics with 
the addition of a general noise term (to be considered in the It\^o sense):
\begin{equation}
\label{eq4:21}
\dot x_i = f_i(x) + \sum_{j=1}^N g_{ij}\xi_j(t)
\end{equation}
where $g_{ij}(x)$ are arbitrary functions and
$\xi_j(t)$ are white noise variables: random Gaussian variables
of zero mean and correlations:
\begin{equation}
\label{eq4:20}
\langle \xi_i(t)\xi_j(t') \rangle = 2 \epsilon \delta_{ij}\delta(t-t')
\end{equation}
$\epsilon$ is the noise intensity, which is included here explicitly.
Due to the presence of the noise terms, it is not adequate to talk
about fixed points of the dynamics, but consider instead the maxima
of the probability distribution function $P(x,t)$. This satisfies
the multivariate Fokker--Planck equation \cite{Risken}:
\begin{equation}
\frac{\partial P(x,t)}{\partial t} = \sum_{i=1}^N \frac{\partial~}{\partial x_i}
\left[-f_i P + \epsilon \sum_{j=1}^N \frac{\partial~}{\partial x_j}
\left( G_{ij}P \right) \right]
\end{equation}
where the matrix $G$ is 
\begin{equation}
G =  g  \cdot  g^T
\end{equation}
For the case of relaxational (gradient or non--gradient)
flows, an explicit solution in the
asymptotic regime of the Fokker--Planck equation can be found if the
relation $G=S$ is satisfied and $S$ is a constant matrix. 
In this case one can show
easily that the stationary solution is given by:
\begin{equation}
\label{eq4:25}
P_{st}(x) = N \exp\left\{-\frac{V(x)}{\epsilon}\right\}
\end{equation}
In some sense, we can say that the stationary distribution is still
governed by the minima of the potential, since the maxima of this
stationary distribution will be centered around these minima, see
Fig. (\ref{fig:4-5}). 
In Statistical Mechanics we encounter a large variety of dynamical
systems that can be described by the stationary distribution 
(\ref{eq4:25}) \cite{HH}. 
The potential $V$ is the free energy $F$ of the system and
the noise terms have a thermal origin and, consequently,
the noise intensity is nothing but the absolute temperature, $\epsilon=k_BT$.
The equality $G=S$ is known as the fluctuation--dissipation 
relation \cite{HH}
and it is the one that ensures that the asymptotic (equilibrium)
distribution of a dynamical model defined by eqs. (\ref{eq4:21})
and (\ref{eq4:20}) is the Gibbs distribution $\exp[-F/k_BT]$. Sometimes,
noise has to be added to the equations of motion in order to reach thermal 
equilibrium. This is, for instance, Cook's modification \cite{cook}
to the Cahn--Hilliard model for spinodal decomposition mentioned above
\cite{Gunton83}.

The situation for non--relaxational flows is somewhat more complicated.
General conditions for the validity of (\ref{eq4:25}) and their
relation with propoerties of detailed balance have 
been discussed in \cite{smc}. We explain now some particular results.
Let us consider a general non--relaxational potential flow with the addition of 
noise terms:
\begin{equation}
\dot x_i = f_i(x) + \sum_{j=1}^N g_{ij}\xi_j(t)
\end{equation}
with
\begin{equation}
f_i = -\sum_{j=1}^NS_{ij}\frac{\partial V}{\partial x_j} +v_i
\end{equation}
The noise variables $\xi_j(t)$ are Gaussian distributed with zero mean
and correlations given by (\ref{eq4:20}).
It can be easily shown that: if (i) $S$ is a constant matrix, 
(ii) the fluctuation--dissipation relation $G=S$ holds, (iii) the 
residual term $v_i$ satisfies the orthogonality condition, eq. (\ref{eq4:10a}),
and (iv) $v_i$ is divergence free:
\be
\label{eq4:25a}
\sum_i \frac{\partial v_i}{\partial x_i} = 0
\ee
then the stationary distribution is still given by (\ref{eq4:25}). In this
case $v_i$ can be easily interpreted as a non dissipative contribution
to the dynamics since it does not contribute to the time variation of $V$
and (\ref{eq4:25a}) implies that it conserves phase space volume.
These conditions are satisfied, for example,
if the non--dissipative contribution
is of the form $v_i= -\sum_j A_{ij}\partial_{x_j} V$ with $A$ an antisymmetric
matrix. 

An important result, due to Graham \cite{gra1,gra3,grahamsitges} states quite generally that:
For a non--relaxational potential flow (i.e. the orthogonality condition
(\ref{eq4:12}) is satisfied) and the fluctuation--dissipation
relation $S=G$ holds, then the
stationary distribution is still governed by (\ref{eq4:25}) 
in the small noise limit. More precisely, we have
\begin{equation}
V(x) = \lim_{\epsilon \to 0} -\epsilon \ln P_{st}(x)
\end{equation}
In those case the potential $V(x)$ is given also 
the name of ``Graham potential",
and the effect of the noise terms on the
asymptotic dynamics is to introduce fluctuations (governed by the
Gibbs distribution)
around the remanent dynamics which occurs in the atractors identified
by the minina of $V$.

\subsection{The K\"uppers-Lortz instability}

When a system displaying the Rayleigh--B\'enard instability 
\cite{HohCross} is put under
rotation, the action of the Coriolis forces induces what is called the 
K\"uppers--Lortz 
(KL) instability \cite{KL}. This is relevant  to many geophysical 
and astrophysical fluids which are subjected to, e.g.,  planet rotation.
The KL instability can be described as follows: for 
an angular rotation speed, $\Omega$, greater than some critical 
value $\Omega_c$, the convective rolls, in the reference frame that rotates
with the system, alternatively change orientations between three preferred
directions. 

Based on the fact that, in a first approximation,
only three directions are relevant to this problem,
Busse and Heikes \cite{BH} have proposed a dynamical model to study the KL
instability. In this model the amplitudes of the three rotating modes,
$A_1,~A_2,~A_3$, follow the evolution equations:
\be
\left\{
\begin{array}{rcl}
\dot A_1 & = & A_1\left(1-|A_1|^2-(1+\mu+\delta)|A_2|^2-
(1+\mu-\delta)|A_3|^2\right) \\
\dot A_2 & = & A_2\left(1-|A_2|^2-(1+\mu+\delta)|A_3|^2-
(1+\mu-\delta)|A_1|^2\right) \\
\dot A_3 & = & A_3\left(1-|A_3|^2-(1+\mu+\delta)|A_1|^2-
(1+\mu-\delta)|A_2|^2\right) \\
\end{array}
\right.
\label{eq4:bh}
\ee
$\mu$ is a parameter related mainly to the temperature gradient and 
$\delta$ is related to the rotation speed $\Omega$ in such a way 
that $\Omega=0$
implies $\delta=0$. We will consider only $\Omega > 0$, or $\delta > 0$; the
case $\Omega < 0$ ($\delta < 0$)
follows by a simple change of the coordinate system.
Finally, although $A_i$ are complex numbers, their phase effectively disappears
from the previous equations and they can be considered real variables. 
A similar set of equation has been proposed to study population 
competition dynamics. If we a have a single biological species, 
the Verhulst or logistic model assumes that 
its population $N(t)$ satisfies the evolution equation: 
\be
\frac{dN}{dt} = r N (1-\alpha N)
\ee
where $r$ is the reproductive growth rate and $\alpha$ is a coefficient denoting
competition amongst the members of the species (for food, for example). If
three species are competing together, it is natural to modify these
equations to a Gause--Lotka--Volterra form \cite{ML}:
\be
\left\{
\begin{array}{rcl}
\dot N_1 & = & r_1 N_1\left(1-N_1-\alpha N_2-\beta N_3\right) \\
\dot N_2 & = & r_2 N_2\left(1-N_2-\alpha N_3-\beta N_1\right) \\
\dot N_3 & = & r_3 N_3\left(1-N_3-\alpha N_1-\beta N_2\right) \\
\end{array}
\right.
\ee
which, after suitable rescaling, are the same that the Busse-Heikes 
equations for the modulus square $a_i \equiv |A_i|^2$ of the amplitudes. 

The fixed point solutions of the Busse--Heikes equations are either of
two types: (a) roll (R) solutions and (b) hexagon (H) solutions.

(a) Roll solutions. There are three of these solutions, each characterized
by a unique non-vanishing amplitude, for instance: $(A_1,A_2,A_3)=(1,0,0)$
is a roll solution with rolls parallel to the $\hat e_1$ direction, and so on.

(b) Hexagon solutions. These are stationary
solutions in which the three amplitudes are different from 0, namely 
$A_1^2=A_2^2=A_3^2= \frac{1}{3+2\mu}$.

\begin{figure}
\begin{center}
\def\epsfsize#1#2{0.82\textwidth}
\leavevmode
\epsffile{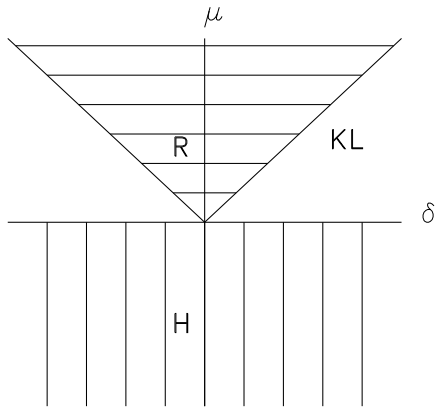}
\end{center}
\caption{\label{fig:4-6}
Stability regions for the Busse-Heikes dynamical system 
(\ref{eq4:bh}). The region H is where the hexagon solution (three equal
amplitudes) is stable. In the R region, the three roll
solutions are stable, and in region KL there are 
no stable fixed points.}
\end{figure}
The stability of the roll and the hexagon solutions can be studied by 
means of  a linear stability analysis.  The result is summarized in 
Fig. (\ref{fig:4-6}). For $\mu < 0$ the only stable situation
is the hexagon solution. For $\mu > 0$, but $\mu > \delta$, the
roll solution is stable and when $\mu < \delta$ however, the roll
solution is unstable. This is the KL instability that occurs at
$0 < \mu = \delta$ and can be described as follows: consider the
roll solution $(A_1,A_2,A_3) = (1,0,0)$, the mode $A_2$ starts growing
and $A_1$ decreasing in order to reach the roll solution $(0,1,0)$. However,
this new roll solution is also unstable, and before it can be reached,
the dynamical system starts evolving towards the
roll solution $(0,0,1)$, which is unstable and evolves towards the
solution $(1,0,0)$ which is unstable ... This is the KL instability that 
shows up in the rotation of the convective rolls.
\be
(1,0,0) \to (0,1,0) \to (0,0,1) \to (1,0,0) \to (0,1,0) \dots
\ee
In brief: we have in the KL region 3 unstable fixed points, each one
of them evolves in the next one.

Further insight into the dynamical system can be obtained
by rewriting the BH equations showing up explicitly that the terms
proportional to $\delta$ do not have a relaxational origin. If we
introduce the potential function $V(A_1,A_2,A_3)$:
\be
\label{eq4:210}
V(A_1,A_2,A_3) = \frac{-1}{2}(A_1^2+A_2^2+A_3^2) +
\frac{1}{4}(A_1^4+A_2^4+A_3^4) +
\frac{1+\mu}{2}(A_1^2A_2^2+A_2^2A_3^2+A_3^2A_1^2)
\ee
we can write:
\be
\label{eq4:200}
\frac{dA_i}{dt} = -\frac{\partial V}{\partial A_i} + v_i
\ee
where
\be
\begin{array}{rcl}
v_1 & = & \delta A_1(-A_2^2+A_3^2) \\
v_2 & = & \delta A_2(-A_3^2+A_1^2) \\
v_3 & = & \delta A_3(-A_1^2+A_2^2) \\
\end{array}
\ee 

If $\delta=0$ the system is relaxational gradient and the corresponding 
stability regions can be obtained also by looking at the minima of the potential
$V$. For the roll solution, the potential takes the value:
\be
V_{R}= \frac{-1}{4} \hspace{2.0cm} {\rm Roll ~ solution}
\ee
whereas for the hexagon solution:
\be
V_H= \frac{-3}{4(3+2\mu)} \hspace{2.0cm} {\rm Hexagon ~ solution}
\ee
According to this potential 
function, and for $\delta=0$, the rolls are more stable than the
hexagons whenever $V_R < V_H$, i.e. $\mu > 0$. Unfortunately, this 
simple criterion does not have an equivalent in the non 
relaxational case, $\delta \ne 0$.

\begin{figure}[h]
\begin{center}
\def\epsfsize#1#2{0.82\textwidth}
\leavevmode
\epsffile{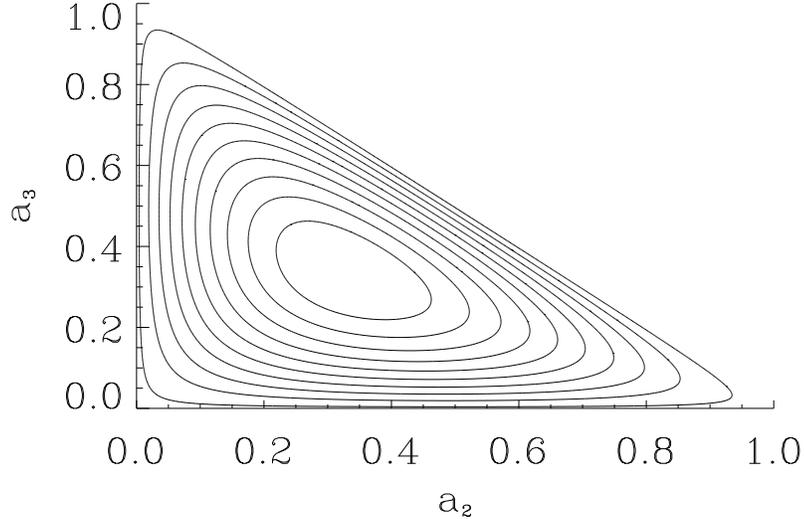}
\end{center}
\caption{\label{fig:4-7} Asymptotic dynamics for $\mu=0$. Here we plot
the level curves $a_2a_3(1-a_2-a_3)=H$ for the Hamiltonian $\hat \cH$
defined in (\ref{eq4:220}). The allowed
values for $H$ range between $H=1/27$ (most inner level curves) and $H=0$
external level curves.}
\end{figure}

The potential $V$ acts as a Lyapunov function in the potential case, 
$\delta=0$. This is obvious from the form (\ref{eq4:200}) of
the dynamical equations. For the case $\delta \ne 0$ we write down
the orthogonality condition (\ref{eq4:10a}) and it is satisfied 
by the same potential function (\ref{eq4:210}) provided the following
condition is satisfied:
\be
\delta \mu = 0
\ee
The case $\delta=0$ is the relaxational gradient case.
The case $\mu=0$ is very interesting because the movement 
appears to be non--relaxational potential with a known potential.
We will explore later this interpretation. Now we 
solve ``essentially" the dynamical equations for $\mu=0$.
By ``essentially" we mean that we
can establish the asymptotic period of movement, after a transient
time of order $1$. First we write down the general 
equations for the variables
$a_i$ (time has been rescaled by a factor of $2$):
\be
\begin{array}{rcl}
\dot a_1 & = & a_1(1-a_1-1+\mu(a_2+a_3)- \delta(a_2-a_3)) \cr
\dot a_2 & = & a_2(1-a_1-1+\mu(a_2+a_3)- \delta(a_3-a_1)) \cr
\dot a_3 & = & a_3(1-a_1-1+\mu(a_2+a_3)- \delta(a_1-a_2)) \cr
\end{array}
\ee
If we introduce
now the variable $x(t)=a_1+a_2+a_3$, it is straightforward to show
that it satisfies the evolution equation:
\be
\dot x = x(1-x)-2\mu y
\ee
where:
\be
y(t)=a_1a_2+a_2a_3+a_3a_1
\ee
In the case $\mu=0$ the equation for $x(t)$ is a closed equation 
whose solution is 
\be
\label{eq4:205}
x(t) = \frac{1}{\left( \frac{1}{x_0}-1 \right) \e^{-t} +1}
\ee
(here $x_0=x(t=0)$). From this expression it turns out that 
$\lim_{t \to \infty} x(t) = 1$ independently of the initial condition.
In practice, and due to the exponential decay of the above expression,
after a transient time of order 1, $x(t)$ already takes its asymptotic
value $x(t)=1$. Therefore, we can substitute $a_1(t)$, say, by
$1-a_2(t)-a_3(t)$ to obtain evolution equations for $a_2(t)$ and 
$a_3(t)$. In this way, the original 3-variable problem is reduced to
a 2-variable one:
\be
\label{eq4:60}
\begin{array}{rcl}
\dot a_2 & = & - \delta a_2(1-a_2-2a_3) \cr
\dot a_3 & = & - \delta a_3(1-2a_2-a_3) \cr
\end{array}
\ee
These equations have a Hamiltonian form:
\be
\begin{array}{rcl}
\dot a_2 & = & \delta \frac{\partial \hat {\cal H}}{\partial a_3} \cr
\dot a_2 & = & -\delta \frac{\partial \hat {\cal H}}{\partial a_2} \cr
\end{array}
\ee
with the Hamiltonian:
\be 
\label{eq4:220}
\hat {\cal H} = a_2a_3(1-a_2-a_3)
\ee
As a consequence of this Hamiltonian structure for the 
dynamics valid for $\mu=0$,
it turns out that the ``energy" $H=\hat {\cal H}(t)$ is
a constant of movement after the transient regime of time (of order
$1$). After this transient, the
Hamiltonian description is valid. The movement in the line ${\cal H}(t)=H$,
see Fig. (\ref{fig:4-7}), is periodic in the variables
$a_2$, $a_3$ (and hence in the $a_1$ variable also).

\noindent The value of $H$ depends only on initial conditions. To check this
let us use the following equivalent (in the asymptotic time limit)
definition of $\cH$:
\be
\cH = a_1 a_2 a_3
\ee
From this definition, and for arbitrary value of $\mu$,
we obtain the following exact equation for $\cH$:
\be
\cH^{-1}\frac{d \cH}{dt} = 3-(3+2\mu)x
\ee
(one can reduce the original Busse--Heikes model to 
dynamical equations for variables $x,~y,~\cH$ but the equation
for $\dot y$ turns out to be too complicated, see  \cite{soward}).
If we substitute the solution (\ref{eq4:205}) 
for $x(t)$ valid in the case $\mu=0$ we obtain:
\be
\cH(t)= \cH_0\left[(1-x_0)e^{-t}+x_0\right]^{-3}
\ee
with $\cH_0=\cH(t=0)= a_1(0)a_2(0)a_3(0)$. 
The asymptotic value for $\cH$ is 
\be
H = \lim_{t \to \infty} \cH(t) = \cH_0 x_0^{-3} = 
\frac{a_1(0)a_2(0)a_3(0)}{(a_1(0)+a_2(0)+a_3(0))^{3}}
\ee
Again, this asymptotic value is reached after a transient time of 
order 1. 

\begin{figure}[h]
\begin{center}
\def\epsfsize#1#2{0.82\textwidth}
\leavevmode
\epsffile{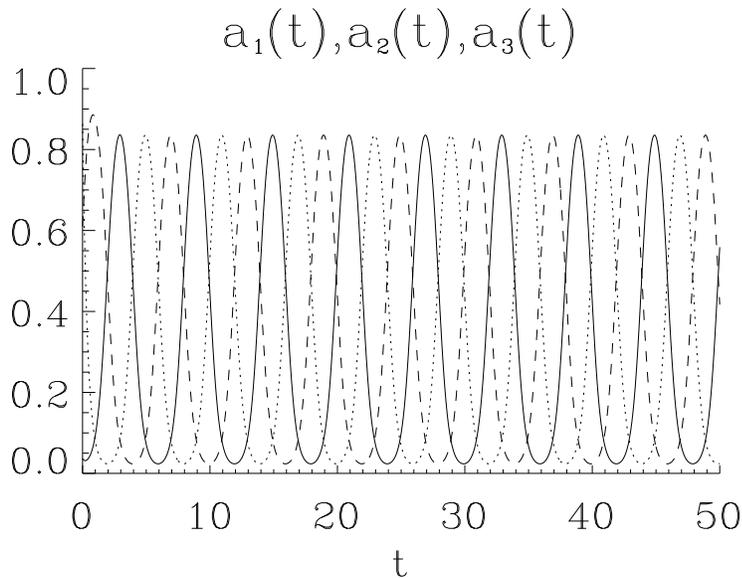}
\end{center}
\caption{\label{fig:4-8}
Periodic trajectories in the case $\mu=0$, $\delta=1.32$. The 
three lines, solid, dotted and dashed, show the
time evolution of $a_1(t)$, $a_2(t)$ and $a_3(t)$, respectively. 
Notice that the
movement is periodic with a well defined period.}
\end{figure}

The behavior of the dynamical system in the case $\mu=0$ 
can now be fully described and interpreted:
after a transient time (or order $1$)
the three variables $a_1$, $a_2$, $a_3$ vary periodically in time
such that $a_1+a_2+a_3=1$. When $a_1$ decreases, $a_2$ increases, etc.
This is characteristic of the Kuppers-Lorz instability. The motion
is periodic because it is a Hamiltonian orbit with a fixed energy.
The orbits can be plotted as closed trajectories in the $a_1+a_2+a_3=1$ plane.
The exact shape of the trajectory depends on the value of
the energy$H$ which, in turn, depends on initial conditions. 
The period of the orbit can also be computed as a function of the energy 
in term of elliptic functions, $T=T(H)$. 
The exact expression is not needed
here, but we mention that when the energy tends to zero the period
tends to infinity, $\lim_{H \to 0} T(H) = \infty$. 
This periodic movement, see Fig. (\ref{fig:4-8}),
with a well defined period
which appears for $\mu=0$ is characteristic of the KL instability.

We now interpret the above characteristic of the movement in terms of the
non--relaxational potential 
flow plus the orthogonality condition: the relaxational terms
in the dynamics make the system evolve towards the 
degenerate minimum of the potential which for $\mu=0$ occurs at
$a_1+a_2+a_3=1$. 
The residual movement is governed by the non--potential part, see
also Fig. (\ref{fig:4-4}). Notice that
this residual movement, equations (\ref{eq4:60}), disappears for $\delta=0$, the
relaxational gradient case.

\begin{figure}
\begin{center}
\def\epsfsize#1#2{0.82\textwidth}
\leavevmode
\epsffile{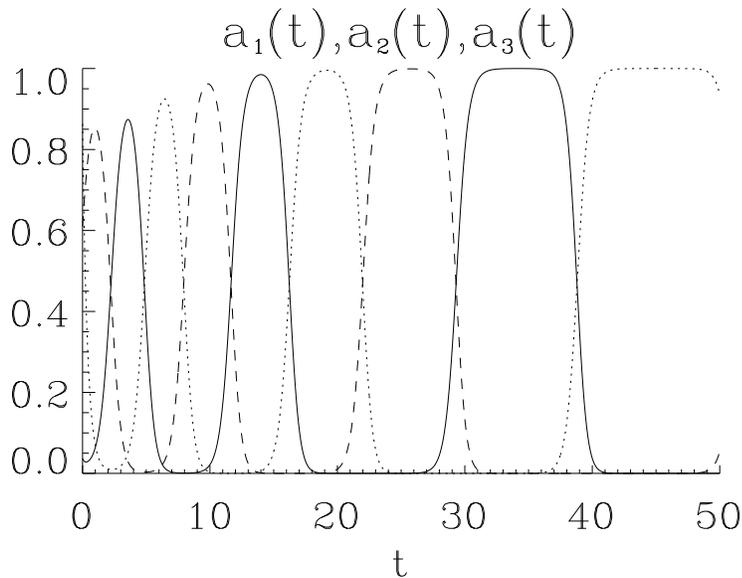}
\end{center}
\caption{\label{fig:4-9}
Trajectories in the case $\mu=1.20$, $\delta=1.32$. Notice that the
movement is such that the time interval between the domination periods
of a given mode increases with time.}
\end{figure}

Once we have understood the case $\mu=0$, we now turn to $\mu > 0$. The 
orthogonality condition is no longer satisfied and we can not get such a nice
picture of the movement as before. 
However, we can understand what is going on in 
the following terms: It turns out that
the energy is no longer a constant of movement after a transient time,
but decreases exponentially
with a characteristics decay time which is of order $\mu^{-1}$ \cite{ML}. 
Consequently, and according to the previous analysis, 
the period of the orbits, 
which is a function of the energy, increases with time. 
We understand in this way a basic feature of the Busse--Heikes model
for the KL instability in the case $0 < \mu < \delta $: 
the increase of the period between successive
alternation of the dominating modes, see Fig. (\ref{fig:4-9}).

\begin{figure}[h]
\begin{center}
\def\epsfsize#1#2{0.82\textwidth}
\leavevmode
\epsffile{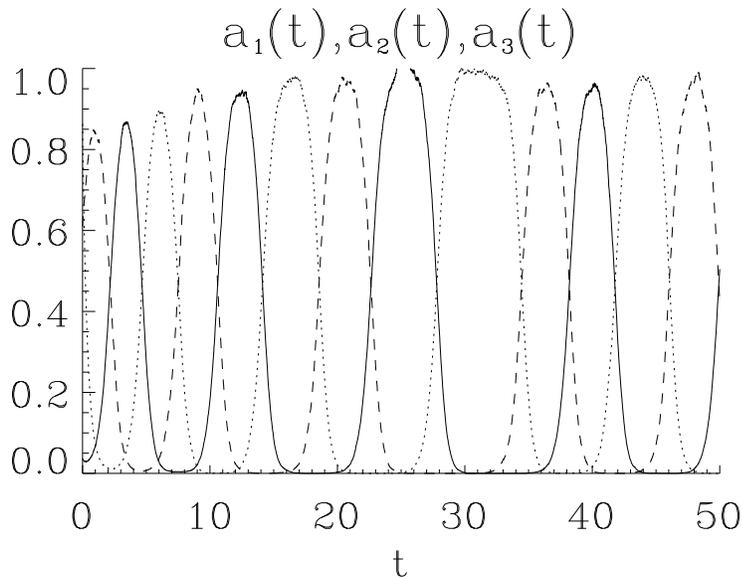}
\end{center}
\caption{\label{fig:4-11} 
Temporal evolution of amplitudes in the case $\delta =1.32$, $\mu=1.20$, 
$\epsilon = 10^{-4}$. In this case, the movement is such that the
time interval between dominations of a single mode fluctuatues around
a mean value.}
\end{figure}

This is, indeed, an unwanted feature, since
the experimental results do not show such an period increase. Busse and Heikes 
were fully aware of this problem and suggested that noise (``small amplitude
disturbances" as they called them), that is present at all times prevents the
amplitudes from decaying to arbitrary small levels and a movement 
which is essentially periodic but with a fluctuating period is 
established.\footnote{It is necessary to point out that Tu and 
Cross \cite{tc} have proposed an 
alternative explanation for the stabilization of the period
without the necessity of the inclusion of the noise term: they modify the 
Busse--Heike equations by considering a two--dimensional amplitude field
and including spatial variation terms.} We modify, then, Busse--Heikes
equations by the inclusion of noise terms:
\be
\begin{array}{rcl}
\dot a_1 & = & a_1(1-a_1-1+\mu(a_2+a_3)- \delta(a_2-a_3)) + \xi_1(t) \cr
\dot a_2 & = & a_2(1-a_1-1+\mu(a_2+a_3)- \delta(a_3-a_1)) + \xi_2(t) \cr
\dot a_3 & = & a_3(1-a_1-1+\mu(a_2+a_3)- \delta(a_1-a_2)) + \xi_3(t) \cr
\end{array}
\ee
where $\xi_i(t)$ are white--noise processes with correlations:
\be
\langle \xi_i(t) \xi_j(t') \rangle = 2 \epsilon \delta(t-t') \delta_{ij}
\ee
Numerical simulations of these equations for small noise amplitude $\epsilon$ 
shows that the role of noise is that 
of stabilizing the orbits around a mean period. We can understand 
this in the following terms \cite{tmsm}: 
the inclusion of noise has the effect of injecting 
``energy" into the system. As a consequence, the energy ${\cal H}$ no longer
decays to zero but it stabilizes 
around the mean value $\langle H \rangle$. In this way a periodic movement with 
a fluctuating period is produced, see Fig. (\ref{fig:4-11}). 
The mean period $\langle T \rangle$ can be
computed from the mean energy $\bar H$ by using 
the same function $\langle T \rangle =T(\langle H \rangle )$ 
that was deduced in the Hamiltonian case.

We might interpret this in terms of the discussion of Sect. (4.2).
For $\mu = 0$ and $\delta > 0$ the function $V$ given by 
(\ref{eq4:210}) is a Lyapunov potential since it satisfies
the orthogonality condition. For $\mu >0 $ this is no longer true
but we should expect that for small $\mu$ a perturbative solution of the
orthogonality condition should give us a potential function $V^{(\mu)}$
that differs from $V$ it terms that vanish for vanishing $\mu$.
In the absence of noise, the dynamics leads the system to the minima of
the potential $V^{(\mu)}$ and the residual movement in this attractor
is one of increasing period between successive alternation of amplitude modes,
see Fig. (\ref{fig:4-9}). When
noise is switched--on, however, fluctuations in the residual motion
stabilize the mean period
to a finite value. The mechanism for this is that fluctuations are amplified
when the trajectory comes close to one of the unstable fixed points of
the dynamics and the trajectory is then repelled towards
another unstable fixed point. The fluctuating period is, hence, sustained
by noise.

\section{Noise effects in spatially extended systems}

\setcounter{equation}{0}
\setcounter{figure}{0}
So far we have considered noise effects in time dependent properties for which spatial
degrees of freedom are not important. However dynamical noise also has important consequences in
pattern formation phenomena \cite{HohCross,AhlersSitges,Walgraef97} occurring in spatially extended systems. We
consider now some of these relevant effects.

\subsection{Symmetry restoring by noise in pattern formation}
Pattern formation is associated with spatial symmetry breaking, but broken 
symmetries might be restored by noise \cite{SanMiguel93}. Equilibrium phase transitions is a well known class of 
problems associated with symmetry breaking. A phase transition takes place when, in the 
thermodynamic limit, thermal fluctuations are not able to mix states with different symmetries. 
This only occurs for a large enough spatial dimensionality, which depends on the type of 
broken symmetry. Noise in phase transitions is discussed in Sect. 6.
 An example of the opposite situation of symmetry restoring by noise is 
given by the laser instability. The description of this instability (Sect. 3) is zero 
dimensional (that is, described in terms of ordinary SDE) and spontaneous emission noise restores the
 phase symmetry of the lasing field 
after a time of the order of the coherence time: In the long time limit the phase noise
$\xi_{\phi}$ implies that there is no preference for any phase $\phi$.
 Pattern formation usually occurs in open 
systems of low dimensionality, which corresponds to situations somehow intermediate between 
the two discussed above. Still, in these systems noise may restore the broken 
symmetry 
implied by the emergence of a pattern, therefore destroying long range order. Pattern formation
can be sometimes described in terms of ordinary SDE for the amplitudes of a few spatial modes,
while sometimes a fully continuous description is needed in terms of partial SDE. The role of
noise in the Kuppers-Lortz instability (Sect. 4) falls within the first category, as well as an
example of optical pattern formation that we discuss next. A prototype of the second category
is the stochastic Swift-Hohenberg
 equation considered later in this section. We mention here that the opposite phenomenon of
symmetry restoring by noise is the one of  symmetry breaking by noise in the large system limit.
Such phenomenon is the one of a noise--induced phase transition considered in Sect. 6.

The study of spatial and spatiotemporal phenomena in nonlinear optics has emerged 
\cite{Transverse} as an interesting alternative to more conventional pattern formation 
studies in fluids. For lasers with high Fresnel number (the equivalent of large aspect ratio in
hydrodynamics) the transverse profile of the laser beam displays pattern formation caused by
diffraction and nonlinearities. As an example of this, it is known that \cite{Lugiato89}
 when the pump 
bias level of an optically pumped laser is increased above a critical value, a spontaneous 
breaking of cylindrical symmetry in the transverse profile of the laser beam occurs. The
emerging state  displays a spatial pattern associated with phase locking of three appropriate
spatial modes. The  angular position of the pattern fluctuates strongly over time intervals
larger than a  few miliseconds, so that symmetry is restored by a phase diffusion mechanism
induced by noise \cite{ColetLugiato}.
 We will describe now this phenomenon. Further instabilities of such pattern (not considered
here) have also been  analyzed \cite{Brambilla91}. 

The slowly varying envelope $E$
of the electric field inside a ring laser with spherical 
mirrors can be expanded in terms of Gauss-Laguerre cavity modes $A_{i} (\rho, \varphi)$.
They describe 
the transverse profile of the beam with $\rho$ and $\varphi$ being the polar transverse
coordinates. In some cavity resonance conditions \cite{ColetLugiato} the electric field can 
be described in terms of just three modes with complex amplitudes $f_1,f_2,f_3$:
\begin{equation}
E ( \rho , \varphi, t) = \sum^{3}_{i = 1} f_i (t) A_{i} (\rho, \varphi),  
\end{equation}
\begin{equation}
d_{t} f_{i} =  - f_{i} + 2 C [ M_{i} f_{i} - \sum_{j, k, l} 
A_{ijkl} f_{j} f_{k} f_{l}^* ] + \xi_{i} (t), 
\end{equation} 
where $C$ is the pump parameter, 
$M_{i},A_{ijkl}$ given coefficients and $\xi_{i}$
 model spontaneous 
emission noise as independent Gaussian white noise processes.
In this description the system becomes effectively zero dimensional 
with 3 relevant degrees of 
freedom. $A_1$ is independent of $\varphi$, 
while $A_2$, $A_3$ have a cosine and sine profile respectively.
In a noise-free framework, the laser instability occurs at 
$C=C_{th}$, leading to a cylindrically 
symmetrical pattern with 
$f_{1} \not= 0$, $f_{2} = f_{3} = 0$. Increasing further 
the pump parameter above a second threshold  $C = C_A$
the modes $f_2,f_3$ become unstable and a new pattern 
emerges. This pattern shown in Fig (5.1a) breaks the 
cylindrical symmetry. The side peaks 
disappear for $C < C_A$ when $f_{2} = f_{3} = 0$. 
A continuous set of solutions with any angular orientation exists 
and the system spontaneously
breaks the symmetry by choosing one of these orientations.
The dynamical system (5.2) is of the class of relaxational gradient dynamical 
systems (Sect. 4):
\begin{equation}
 d_{t} f_{i}^{R,I} = - {\partial V \over \partial f_{i}^{R,I}} + 
\xi_{i}^{R, I}, 
\end{equation}
where $f_{i}^{R,I}$ are the real and imaginary parts of $f_i$.	Introducing new mode variables  
$g_i=|g_i| e^{\beta_i}$, $g_1=f_1$, $g_{2, 3} = (f_2 \pm i f_3)/\sqrt{2}$, the
angular orientation of the pattern is given by the variable 
$\eta=(\beta_3-\beta_2)/ 4$. The
deterministic stationary solutions of the problem are 
given by the degenerate  minima of the potential $V$ for which
at $\vert f_{2} \vert = \vert f_{3} \vert$, 
$\delta = \pm \pi / 4$, $ \pm {3
\pi/ 4}$, with $\delta=(\beta_2 +\beta_3)/4 - \beta_1/2$. 
The discrete values of $\delta$ evidentiate phase locking.

Symmetry restoring by 
noise manifests itself in the fact that the potential $V$, and therefore the 
stationary solution
$P_{st}$ (4.36) of the Fokker-Planck equation associated 
with (5.3) is independent of the variable 
$\eta$. Dynamically, the effect of noise is 
to restore the cylindrical symmetry by driving the system 
through different values of $\eta$. This 
effect of noise can be visualized by a numerical integration 
of the stochastic equations for the 
modal amplitudes \cite{ColetLugiato}. Taking as the initial condition 
a deterministic solution 
with arbitrary 
orientation, Fig. (5.1a), fluctuations of the modal 
amplitudes induced by noise lead to a
changing  pattern in time with global rotation 
and distortion, Fig (5.1b). A long time average,
or an  ensemble average over many realizations 
(as given by $P_{st}$), results in an averaged
cylindrically symmetric  pattern.
\begin{figure}
\begin{center}
\def\epsfsize#1#2{0.80\textwidth}
\leavevmode
\epsffile{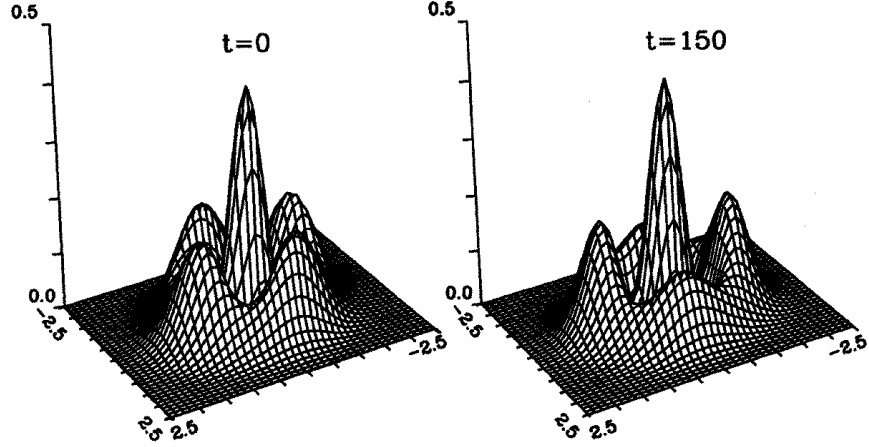}
\end{center}
\caption{ (a) Noise free stationary pattern as obtained from (5.1)-(5.2) and a stationary 
solution of
(5.2) for $C>C_A$. (b) Configuration at time t=150 obtained from the stochastic evolution of
the configuration in (a) with noise level $\epsilon=0.001$}
\label{pattgigi}
\end{figure}
The dynamics of symmetry restoring is described by the stochastic equation for $\eta$:
\begin{equation}
d_{t} \eta = c I_{1} {I_{2} - I_{3} \over \sqrt{I_{2} 
I_{3}}} \sin 4 \delta + {1 \over 4} ( {1 \over \sqrt{I_{3}} } 
\xi^{'}_{3} - {1 \over \sqrt{I_{2}} } \xi^{'}_{2} ),
\end{equation}
where $c$ is a constant proportional to the pump $C$, $I_i$=$|f_i|^2$ and
$\xi^{'}_{2,3}$
are real and independent Gaussian random processes.  The drift 
part vanishes in the deterministic steady states. For those states, (5.4)
describes a diffusive motion of $\eta$ (Wiener process) which is the basic physical mechanism of
symmetry restoring. We also note that the pattern shown in Fig. (5.1) has four phase
singularities in  which $E ( \rho , \varphi) =0$. These singular points also exhibit a
diffusive motion \cite{ColetLugiato}.

The mechanism of symmetry restoring by phase diffusion also holds in systems which require a fully 
continuous description. If we consider a one dimensional system with a stationary pattern forming
instability, the relevant pattern forming variable $\psi$ can be expressed, close to the
instability, as \cite{HohCross}
\begin{equation}
\psi(x,t) = A(x,t) e^{i q_{M}x} + {\rm c.c.},
\end{equation}
where $A=R e^{i \varphi}$ is the slowly varying amplitude of the most unstable mode
$q_M$. Invariance under continuous space translations of the original problem
 ($x \rightarrow x + x_0$) implies gauge invariance under a change of a constant
phase of the amplitude  ($\varphi \rightarrow \varphi + \varphi_{0}$). Spontaneous symmetry 
breaking associated with pattern formation is here the choice of a particular phase of $A$, which
means fixing a global position of the pattern by choosing among a continuous family of pattern 
forming solutions. The phase is
the Goldstone mode associated with the broken symmetry, which generally satisfies a
diffusion equation \cite{Pomeau79}: The linear evolution of longwavelength modulations of a
pattern with dominant wave number $q+q_M$ is given by
\begin{equation}
\partial_{t} \varphi_q(x,t) = D (q) \partial^{2}_{x} \varphi_q(x,t) + \xi (x,t).
\end{equation}
The Eckhaus instability \cite{HohCross,Fauve} is given by the wavenumber $q_E$ for which $D(q_E)=0$.
The condition
$D(q)>0$ identifies the range of wave numbers around $q_M$ which give linearly stable patterns.
Within the range of Eckhaus stable patterns, we have pure stochastic diffusion (no drift term) of 
the $k=0$ Fourier mode component of the phase $\varphi_q(x,t)$. 
For other $k$-modes we have damped phase fluctuations driven by the noise $\xi$. Such stationary phase
fluctuations scale as  $\langle \varphi^{2}(k) \rangle \sim k^{-2}$. The integral of 
$\langle \varphi^{2}(k) \rangle$
over all wave numbers diverges in spatial dimension $d < 2$. Dimension $d=2$ is the critical dimension
below which fluctuations associated with a broken continuous symmetry destroy long range order, or
in other words symmetry is restored by noise. In our case this means that fluctuations in the
local phase 
of the pattern preclude the existence of a rigidly ordered pattern with well defined 
wavenumber in a large system. We will now be more specific with these ideas in the example provided
by the Swift-Hohenberg equation.

\subsection{The stochastic Swift-Hohenberg equation: Symmetry restoring, pattern
selection and noise modified Eckhaus instability.}
The Swift-Hohenberg equation (SHE) was originally introduced to model the onset of a 
convective instability in simple fluids \cite{SwiftHoh77}. It describes the dynamical evolution of a scalar real 
variable $\psi (x, t)$,
\begin{eqnarray} 
& \   \partial_{t} \psi (x, t) &= [ \mu - (1 + \partial^{2}_{x} )^{2} ] \ \  
\psi (x, t) - \psi^{3} (x, t) + \sqrt{\epsilon} \xi (x, t) = \cr 
&\  &= - {\delta F \over \delta \psi} + \sqrt{\epsilon} \xi (x, t), \ \
\end{eqnarray} 
where we have explicitly indicated that the SHE has relaxational gradient dynamics determined by  
a Lyapunov functional $F$, and where $\xi(x,t)$ models thermal fluctuations as a Gaussian
process with zero mean and \begin{equation}
\langle \xi (x, t) \xi (x', t') \rangle = \delta (x - x') \delta (t - t').  
\end{equation}
The study of the effect of thermal fluctuations in the onset of pattern formation $(\mu =0)$ was
one of the original motivations in the study of the SHE \cite{SwiftHoh77,Graham74}, but 
they were commonly
neglected for some time because of its smallness for simple fluids. However, in several more 
recent experiments, in particular in the electrohydrodynamic 
instability in nematic liquid crystals, these fluctuations have been either directly or 
indirectly observed \cite{AhlersSitges,Ahlers}.

In the absence of noise ($\epsilon = 0$) 
and above threshold ($\mu >0$), the SHE admits stationary
periodic solutions $\psi_{q} (x)$
characterized by a wavenumber $q$ 
\begin{equation}
\psi_{q} (x) = 
\sum_{i} A_{i} (q) \sin [(2i + 1) qx]\ , \ \ \ q_{-L} < q < q_{L}.
\end{equation}
These solutions exist for a band of wavenumbers centered around $q = q_M =1$ with $q_{\pm L}=
\sqrt{1 \pm \sqrt{\mu}}$. These are
linearly unstable wave numbers for the ©©no pattern" solution $\psi= 0$, being $q = q_M$ the
most unstable mode. We emphasize that this mode does not coincide with the mode $q_{min}$
associated with the periodic solution
which minimizes the potential $F$. Close to threshold one finds \cite{Pomeau79}
\begin{equation}
q_{min} \approx  1 - {\mu^{2} \over 1024}.
\end{equation}

\begin{figure}
\begin{center}
\def\epsfsize#1#2{0.70\textwidth}
\leavevmode
\epsffile{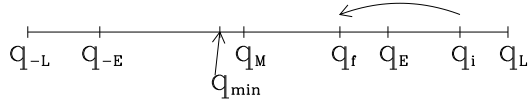}
\end{center}
\caption{\label{rectaSH}
Schematic position of the different wavenumbers mentioned in the text. The decay
from an Eckhaus unstable wavenumber $q_i$ to a final wavenumber $q_f$ is also indicated}
\end{figure}
\noindent Stable periodic solutions only exist in a restricted band 
of wave numbers $q_{-E} < q < q_{E}$,
where $q_{\pm E}$ are the critical wave numbers for the
Eckhaus instability described above \cite{Kramer85}. For small $\mu$, 
$q_{\pm E} \sim 1 + (q_{\pm L} -1) / \sqrt{3}$, Fig. (5.2).
An important fact about the Eckhaus instability in the SHE is that it is a subcritical bifurcation.
Defining $V(q)= F[\psi_q(x)]$, the Eckhaus boundary is given by the wavenumbers such that
${\partial^{2} V(q) \over \partial q^{2}}=0$, so that $q_E$ plays the role of a spinodal
line in first order phase transitions \cite{Gunton83} separating unstable $(q_L > q > q_E, q
_{-L} < q < q_{-E})$ from metastable states $(q_{-E} < q < q_E)$. The only globally stable
state, in the presence of noise, is the one minimizing
 $F$. For further reference we also note that for the decay of a solution with wavenumber
$q_i$ outside the Eckhaus 
stable range, the initially fastest growing mode is determined by a linearized analysis around a
periodic solution $\psi_{q_i}(x)$ (Bloch eigenvalue problem). This fastest growing mode is no longer
$q_M=1$ but $q_M(q_i)$, a function of $q_i$.

We will now address three questions regarding the role of noise in the SHE. First is how noise destroys long range
order, second the influence of noise in the dynamical process of pattern selection from a given 
initial condition, and finally how noise modifies the Eckhaus instability. Answers to these 
questions \cite{Vinals91,HernandezGarcia92,HernandezGarcia93} are
mostly found from extensive numerical simulations in which a system of size $L$, with periodic
boundary conditions, is discretized in 
a grid of $N$ points $x_i$, with $N$ ranging from $1024$ to $8192$ points and a discretization step 
$\Delta x = L/N = 2 \pi / 32$. For these sizes the number of changes of sign (NCS) of a periodic
solution with $q=q_M$ ranges from $64$ to $512$, which identifies the number of convective rolls
in  the system. 

Our general arguments above imply that there is no long range order in this one dimensional system
when any amount of noise is present ( a discussion of the $d=2$
case is given in \cite{Elder}). A
quantitative measure of this statement can be given in terms of a stationary structure factor defined
as
\begin{equation}
\lim_{t \rightarrow \infty}  \ P (q, t) = {1 \over N} \vert 
\sum^{N}_{i = 1} \psi (x_{i}, t) e^{- i q x_{i}} \vert^{2}.
\end{equation}
Alternatively one can look at the normalized correlation function
\begin{equation}
 G (r) = \lim_{t \rightarrow \infty} {\langle \psi (x + r, t) \psi (x, t) \rangle 
 \over \langle \psi (x, t) \rangle^{2} }.
\end{equation}
The numerical results make explicit the absence of long range order in several ways. Consider first
a single realization of the process, that is the stochastic evolution of a configuration
$\psi (x, t)$. The long time dynamics is characterized by mode competition with no single dominant mode,
so that no periodic solution is established. For small systems ($N=1024$) this is evidentiated by
a power spectrum $P (q, t)$ which evolves in time with a hopping maximum wave number and a few
modes being permanently excited by noise. The absence of a global wave number is also evidentiated by considering a local
structure factor at a fixed time. Such local structure factor centered in $x_0$ 
is defined replacing in (5.11) 
$\psi (x_{i}, t)$ by a filtered value around $x_0$:  
$\psi_{x_{0}} (x_{i}, t) = e^{- (x_{i} - x_{0})^{2} / \beta^{2} } \psi (x_{i}, t)$.
The local structure factor $P_{x_0}(q,t)$, shown in Fig. (5.3), shows a well defined peak at a
wave number  that is a function of $x_0$. Therefore, $\psi (x, t)$ is locally periodic, but the
local wave number varies from point to point. For large systems ($N=8192$) the structure factor
of a configuration at a fixed time is broad, with a typical width of around 40 competing 
Fourier
modes, which gives an  idea of the range of coherence. In summary, configurations $\psi (x, t)$
cannot be characterized by a single  wavenumber. 

\begin{figure}
\begin{center}
\def\epsfsize#1#2{0.70\textwidth}
\leavevmode
\epsffile{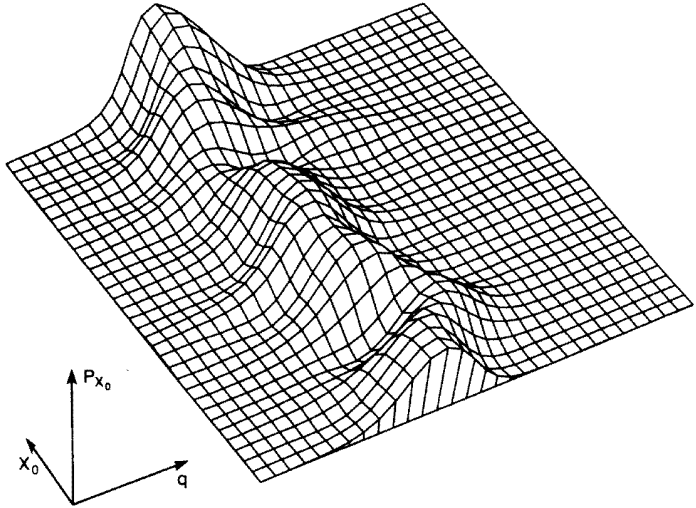}
\end{center}
\caption{Local power spectrum $P_{x_0}(q,t)$ as defined in the text for a system of
N=1024 points. Parameter values: $\mu=0.25$, $\beta=10$ and $\epsilon= 0.2 \Delta x$}
\label{localsf}
\end{figure}
\begin{figure}
\begin{center}
\def\epsfsize#1#2{0.70\textwidth}
\leavevmode
\epsffile{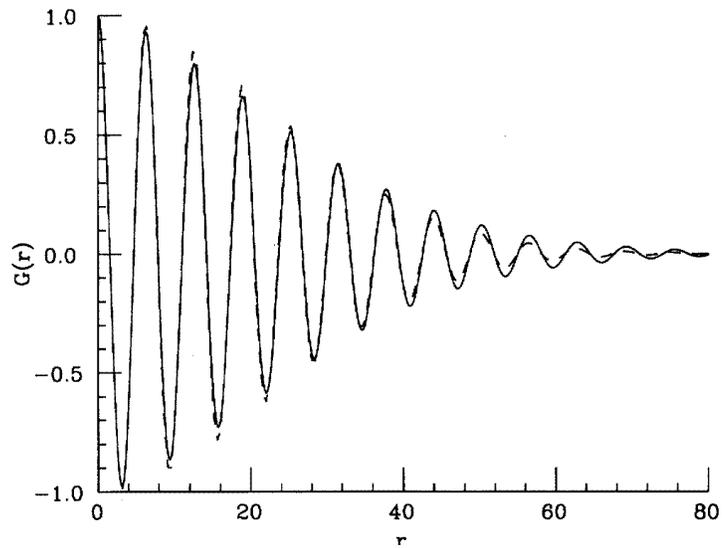}
\end{center}
\caption{Solid line: Correlation function $G(r)\over{G(0)}$ obtained numerically 
as an average over 20 realizations
for a system of N=8192 with parameter values $\mu=0.25$ and $\epsilon= 0.1 \Delta x$.
Dashed line: fit to Eq. (5.13) with $q_M=1$ and $r_0=32$}
\label{corrfct}
\end{figure}
Turning now to ensemble averages over several
realizations, the stationary correlation function shown in  Fig. (5.4), is well fitted by 
\begin{equation}
G (r) = G(0) e^{- ( {r \over r_0})^2} cos (q_M r),
\end{equation}
which identifies a correlation length $r_0$. For the typical parameter values of Fig. (5.4)
we find $L/r_0 \sim 50$ indicating that coherence in the pattern is only maintained in
 $1/50$ of the system size. The Fourier transform of the correlation function, or ensemble
averaged power spectrum, requires a different fitting for the wings of the spectrum
\cite{Vinals91} and its width is found to grow linearly with noise intensity.
We note that despite the absence of long range order, and therefore of a wavenumber for the system,
still a conservation of the NCS holds during the stochastic steady state dynamics. Such conserved
NCS can be used to define a global wave number or a mean periodicity of the system.

The question of pattern selection in this problem can be stated as which of the 
stable solutions $\psi_{q} (x)$ within the Eckhaus stable band is chosen from a given initial
condition. Commonly proposed answers to this question are two: a) An argument based on
relaxational dynamics on $F$ favor the selection of $q_{min}$. 
This is a potential argument
based on the role of  noise along the dynamical path. 
b) The argument of fastest linear growth favors
the selection of $q_M$ since this solution is linearly stable, dominates in the early time
dynamics, and there is conservation of NCS after the initial transient. Two simple comments are here in order. First is that, from a deterministic point of
view, the selected state is entirely determined by initial conditions and any linearly stable
state is reachable from an appropriate initial condition. Secondly is that, from the stochastic
point of view, the question is ill-posed since we have shown that no single wavenumber
characterizes the long time configurations. Still, the question makes sense 
if we ask for the
global wave number corresponding to a selected  NCS. 

From this stochastic point of view one
asks for a distribution of selected wavenumbers obtained as an ensemble average 
from a statistically distributed set of typical initial conditions. A natural situation to consider is the
growth of the pattern from random initial conditions around $\psi=0$. This corresponds
again to the case of the decay of an unstable state (discussed in Sect. 3), created by
changing the control parameter $\mu$ from below ($\mu < 0$) 
to above ($\mu > 0 $) the pattern forming instability.
In this case the numerical solutions indicate that noise along path washes out the influence of
the initial configuration. The final states are statistically distributed around the NCS
corresponding to the global wavenumber $q_M$. However, the distribution is significantly
narrower than the range of Eckhaus stable wavenumbers. Therefore the selected states are not
uniformly distributed among all possible linearly stable states. The issue of selection of
$q_M$ versus selection of $q_{min}$ cannot be clearly answered numerically because the two
wavenumbers are numerically very similar.  An alternative situation to address this question is
to study the decay from a solution with wavenumber $q_i$ which is Eckhaus unstable, Fig. (5.2).
Noise drives the system away from the initial condition and there is a 
transient in  which the NCS
changes, leading to a configuration with a final global wavenumber $q_f$.
 The time dependence of the NCS for a fixed $q_i$ and 
averaged over 20 realizations of the noise 
is shown in Fig. (5.5). 
It is seen that the evolution gets trapped in a metastable state after the 
disappearance of 72 rolls. The final observed NCS is consistent with the global wavenumber 
associated with fastest growth $q_M(q_i)$, 
and it differs from $q_{min}$, which minimizes $F$, in another 
36 rolls. 
Therefore, noise of moderate strength along the dynamical path is not able to drive the 
system, in the explored time scale, to the configuration which minimizes the Lyapunov 
functional of the problem, or equivalently to the most probable configuration in terms of the
stationary solution of the Fokker-Planck equation associated with (5.7). The dynamical mechanism
of pattern selection seems to be by the mode of fastest 
growth and this is robust against moderate fluctuations.  Eventually, and for extremely 
long time scales, it is a rigorous mathematical statement that noise will induce jumps to 
configurations with a lower value of $F$. Of course, larger noise intensity would induce earlier
jumps, but when the noise becomes larger, the conservation of NCS does not hold anymore, and the
question of pattern selection becomes meaningless. 

\begin{figure}
\begin{center}
\def\epsfsize#1#2{0.70\textwidth}
\leavevmode
\epsffile{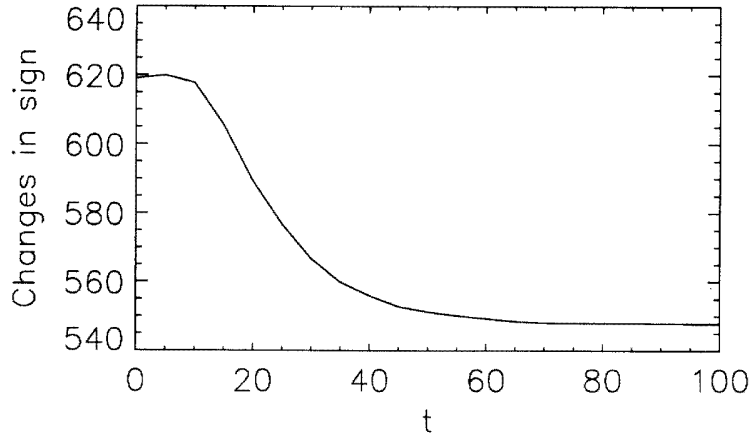}
\end{center}
\caption{Time dependence of the averaged NCS in the decay of an Eckhaus unstable configuration
for a system of N=8192 with $\mu=0.56$ and $\epsilon= 0.1 \Delta x$.  NCS=512 
corresponds to $q=1$}
\label{decayEckhaus}
\end{figure}
The fact that $q_f$ is given by the growth of fastest mode $q_M(q_i)$ associated with the decay from $q_i$
breaks down when $q_i$ becomes close enough to the Eckhaus stability boundary $q_{\pm E}$. In
this case one enters a fluctuation dominated regime in which $q_f$ can be identified neither
with $q_M(q_i)$ nor with $q_{min}$. We already mentioned that the Eckhaus boundary is similar to
a spinodal line in the theory of phase transitions. It is well known in that context that such line
is only a meanfield concept, so that it is not well defined when fluctuations are present
\cite{Gunton83}. A similar results hold here, and the Eckhaus boundary, when fluctuations are
present, is only defined as  an effective stability boundary for a time scale of observation
\cite{HernandezGarcia93}. 

To clarify this point we need first to characterize the range of
$q_i$ for which we are in a fluctuation dominated regime. We can give a criterion which, being
similar to the Ginzburg criterion of phase transitions \cite{huang}, 
is based on the passage time ideas
discussed in Sect. 3. The linearized analysis of the dynamics around a configuration
$\psi_{q_i} (x)$ identifies a most unstable  Bloch eigenmode $q_M(q_i)$
with eigenvalue or linear growth rate $\lambda_m(q_i)$. 
The amplitude $u$ of  this mode satisfies during the early stages of
decay a linear SDE
\begin{equation}
\dot u(t) = \lambda_m u(t) + \sqrt{\epsilon}\xi (t),
\end{equation}
where $\xi(t)$ is a white Gaussian random process which 
results from the projection of $\xi(x,t)$ onto the most unstable Bloch eigenmode.
Equation (5.14), that is the linear regime, is valid up to a time $T$ at which the 
amplitude $u$ becomes large enough, so that nonlinearities
become important. This time is the  MFPT calculated in Sect. 3:
\begin{equation}
T \sim {1 \over 2\lambda_m} \ln ({u_{0}^2 \lambda_m \over \epsilon}),
\end{equation}
where $u_0$ is a reference value for the amplitude $u$.
We should now remember that the calculation of $T$ was based on the replacement of a
stochastic  process $h(t)$, that played the role of an effective initial condition, by a
stationary random variable $h(\infty)$. This replacement was valid after a time $t_{m} \sim
\lambda_m^{-1}$. The time scale given by $t_{m}$ is the fluctuation dominated regime in which
$u(t)$ stays fluctuating around $u=0$. The existence of a linear growth regime invoked in our
earlier discussion requires a separation of time scales between $t_{m}$ and $T$. If this
is not satisfied we enter  directly from a fluctuation dominated regime into
the nonlinear regime, and the argument of fastest growing mode does not make further sense.
The separation of time scales  $t_m \ll T$ breaks down for
\begin{equation}
\lambda_m \approx {\epsilon \over u_0^2}.
\end{equation}
Since it is possible to show that $\lambda_m \sim (q_i - q_E)^2$, eq. (5.16) determines the range
of initial wavenumbers $q_i$ close to $q_E$ for which a linear regime does not exist in the 
decay process.
This is the regime dominated by fluctuations for which the definition $q_E$ of the Eckhaus boundary 
based on a linear deterministic analysis is not meaningful. A clear example of the consequences
of this statement is provided by the numerical evidence that, for small noise amplitudes,
periodic solutions with $q_i<q_E$ (within the linear deterministic Eckhaus stable band) are not
maintained in time and decay to configurations with a global $q_f$ closer to 
$q=1$. In
addition, for initial configurations within the fluctuation dominated regime in the Eckhaus
unstable range one finds that the result $q_f=q_M(q_i)$ does not longer hold.

Having argued that the Eckhaus boundary is not properly defined in the presence of
fluctuations, we still can characterize a fluctuation shifted fuzzy boundary $\tilde{q}_{E}$
separating unstable from metastable states. Such boundary is defined with respect to a given
long time scale of observation of the system. Within that scale of observation, it is defined as
the value of $q_i$ for which a relaxation time diverges. Of course this is based on
extrapolation of data for a range $\tilde{q}_{E}<q_i$ in which decay is observed. Numerical
results indicate  that $q_f$ becomes a linear function of $q_i$ so that $\tilde{q}_{E}$ can be
identified by extrapolation to $q_i=q_f$. We further note that the nonlinear relaxation process
in this fluctuation dominated regime exhibits a form of dynamical scaling
\cite{HernandezGarcia93}.

We close this section with a final remark on pattern selection on 
nonrelaxational systems.
There are pattern forming models that, as described in Sect. 4, do not follow a 
relaxational
dynamics, but still can have a Lyapunov functional. An example recently 
considered is the
Greenside-Cross equation in d=2 \cite{Kurtze96}. Our general discussion on pattern
selection dynamics makes clear that the configuration minimizing a Lyapunov 
functional
might not be the selected configuration in many cases. This has been shown here 
in the case 
of a purely relaxational dynamics and, a fortiori, will also be true for
nonrelaxational dynamics.

\subsection{Noise amplification in convective instabilities: Noise sustained structures}
Generally speaking, we call a state of a system absolutely unstable if a perturbation
localized around a spatial position grows in time at that spatial position. A state is
absolutely stable  if the perturbation decays in time. An intermediate situation occurs at a
convectively unstable state: A local perturbation grows in time but traveling in the system, so
that the perturbation decays in time at fixed space positions but, being convected away from
the place where it originated, grows in time in the frame of reference that moves with the
perturbation. The role of noise in a convectively unstable situation is very important
\cite{Deissler} because, as it happened
 in the decay
of an unstable state, there is a mechanism of noise amplification: If what grows at the instability
is some sort of spatial structure (wave or pattern), this structure will not be seen at steady
state in the absence of noise because any perturbation is convected away and leaves the system.
However, if noise is continuously present in the system, it will be spatially amplified and 
a persistent structure
sustained by noise will be observed. Important questions in this context are the determination
of conditions for the existence of a convectively unstable regime, the characterization of
a noise sustained structure and the characterization of the threshold for the 
transition between a noise sustained
structure and a deterministically sustained structure. 

As an example of this situation we consider the complex Ginzburg Landau Equation (CGLE) which is the 
amplitude equation for the complex amplitude $A$ of the dominating mode at a Hopf bifurcation in a 
spatially extended system \cite{HohCross}. Using a different notation than in
Sect. 4,
\begin{equation}
\partial_t A(x,t) - v \partial_x A(x,t) =
\mu A(x,t) + (1+i \alpha)\partial_x^2 A(x,t)  
- (1+i\beta)\vert A(x,t)\vert^2 A(x,t) + \sqrt{\varepsilon}\xi(x,t) 
\end{equation}
For $v=0$ the state $A=0$ changes from absolutely stable to absolutely unstable when the
control parameter changes from $\mu <0$ to  $\mu >0$. Linearly stable traveling waves exist as solutions
of the CGLE for $\mu>0$ and $1 +\alpha \beta>0$. The convective term $v \partial_x A(x,t)$ 
becomes important when
boundary conditions are such that it can not be eliminated by a change of frame of reference. 
To understand the role of this term we linearize  (5.17) around $A=0$. The complex dispersion
relation $\omega$ for a disturbance of wavenumber $K$, that thus
behaves as $e^{\omega t + K x}$, becomes:
 
\begin{equation}
\omega=\mu +Kv+(1+i\alpha)K^2 \quad , \quad K=k+iq \quad,
\end{equation}

\noindent and the growth rate of such a perturbation is
given by $Re \omega (K)$. Using the method of steepest descent, the
long-time behavior of the system along a ray defined by fixed $x/t$, i.e. in
a frame moving with a velocity $v_0 = x/t$, is governed by the saddle point
defined by :

\begin{equation}\label{reim}
Re\left(\frac{d\omega}{dK}\right)= v_0 \quad , \quad
Im\left(\frac{d\omega}{dK}\right) = 0.
\end{equation}

\noindent Since absolute instability occurs when perturbations grow at fixed locations, one
has to consider the growth rate of modes evolving with zero group velocity,
which are defined by:

\begin{equation} \label{reim2}
Re\left(\frac{d\omega}{dK}\right)=Im
\left(\frac{d\omega}{dK}\right)=0
\end{equation}

\noindent These conditions define the following wave number

\begin{equation}
\label{wavevekt} q=-\alpha k \quad , \quad k=-\frac{v}{2(1+\alpha^2)}\quad .
\end{equation}

\noindent The real part of
$\omega$, which determines the growth rate $\lambda$ of these modes is then:
\begin{equation} 
\lambda=Re (\omega)=\mu-\frac{v^2}{4(1+\alpha^2)}\quad.
\end{equation}
Therefore, the uniform reference state $(A = 0)$ is absolutely
unstable if $\lambda>0$. This condition determines a critical line in the
parameter space which can be expressed for the group velocity $v$ or the
control parameter $\mu$ as 

\begin{equation} v_c = 2\sqrt{\mu(1+\alpha^2)}
\qquad
 \mbox{or} \qquad
\mu_c =\frac{v^2}{4(1+\alpha^2)}\quad .
\end{equation}

\noindent Hence, for $0 < \mu < \mu_c$, the uniform reference state is
convectively unstable, and wave patterns are convected away in the
absence of noise. For $\mu> \mu_c$, wave patterns may grow and are
sustained by the dynamics, even in the absence of noise. 

This analysis of the convective instability in the CGLE has been used 
\cite{Ahlers94} to account for the corresponding experimental
situation in the case of a Taylor-Couette system with through flow, where 
the transition from  convective to absolute instability and
 noise sustained structures are observed. Noise sustained structures in the
convectively unstable regime $\mu<\mu_c$ can be characterized by the stochastic
trajectory of $A$ at a fixed space point. It is seen that the statistical
properties can be explained in terms of a stochastic phase diffusion for the phase
of $A$, while $|A|$ shows small fluctuations. This  identifies again a structure with 
no long range order
since it is being excited by noise. An equivalent quantitative measure of the phase
diffusion wandering induced by noise amplification is given by the width of 
the frequency power spectrum of $A$ at 
a fixed point. This spectrum is broad in the convectively unstable regime. The 
threshold at $\mu=\mu_c$ can be characterized by the way in which the 
width of the spectrum vanishes (divergence of
a correlation time) as one moves from the convectively unstable regime to the
absolute
unstable regime. In the latter regime the structure persists for vanishing noise and the
spectrum is essentially noise free.

\begin{figure}
\begin{center}
\def\epsfsize#1#2{0.60\textwidth}
\leavevmode
\epsffile{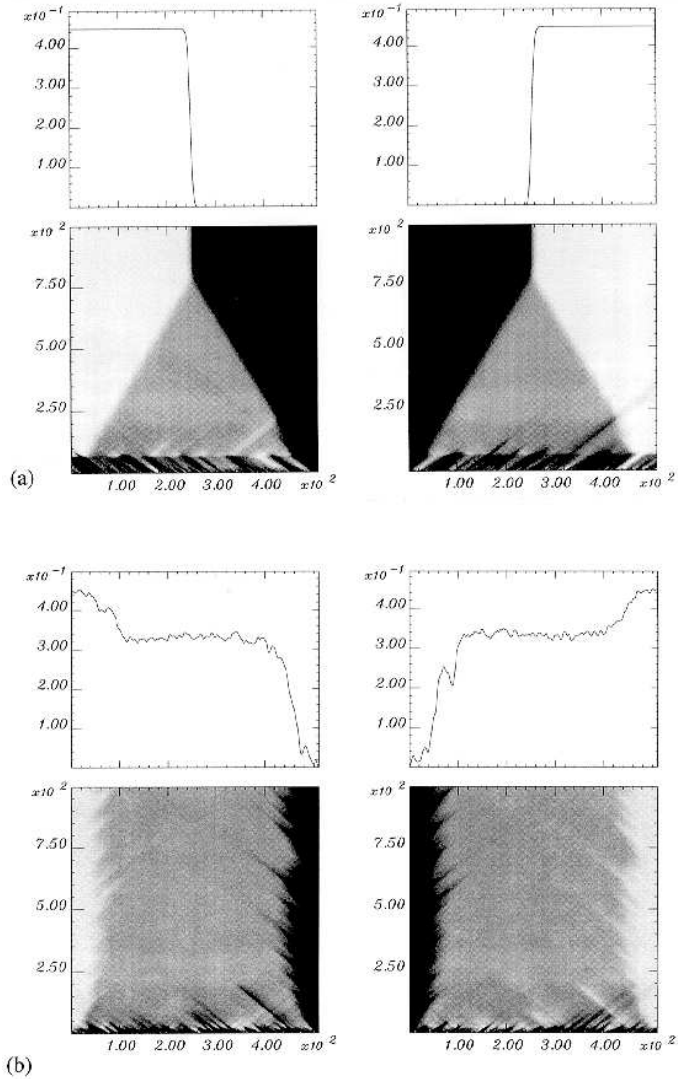}
\end{center}
\caption{Space (horizontal axis)-time (vertical axis) grey scale plot of the moduli of the amplitudes
$A(x,t)$ (left) and $B(x,t)$ (right) for $\gamma=0.8$, $\mu=0.408$, $v=1$, $\alpha=0.02$, $\beta=0.04$,
$\delta=0.05$. The upper diagrams show the spatial dependence of $|A|$
and $|B|$ at the end of the space-time plots. (a) Deterministic case ($\epsilon=0$):
the disturbances of the initial random pattern create initially a standing wave
pattern which is replaced, due to front propagation, by a symmetric traveling 
wave pattern.
(b) Stochastic case ($\epsilon=10^{-4}$): the spatially distributed noise gives 
rise, in the
bulk of the system, to a noise sustained wave structure fluctuating 
around the deterministic 
value}
\label{marc}
\end{figure}
These ideas have been recently extended \cite{Neufeld96} to the case of 
two coupled CGLE which describe a system
undergoing a Hopf bifurcation at a finite wave number. 
The amplitudes $A$ and $B$ of the two emerging
counterpropagating waves satisfy
\begin{eqnarray}
\partial_t A(x,t) - v \partial_x A(x,t) & = &
\mu A(x,t) + (1+i \alpha)\partial_x^2 A(x,t) 
- (1+i\beta)\vert A(x,t)\vert^2 A(x,t) \nonumber \\
& - & (\gamma +  
i\delta)\vert B(x,t)\vert^2 A(x,t) + \sqrt{\varepsilon}\xi_A(x,t) \\
\partial_t B(x,t) + v\partial_x B(x,t) & = &
\mu B(x,t) + (1+i \alpha)\partial_x^2 B(x,t) 
- (1+i \beta)\vert B(x,t)\vert^2 B(x,t) \nonumber \\
& - &(\gamma+i\delta)\vert A(x,t)\vert^2 B(x,t) + \sqrt{\varepsilon}\xi_B(x,t).
\end{eqnarray}
For $\gamma>1$ there is strong coupling of the two waves, so that only one
of them survives,
and the situation is the same than for the single CGLE. For $\gamma<1$ a stability 
analysis analogous to the one discussed for the single CGLE shows that the Traveling Wave
solution  (TW), $A \not =0$, $B=0$, is convectively unstable for
$\mu_c<\mu<\frac{\mu_c}{1-\gamma}$, while it becomes absolutely unstable for
$\frac{\mu_c}{1-\gamma}<\mu$. Therefore, this deterministic analysis would imply that TW
solutions are observed while they are 
convectively unstable and  Standing Wave solutions (SW),
$A=B\not=0$, emerge when crossing the line 
\begin{equation}
\gamma_c =1- {\mu_c \over \mu} = 1- \frac{v^2}{ 4 \mu ( 1+ \alpha^2)},
\end{equation}
and getting into the region in which the TW solution is absolutely unstable. However, from 
a stochastic point of view one expects that for $\gamma>\gamma_c$ (TW convectively unstable)
noise will be amplified so that the two amplitudes $A$ and $B$ become nonzero and a noise
sustained SW emerges. Crossing the line $\gamma=\gamma_c$ would then lead from a noise sustained
SW to a deterministically sustained SW. Evidence for the noise sustained SW for $\gamma>\gamma_c$
is shown in Fig. (5.6), where the results of a numerical simulation with and without noise 
are shown. Initial conditions are random fluctuations around the state $A=B=0$
and the  
boundary conditions used in a system of size $L$ are
\begin{equation}
 A(L,t)=0\quad,\quad B(0,t)=0\quad, \quad \partial_x A(0,t) =0\quad,\quad  \partial_x B(L,t) = 0\quad .
\end{equation}

The transition at $\gamma=\gamma_c$ can be characterized in terms of the width of the
frequency power spectrum $|A(x,\omega)|^2$ of $A$ at a fixed space point 
(inverse correlation time) or by
the width of the time average of the power spectrum $|A(k,t)|^2$ (inverse correlation length).
The two power spectra are broad in the region of noise sustained SW. The onset of the deterministic
SW is identified by a divergence of the correlation time and a correlation length becoming of the order
of the system size (for the relatively small sizes explored). However the transition does not
occur at $\gamma=\gamma_c$ but at a slightly noise-shifted value \cite{Neufeld96}. This is just
another  example of instability point shifted and modified by noise as the one described above
for the Eckhaus instability and the shifting of the phase transition point discussed in Sect 6.


\section{Fluctuations, phase transitions and noise--induced transitions}
\setcounter{equation}{0}
\setcounter{figure}{0}
In many cases, 
as we have encountered in previous examples, noise has a disordering
effect. In this section, however, we consider a class of
problem in which noise acts
in a non trivial and unexpected way inducing a phase transition to an ordered
state in spatially distributed system \cite{VPT,VPT2,GOPSB}. 
We first review the transitions induced by noise in the zero--dimensional
case and then turn to the extended systems.

\subsection{Noise--induced transitions}
\begin{figure}[t]
\begin{center}
\def\epsfsize#1#2{0.82\textwidth}
\leavevmode
\epsffile{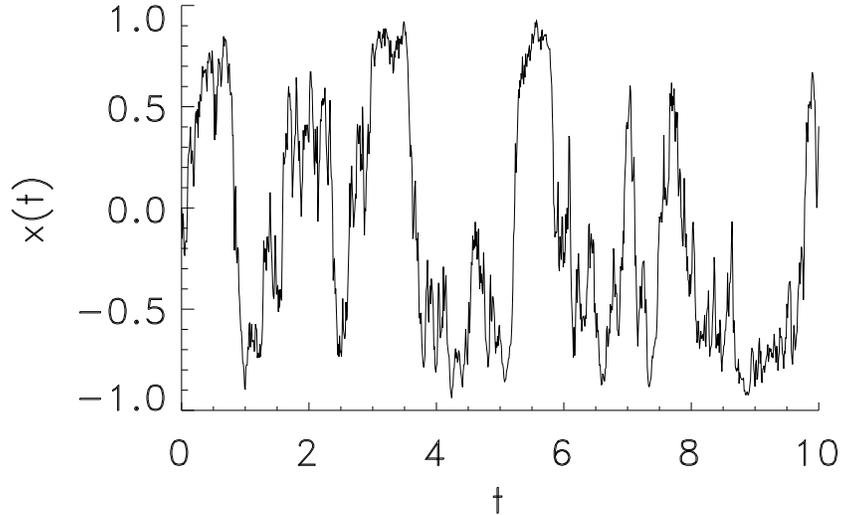}
\end{center}
\caption{\label{fig:6-1}
Typical trajectory coming from a numerical solution of the SDE
(\ref{eq6:2a}). We see that, although there are two preferred values, $\pm x_0$,
for the variable $x$ (i.e. the stationary distribution is bimodal), there
are many jumps between these two preferred values. If one were
to interpret $+x_0$ and $-x_0$ as two ``phases" of the system, we would
conclude that noise is capable of restoring the symmetry between these
two phases. Values of the parameters are $\lambda=0.5$, $\sigma^2=2$. }
\end{figure}

Noise induced transitions have been known for some time now 
\cite{hor84}. Let us start with a
simple example: consider the following SDE with multiplicative noise 
(Stratonovich sense)
\be
\label{eq6:1}
\dot x = f(x) + g(x) \xi (t)
\ee
$\xi(t)$ is a white noise process with mean value zero and correlations:
\be
\langle \xi(t)\xi(t')\rangle = \sigma^2 \delta(t-t')
\ee
The parameter $\sigma$ is called the noise intensity and, although it could be
absorbed in the function $g(x)$, is included here to stress the fact that when
$\sigma=0$ the noise effectively disappears from the problem. By solving the
Fokker--Planck equation, one can check that the stationary distribution is
given by: 
\be 
P_{st}(x) = C \exp\left\{\int_0^x dy \frac{f(y)-{\sigma^2 \over 2}
g(y)g'(y)} {{\sigma^2 \over 2} g^2(y)}\right\} 
\ee
Consequently, the extrema $\bar x$ of the stationary distribution are given by:
\be
\label{eq6:2}
f(\bar x) - \frac{\sigma^2}{2} g(\bar x)g'(\bar x) =0
\ee
And, quite generally, they will be different from the fixed
points of the deterministic dynamics,
$\sigma=0$. It is possible, then, that new stationary states appear as a
consequence of the presence of the noise term. Let us consider a specific case:
\be 
\label{eq6:2a}
\dot x = -x + \lambda x (1-x^2) + (1-x^2) \xi(t)
\ee
for $|x| < 1$ and $\lambda < 1$. It can be put in the form:
\be
\dot x = - \frac{\partial V}{\partial x} + (1-x^2)\xi(t)
\ee
with 
\be
V(x) = \frac{1-\lambda}{2} x^2+ \frac{\lambda}{4} x^4
\ee
In the absence of noise, $\sigma=0$, the deterministic problem has a unique
fixed stable point at $x=0$. For small $\sigma$ the situation is
somewhat similar, the stationary probability distribution is peaked
around the value $x=0$. However, by increasing the noise intensity, namely for
$\sigma^2 > 2 \lambda$,  the stationary probability distribution changes shape
and becomes bimodal. This change from a unimodal to a bimodal distribution
obtained by increasing the noise intensity has been called a noise--induced 
transition \footnote{A different situation recently studied in \cite{nittira}
is that of noise induced transition in the limit of weak noise}.
Of course, this can not be identified as a true ``phase transition", because
during the stochastic evolution the system will jump many times  from one
maximum of the probability to the other, thus not breaking ergodicity, which is
a clear signature of a phase transition. We show in Fig.(\ref{fig:6-1}) how
noise is capable of restoring the symmetry and, on the average the
``order parameter", $m\equiv \langle x \rangle =0$.
One would think that, in similarity to what happens in other models of
statistical mechanics, it should be possible to obtain a phase transition by
coupling many of these systems. The coupling should favor the ordering of
neighboring variables to a common value. The system would, then, choose between
one of the two maxima displaying a macroscopic not vanishing order parameter.
However, it turns out that, although the noise is effectively capable of
inducing a phase transition, it does so in a different way.

\subsection{Noise--induced phase transitions}

Remember that for relaxational systems
noise has the effect of introducing fluctuations around the minima
of the potential. If the potential is bistable and 
the noise intensity becomes large enough, the system is brought out 
of the minima in a short time scale
and the order parameter vanishes. One can interpret these
results in terms of a shift in the transition. Let us be more specific:
consider the Ginzburg--Landau model \cite{amit,TC1}
(model $A$ in the taxonomy of
\cite{HH}, also called the $\varphi^4$ model). We consider the lattice
version: the variables $x_i$ follow the relaxational gradient
evolution equations:
\be
\label{eq6:a1}
\dot x_i = -\frac{\partial V(x)}{\partial x_i} + \xi_i(t)
\ee
The index $i$ runs over the $N=L^d$ sites of a $d$--dimensional
regular lattice (of lattice spacing $\Delta r=1$). 
The potential is:
\be
\label{eq6:a2}
V(x) = \sum_{i=1}^N \left[ -\frac{b}{2}x_i^2 + \frac{1}{4}x_i^4
+\frac{1}{2}|\vec \nabla x_i|^2 \right]
\ee
We use the simplest form for the lattice gradient $\vec \nabla x_i$. It 
is such that its contribution to the dynamical equations is the 
lattice--Laplacian:
\be
-\frac{\partial}{\partial x_i} \left[\sum_j|\vec \nabla x_j|^2\right] =
2\sum_{j \in n(i)} (x_j-x_i)
\ee
the sum over the index $j$ runs over the set $n(i)$ of the $2d$ lattice
sites which are nearest neighbors of site $i$.

As discussed in Sect. 4.2, if the noise term satisfies the 
fluctuation--dissipation relation:
\be
\label{eq6:a3}
\langle \xi_i(t) \xi_j(t') \rangle = 2 k_B T \delta_{ij}\delta(t-t')
\ee
the stationary distribution  of (\ref{eq6:a1}) is given by:
\be
\label{eq6:a4}
P_st(x) = Z^{-1}\exp[-V(x)/k_BT]
\ee
$Z$ being a normalization constant (the partition function).
The parameter $b$ also depends on temperature $b=b(T)$ and
is such that changes sign at $T=T_0$, being positive for $T < T_0$
and negative for $T > T_0$. If we neglect thermal fluctuations,
the stationary distribution becomes a sum of delta functions centered
around the minima of 
the potential $V(x)$. If $b<0$, the potential has a single minimum
and if $b>0$ the potential has two minima. In these minima
all the variables $x_i$ adopt a common value $m$. The order parameter is:
\be
\label{eq6:a5}
m=\left\{
\begin{array}{lll} 0 & b < 0 &, T > T_0 \\
\pm \sqrt{b} & b > 0 & , T < T_0 \end{array} \right.
\ee
This is nothing but a simple example of
Landau theory of phase--transitions \cite{landau}.
The two minima, corresponding to the $\pm$ signs in the order parameter,
are the two phases whose symmetry is broken by the time evolution.
In the absence of noise, there is no mechanism to jump from one minimum
to the other. 

\begin{figure}[t]
\begin{center} \def\epsfsize#1#2{0.82\textwidth}
\leavevmode
\epsffile{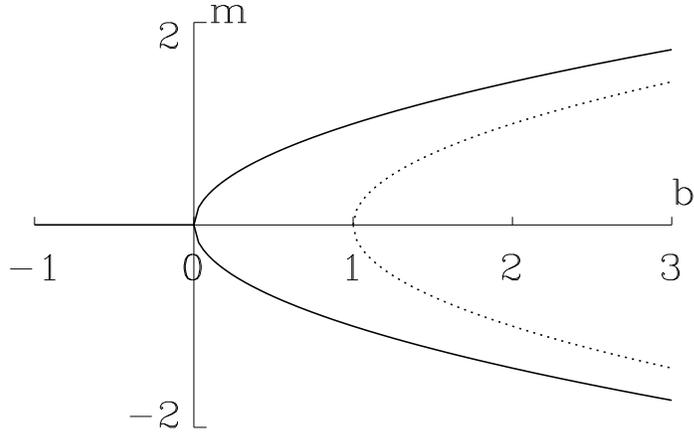}
\end{center}
\caption{\label{fig:6-1b}
Order parameter $m$ for the $\varphi^4$ model of phase transitions as 
a function of the parameter $b$. The continuous line is Landau's solution,
eq. (\ref{eq6:a5}); it has a critical point at $b=0$. The dotted line
is the (schematic) exact result (when including thermal fluctuations)
for spatial dimension $d > 1$. Notice
that there is a shift in the transition when including fluctuations towards
larger values for $b$, smaller temperatures.}
\end{figure}

If we now go beyond Landau's mean field approximation, we include thermal
fluctuations by explicitly considering the noise term. Without going
into details \cite{amit,TC1,cowley}, 
we can understand that fluctuations make it possible
for the system to jump from one minimum to the other, hence restoring
the symmetry. For this to happen, the thermal energy $k_BT$ must be of
the order of the energy barrier separating the two minima. Put in
another way: when thermal fluctuations are taken into account, one 
needs a deeper potential in order to keep the broken symmetry. Or
still in other words: when fluctuations are included, there is a shift in 
the location of the phase transition from $b=0$ to $b >0$, what implies
that the new critical temperature is smaller than the Landau mean--field
temperature, see Fig. (\ref{fig:6-1b}). We conclude that the
probability distribution for the order parameter changes from bimodal
to unimodal when increasing the fluctuations (noise intensity). In fact,
for this change to happen at temperature greater than zero, one needs
the spatial dimension to be strictly greater than $1$, $d > 1$
(lower critical dimension of the $\varphi^4$ model).
We have witnessed in the case of zero--dimensional (one variable)
models, how
fluctuations can actually go the other way and induce a bimodal distribution
when increasing the noise intensity. We called this a noise--induced 
transition. In order to have a real noise--induced
{\sl phase} transition, we guess
that we need to go a spatially extended system. The role of
multiplicative noise in spatially extended systems in the context of
phase transitions \cite{kra95,VPT,VPT2,GOPSB} and pattern--forming
instabilities \cite{sancho1,sancho2,sancho4} has been thoroughly studied.
We consider here the $d$--dimensional version of (\ref{eq6:1}) with
the inclusion of nearest neighbor coupling terms:
\be
\label{eq6:5}
\dot x_i = f(x_i) + \frac{D}{2d}\sum_{j\in n(i)} (x_j-x_i) +g(x_i)\xi_i(t)
\ee
The parameter $D>0$ measures the strength of the coupling between
neighboring sites. 
The noise variables $\xi_i(t)$ are Gaussian distributed of zero mean and
correlations given by:
\be
\langle \xi_i(t)\xi_j(t') \rangle = \sigma^2 \delta_{ij}\delta(t-t')
\ee
We skip now some of the mathematical details. Although it is possible to
write down the 
Fokker--Planck equation  for the joint probability distribution function
$P(x_1,\dots,x_N;t)$, corresponding to this set of
stochastic differential equations, it is not possible now, as it was in the
zero--dimensional case, to solve it in order to find the
stationary distribution $P_{st}(x_1,\dots,x_N)$. What one does, though, is to
obtain an exact equation for the one--site 
probability $P_{st}(x_i)$: 
\be
\label{eq6:3}
\frac{\partial}{\partial x_i} \left[-f(x_i)+
D[x_i-\langle x_{n(i)} | x_i\rangle]+\frac{\sigma^2}{2} g(x_i) \frac{\partial}
{\partial x_i}g(x_i)\right] P_{st}(x_i) =0
\ee
In this equation appears the conditional mean value $\langle x_{n(i)} |
x_i\rangle$: the mean value of a neighbor site of $i$ given that the
random variable $x_i$ takes a fixed value. In order to proceed, we make now a
mean--field type assumption. This approximation, close in spirit to Weiss
mean--field theory for ferromagnetism \cite{pathria}, 
consists in the assumption that the above
conditional mean value is equal to the mean value of $x_i$ itself:
\be
\label{eq6:8}
\langle
x_{n(i)} | x_i  \rangle = \langle x_i \rangle 
\ee
It is possible now to integrate out all the variables except a given
one, say $x_1$, from
equation (\ref{eq6:3}) to obtain a closed equation for $P_{st}(x_1)$ whose
solution is (we have dropped the subindex $1$ for the sake
of simplicity and because similar equations are satisfied for any $x_i$):  
\be
\label{eq6:9}
P_{st}(x) = Z^{-1} \exp \left[ \int^x dy \frac{f(y)-\frac{\sigma^2}{2} g(y)g'(y)
- D(y-\langle x \rangle)}{{\sigma^2 \over 2}g^2(y)} \right] 
\ee 
In this equation, the unknown value
$\langle x \rangle$is then obtained by the consistency relation: 
\be
\label{eq6:10} 
\langle x \rangle = \int dx P_{st}(x)
\ee
The solutions of equations (\ref{eq6:9}) and (\ref{eq6:10}) are to 
be considered the order parameter, $m=\langle x \rangle$. 
These  equations can be
solved numerically, but a first clue about the possible solutions can be
obtained in the limit $D \to \infty$ by means of a saddle--point integral, which
yields:  
\be
\label{eq6:20}
f(\langle x \rangle) + \frac{\sigma^2}{2} g(\langle x \rangle) g'(\langle x
\rangle) = 0
\ee
If by varying the noise intensity, $\sigma$, the order
parameter changes from $m=0$ to a non--zero value, we interpret this as a
phase transition. In the example given below, we will show that the main
features of a true phase transition hold, namely: 
the breaking of ergodicity, the
divergence of correlations, the validity of scaling relations, 
etc. \cite{stanley}. Finally,
note the difference in sign of this equation with equation (\ref{eq6:2})  giving
the maxima of the probability distribution in the case of
noise--induced--transitions. This means that, at least in the limit of very
strong coupling in which equation (\ref{eq6:20}) holds, those systems displaying
a noise--induced transition will not show a noise--induced phase transition
and viceversa.

\begin{figure}[h]
\begin{center} \def\epsfsize#1#2{0.62\textwidth}
\leavevmode
\epsffile{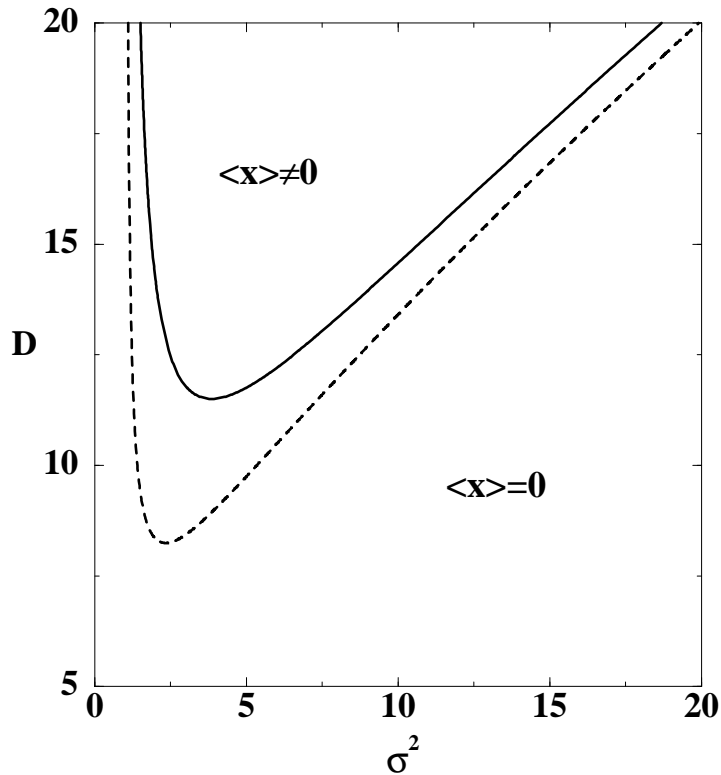}
\end{center}
\caption{\label{fig:6-2}
Numerical solution of equations (\ref{eq6:9}) and (\ref{eq6:10}) for the model
(\ref{eq6:30}) showing the regions of
order and disorder (solid line). The dotted line is the result of a higher
order approximation (not explained in the main text)
which goes beyond the approximation (\ref{eq6:8}) by
including correlations amongst nearest neighbors variables.} 
\end{figure}

A simple model is obtained by choosing \cite{VPT,VPT2}: 
\be
\label{eq6:30}
\begin{array}{rcl}
f(x) & = & -x (1+x^2)^2 = -\frac{\partial V}{\partial x}, \hspace{1.0cm} V(x) =
\frac{1}{3} (1+x^2)^3 \\
g(x) & = & 1+x^2
\end{array}
\ee
Let us summarize some properties of this model:\\
(i) For the deterministic
version (no noise, $\sigma=0$) it is a relaxational gradient system with
a potential $V(x)$ which displays a single minimum. Therefore, the
stationary solution  is $x_i=0$ for all $i$.\\ 
(ii) For the
zero-dimensional version, in which there is no coupling between the different
lattice points, $D=0$, each variable becomes independent of the others. By
looking at the solutions of (\ref{eq6:2}), one can check that this system does
not present a noise--induced transition. Therefore, the stationary probability
distribution $P_{st}(x_i)$ has a maximum at $x_i=0$ again for all values of $i$,
and is symmetric around $x=0$, hence, $m= \langle x_i \rangle =0$.\\
(iii) When both coupling and noise
terms are present, $D\ne 0$, $\sigma \ne 0$, the stationary distribution has
maxima at non--vanishing values of the variable. Due to the symmetry $x \to
-x$, these maxima are at values $\pm x_0$, for some $x_0$. In fact, in the
infinite coupling limit, equation (\ref{eq6:20}) yields for the (common) average
value:
\be  
m=\langle x \rangle = \pm \sqrt{\sigma^2-1}
\ee
This existence of a non--vanishing order parameter is also maintained at
finite values for the coupling $D$, although then the consistency relation
(\ref{eq6:10}) must be solved numerically. In Fig. (\ref{fig:6-2}) 
we show the regions in
parameter space for which this mean--field approach predicts a non--zero
value of the order parameter. When the only solution is $\langle x \rangle =0$
we talk about a disordered phase, whereas when non--null solutions $\langle x
\rangle \ne 0$ appear, we say that the system is in an ordered phase. Notice
that, according to Fig. (\ref{fig:6-2}), an ordered phase can appear only for
a sufficiently large value of the coupling parameter $D$ and within a range of
values of the noise intensity $\sigma$, i.e. the transition is reentrant
with respect to the noise intensity. 

\begin{figure}[h]
\begin{center} 
\def\epsfsize#1#2{0.62\textwidth} 
\leavevmode
\epsffile{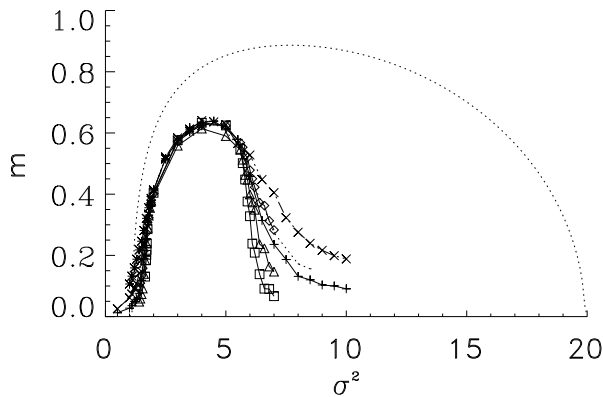} 
\end{center} 
\caption{\label{fig:6-3} 
Phase diagram of the system defined by equations (\ref{eq6:5}) and
(\ref{eq6:30}) for spatial dimension $d=2$. The dotted line is the result of the
mean field approximation (independent of dimension). The symbols are the results
of numerical simulations for different system sizes (from $L=16$, crosses, to
$L=128$, squares). In this case the order parameter is defined as $m \equiv
\langle | N^{-1}\sum_{i=1}^N x_i | \rangle$. 
 } 
\end{figure}
This prediction of the existence of a phase transition in the model of
equations (\ref{eq6:30}) has been fully  confirmed by numerical simulations of
the coupled set of Langevin equations (\ref{eq6:5}) in the two--dimensional,
$d=2$, case. The numerical solution has used appropriate versions of the
Milshtein and Heun algorithms, see Sect. (2.4.4). 
We now describe some observed features of the
transition as obtained from the computer simulations. 

(i) First of all, there is the breaking of ergodicity. Due to the above
mentioned symmetry $x \to -x$ and depending on initial conditions and white
noise realizations, some realizations will go to positive values of the order
parameter and some other to negative values. However, and this is the key
point, the jumps between positive and negative values do not occur very often,
and the frequency of the jumps strongly decreases with system size $N$,
suggesting that in the limit $N \to \infty$ noise is not capable of restoring
the symmetry.  
\begin{figure}[t]
\begin{center}
\def\epsfsize#1#2{0.72\textwidth}
\leavevmode
\epsffile{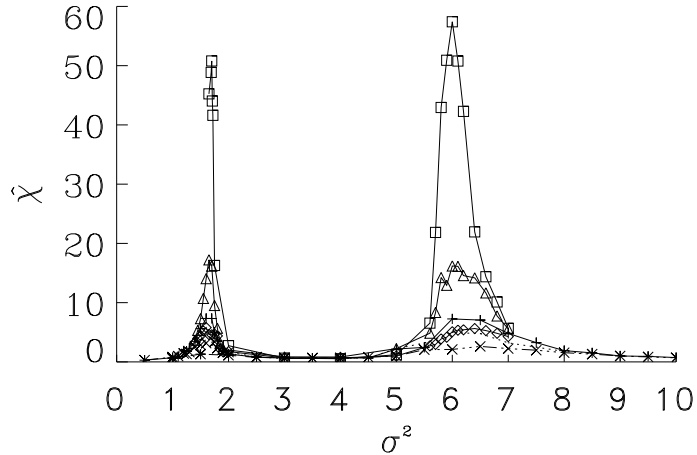}
\end{center}
\caption{\label{fig:6-4}
Susceptibility $\hat \chi$ plotted as a function of the noise
intensity $\sigma ^2$. The symbols are the results of numerical simulations for
different system sizes (same symbols meaning as in Fig. (\ref{fig:6-3}).
Notice the sharp increase of the susceptibility as the system size increases,
as usual in second--order phase transitions.} 
\end{figure}

(ii) The second point is the existence of the transition. This is shown in
Fig. (\ref{fig:6-3}) where we plot the order parameter as a function of
noise intensity $\sigma$ for a value $D=20$ of the coupling parameter.
According to Fig. (\ref{fig:6-2}) there should be a non--zero value for the
order parameter for $1.1 < \sigma^2 < 19.9$, approximately. Although the
reentrant transition occurs at a value $\sigma^2 \approx 6.7$, lower than the
predicted in mean field, $\sigma^2 \approx 19.9$, it is clear the increase of
the order parameter from $m=0$ to a non--zero value and the further decrease
towards $0$ through what appears to be two genuine second--order phase
transitions. Of course, calling this a second--order phase transition is an
abuse of language, since there is no free energy whose second derivatives show
some degree of non--analiticity. Still we keep this name because, as we will see
in the next paragraphs, many of the characteristics associated to a
thermodynamic second--order phase transition are also present in this model.

(iii) In a second--order phase transition, the susceptibility diverges. Here
we have defined a ``susceptibility", $\hat \chi$, 
as the following measure of the fluctuations of the order parameter:
\be
\label{eq6:40}
\hat \chi \equiv \frac{\langle m^2\rangle - \langle m \rangle^2}{N \sigma^2}
\ee
(the presence of the $\sigma^2$ term is reminiscent of the statistical
mechanics relation 
$\chi \equiv [\langle m^2\rangle - \langle m \rangle^2]/(N k_BT)$ since
in our model the equivalent of temperature $T$ is the noise intensity
$\sigma^2$). In Fig. (\ref{fig:6-4}) we can see how the susceptibility
develops maxima at the two transitions, maxima that increase with
increasing system size, suggesting again a true non-analytical
behavior (divergence) in the thermodynamic limit. It is possible to analyze
these results in terms of finite size scaling relations, as shown later.

\begin{figure}[t]
\begin{center}
\def\epsfsize#1#2{0.72\textwidth}
\leavevmode
\epsffile{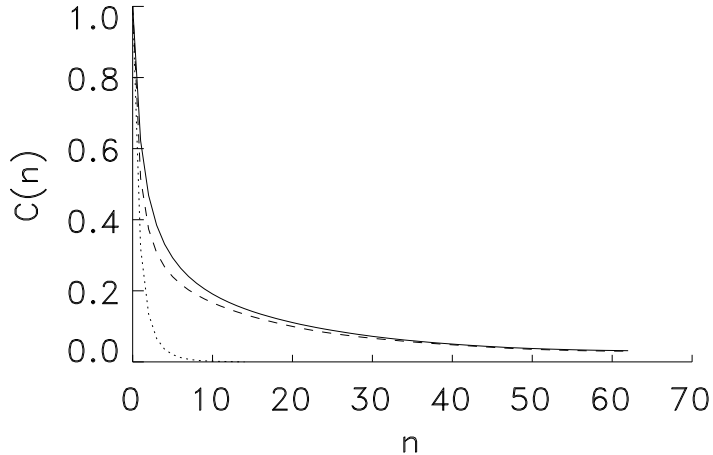}
\end{center}
\caption{\label{fig:6-5}
Correlation function $C(n) \equiv \langle (x_i-\langle x_i
\rangle ) (x_{i+n}-\langle x_{i+n}\rangle ) \rangle$ for $D=20$ and three
different values of the noise intensity $\sigma^2=1.65$ (solid line),
$\sigma^2=4$ (dotted line), $\sigma^2=6.7$ (dashed line). Note that at the two
critical points, the correlation function decays very slowly.} 
\end{figure}

(iv) Another feature of a second--order phase transition is that of the
long--range correlations. This is shown in Fig. (\ref{fig:6-5}) in which we
plot the spatial correlation function $C(n) \equiv \langle (x_i-\langle x_i
\rangle ) (x_{i+n}-\langle x_{i+n}\rangle ) \rangle$. Far from the two
transition points, the correlation function decays very fast, but near both
critical points, correlations decay very slowly. The same phenomenon has been
observed for the correlation times (critical slowing down).

\begin{figure}[t]
\begin{center}
\def\epsfsize#1#2{0.82\textwidth}
\leavevmode
\epsffile{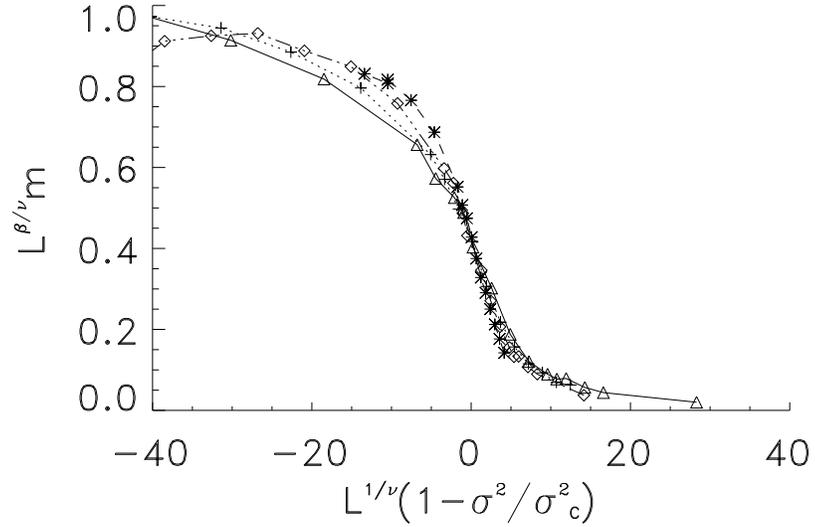}
\end{center}
\caption{\label{fig:6-6}
Plot of $L^{\beta/\nu}m(\sigma,L)$ versus
$L^{1/\nu}(1-\sigma^2/\sigma_c^2)$ with $\sigma_c^2=1.65$ 
for different system sizes (same symbols
than in Fig. (\ref{fig:6-2})), to show that the finite--size scaling
relation (\ref{eq6:45}) is well satisfied.} 
\end{figure}

(v) Finally, we show now that the transition also satisfies the usual scaling
relations valid in second order transitions. We have focused just in
finite--size--scaling relations \cite{cardy}
(so useful for analysing the results of
computer simulations \cite{fs2,fs3}). Although these finite--size scaling relations hold very
generally for the model introduced in this section, we present results for the
order parameter. According to finite size scaling theory, the order parameter
should behave close enough to the critical points as:
\be
\label{eq6:45}
m(\epsilon,L) = L^{-\beta/\nu} m(\epsilon L^{1/\nu})
\ee
where $\epsilon = 1-\sigma^2/\sigma_c^2$ 
(for thermal phase transitions, this
is defined as $\epsilon = 1-T/T_c$, $T$ being the system temperature). 
The meaning of $\beta$ and $\nu$ is that of the usual scaling exponents. In
Fig. (\ref{fig:6-6}) we plot $L^{\beta/\nu}m(\sigma,L)$ versus 
$L^{1/\nu}(1-\sigma^2/\sigma_c^2)$ to show that curves corresponding to
different values of $L$ and $\sigma$ fall onto a master curve, so confirming
the validity of scaling. In this curve we have used the values of the
$2$--$d$ Ising model exponents, $\nu=1$, $\beta=1/8$. 
Our data, then, suggests that this transition belongs to the $2$--$d$ Ising
model universality class, although more precise data are needed in order
to clarify this important point.

As a conclusion, we have shown in this section how the noise can have a
different effect that it is usually accepted for
Statistical Mechanics systems: it can induce long--range order
in a spatially extended system. Recent extensions of this work 
have considered the effect of colored noise \cite{wio}. This is motivated
because one expects that the kind of fluctuations leading to multiplicative
noise will show a finite correlation time. The role of colored noise in 
the $\varphi^4$ model (with additive noise) has been studied in
\cite{sancho3,sancho5} showing that a non--equilibrium phase transition
can be induced by varying the correlation time $\tau$ of the noise.
For zero--dimensional systems, a reentrant transition has been found
in a noise--induced transition as a consequence of the color \cite{castro}.
When multiplicative colored noise is included in an extended system, several
new effects can appear. One of the most counter--intuitive
ones being that of the disappearance of the order when increasing the
coupling between nearest neighbor fields \cite{wio}. It is hoped that 
these recent theoretical developments can boost experimental work
in the search for those predicted phenomena.

\espacio
\espacio
ACKNOWLEDGEMENT: We acknowledge partial financial support from DGICYT (Spain)
projects PB94-1167 and PB94-1172 and from the European Union TMR project 
QSTRUCT (FMRX-CT96-0077). Our knowledge on this subject has been built through research
work in collaboration with many colleagues and collaborators
reported in the references below.


\end{document}